\PassOptionsToPackage{sort&compress}{natbib} 
  \documentclass[3p]{elsarticle}
  \journal{Int.~J.~Mech.~Sci.}
  \usepackage[utf8]{inputenc}
  \usepackage[dvipsnames]{xcolor}
  \usepackage{graphicx}
  \usepackage{amsmath,amssymb}
  \usepackage{hyperref}
  \usepackage{pdflscape} 

  \usepackage{graphicx}
  \usepackage{siunitx}
  \usepackage{subcaption}
  \usepackage{caption} 
  \usepackage{hhline}
  \usepackage{longtable}
  \usepackage{xltabular} 
  \usepackage{booktabs}
  \usepackage{array}
  \usepackage{pifont} 
  \usepackage{tikz}

\newcommand{\cmark}{\ding{51}} 
\newcommand{\figref}{Figure~\ref}
\newcommand{\pdv}[2]{\frac{\partial #1}{\partial #2}}

\usepackage[normalem]{ulem}

\newcommand{\thetitle}{Careful finite element simulations of cold rolling with accurate through-thickness resolution and prediction of residual stress}
\newcommand{\thekeywords}{finite element \sep plasticity \sep elasticity \sep cold rolling \sep residual stress.}

\begin{document}
\begin{frontmatter}

  \title{\textbf{\thetitle}}

\author[macsi]{Francis~Flanagan}\ead{Francis.Flanagan@ul.ie}
\author[csis,lero]{Alison~N.~O'Connor}\ead{Alison.OConnor@ul.ie}
\author[wmi]{Mozhdeh~Erfanian}\ead{Mozhdeh.Erfanian@warwick.ac.uk}
\author[cued,ted]{Omer~Music}\ead{om236@cantab.ac.uk}
\author[wmi,wmg]{Edward~J.~Brambley\corref{cor1}}\ead{E.J.Brambley@warwick.ac.uk}\cortext[cor1]{Corresponding Authors}
\author[macsi]{Doireann~O'Kiely\corref{cor1}}\ead{Doireann.OKiely@ul.ie}

 \affiliation[macsi]{organization={MACSI, Department of Mathematics \& Statistics},
             addressline={University of Limerick},
             city={Limerick},
             postcode={V94 T9PX},
             country={Ireland}}

  \affiliation[csis]{organization={Computer Science and Information Systems (CSIS)},
             addressline={University of Limerick},
             city={Limerick},
             postcode={V94 T9PX},
             country={Ireland}}

  \affiliation[lero]{organization={Lero, The Science Foundation Ireland Centre for Software Research},
             addressline={University of Limerick},
             city={Limerick},
             postcode={V94~T9PX},
             country={Ireland}}

\affiliation[wmi]{organization={Mathematics Institute},
             addressline={University of Warwick},
             city={Coventry},
             postcode={CV4 7AL},
             country={UK}}
             
\affiliation[cued]{organization={Department of Engineering},
             addressline={University of Cambridge},
             city={Cambridge},
             postcode={CB2 1PZ},
             country={UK}}
             
\affiliation[ted]{organization={Mechanical Engineering Department},
             addressline={TED University},
             city={Ankara},
             country={Turkey}}

  \affiliation[wmg]{organization={WMG},
             addressline={University of Warwick},
             city={Coventry},
             postcode={CV4 7AL},
             country={UK}}


\begin{abstract}
Metal rolling is a widespread and well-studied process, and many finite-element (FE) rolling simulations can be found in the scientific literature. 
However, these FE simulations are typically limited in their resolution of through-thickness variations.
In this paper, we carefully assess the accuracy of a number of FE approaches, and find that at least 60 elements through-thickness are needed to properly resolve through-thickness variation; this is significantly more than is used elsewhere in the metal rolling literature.
In doing so, we reveal an oscillatory stress pattern, which is not usually observed in simulations but which we can validate by comparison with recent analytical work, and which is completely deterministic, not arising from numerical noise or error.
We show that these oscillations contribute to the formation of residual stress and may help predict curvature in asymmetric rolled sheets, a phenomenon which is currently not well understood.
Accurate through-thickness variation of stress and strain would also have implications for modelling microstructure evolution, damage, and surface finish.

\end{abstract}


\end{frontmatter}

\section{Introduction}
\label{intro}

\begin{figure}%
\centering%
\includegraphics[width=\linewidth]{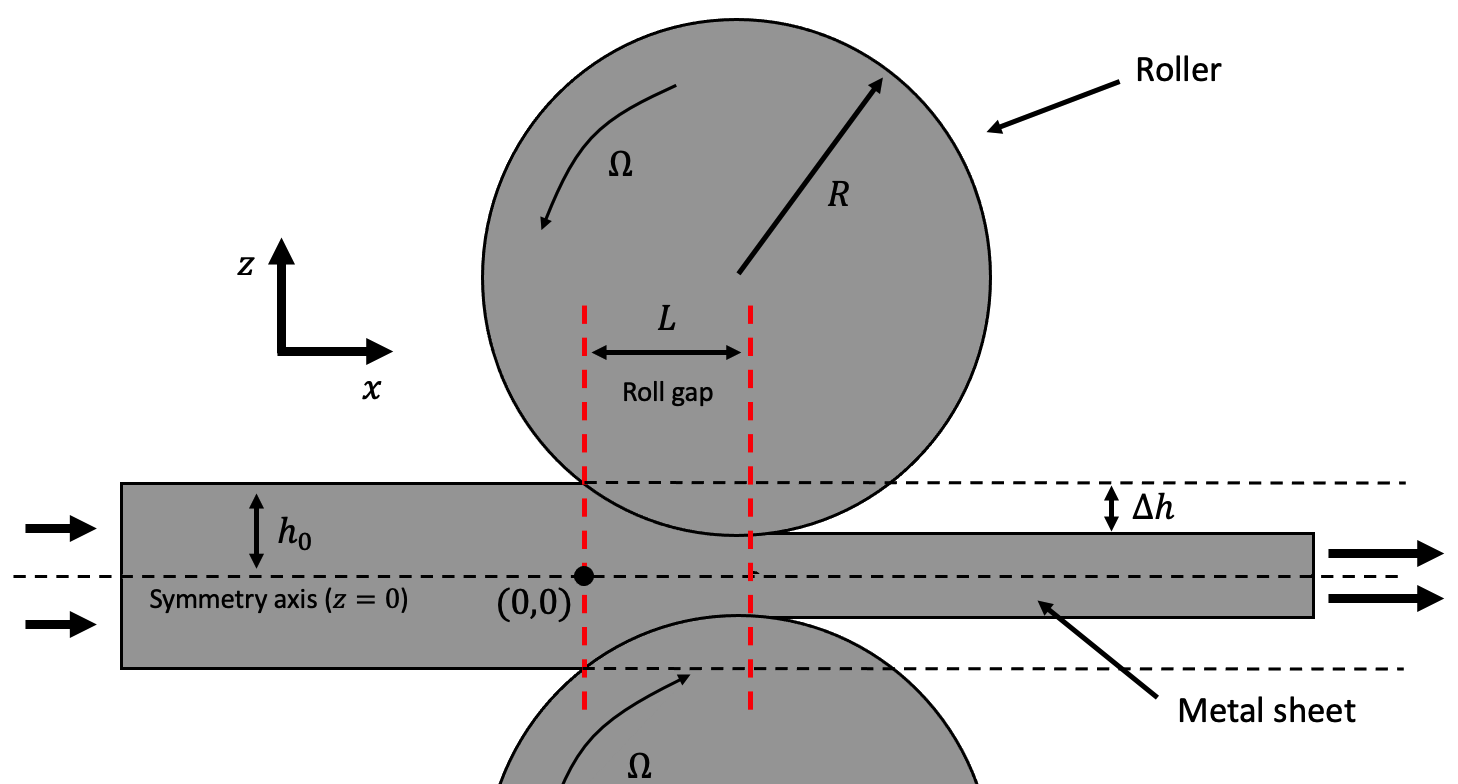}%
\caption{Schematic of the metal rolling process. The sheet moves from left to right and its thickness is reduced by the roller. Symmetry about the centre of the sheet ($z=0$) is assumed, so only the top roller and top half of the sheet are modelled in this work.
}%
\label{fig:rolling_pic}%
\end{figure}

In cold rolling, a sheet of metal, whose temperature is below the material recrystallisation temperature, is gradually fed between two rotating rollers.
The rollers are held at a fixed separation that is less than the initial thickness of the sheet and act to permanently deform the sheet so that its thickness is reduced as illustrated in \figref{fig:rolling_pic}.
Mathematical models of the cold-rolling process have been used for many years to make predictions, optimise system design and tighten process specification~\citep{orowan1943calculation,horton2017yield,wehr2020online}. 
Waste reduction is also an important motivator in this context as the production of steel and aluminium grows year-on-year and is a major contributor to global energy consumption and CO$_2$ emissions~\citep{milford2011assessing,allwood2016closed,WORLDSTEELASSOCIATION2024}.

In this paper we re-visit the task of modelling cold rolling using finite element (FE) analysis, with a particular focus on through-thickness variations.
We show that rolled sheets contain a rich and detailed stress pattern that varies repeatedly across the sheet thickness and that has not been properly resolved previously.
These oscillations do not arise from noise, vibrations or uncertainty, but rather are completely deterministic.
From a modelling point of view, we assess the accuracy of different FE approaches for modelling these through-thickness variations and recommend an approach that can fully resolve stress patterns.
From a physics and process engineering point of view, we fully resolve and validate these patterns and outline their implications for downstream effects like residual stress and curvature.

\begin{figure*}[ht]%
\centering%
\begin{subfigure}[b]{0.48\linewidth}%
   \centering%
   \begin{tikzpicture}
   \node at (0,0) {\includegraphics[width=\linewidth]{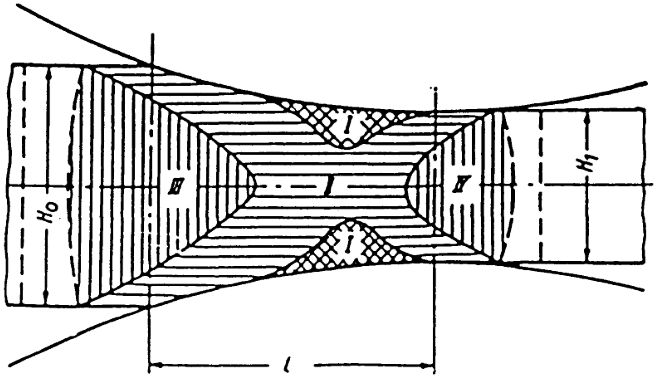}};
   \node[fill=white,font=\small,inner sep=2pt] at (0,0) {II};
   \node[fill=white,font=\small,inner sep=2pt] at (0.171,0.7) {I};
   \node[fill=white,font=\small,inner sep=2pt] at (0.17,-0.7) {I};
   \node[fill=white,font=\small,inner sep=2pt] at (-1.72,0) {III};
   \node[fill=white,font=\small,inner sep=2pt] at (+1.6,0) {IV};
   \end{tikzpicture}%
   \caption{}%
   \label{fig:hartley}%
\end{subfigure}%
\begin{subfigure}[b]{0.48\linewidth}%
        \centering%
        \includegraphics[width=\linewidth]{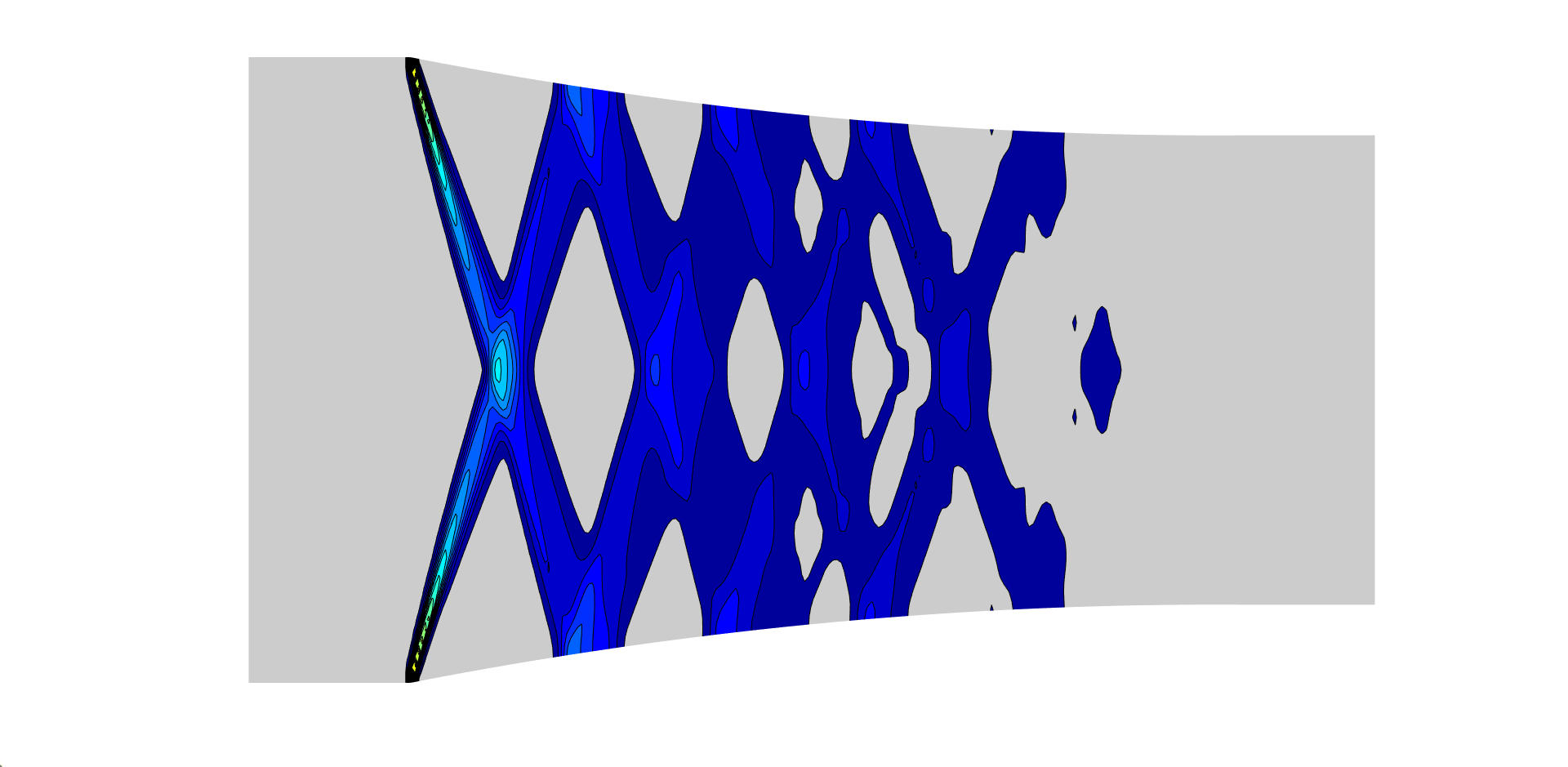}%
        \caption{}%
        \label{fig:peeqintro}%
\end{subfigure}\par%
\caption{Deformation zones in a metal sheet during rolling: (a) deformation zones reported by~\cite{hartley1989experimental} using data from~\cite{tarnovskii2013deformation}.  Region I has ``restricted deformation'', II has thinning, and III and IV have thickening.  (b) PEEQ rate predicted here. Grey zones are not flowing, even if they are at yield, while darker zones are shearing.}%
\label{fig:intro2}%
\end{figure*}
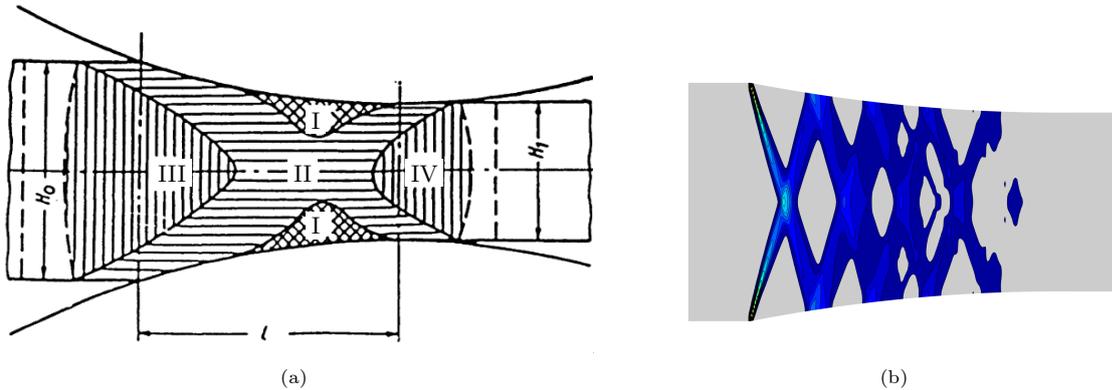

Early mathematical descriptions of cold rolling used a ``slab'' approach pioneered by~\citet{orowan1943calculation}.  
Slab models assume that the sheet can be thought of as a series of vertical slabs which remain vertical throughout the rolling process, and that stresses are uniform across the sheet thickness.  
While this simple mathematical description can yield reasonably accurate predictions of the roll force and torque required for a given thickness reduction, they automatically disregard any through-thickness effects~\citep{minton2016asymptotic}.
Experimental observations of deformation led to pictures like the one shown in \figref{fig:hartley}, with zones of thinning, thickening and restricted deformation.
Through-thickness variations in stress and strain can in principle be described with more advanced modelling techniques such as FE  analysis~\citep{thompson1982inclusion,mori1982simulation,yarita1985stress,lau1989explicit,yoshii1991analysis,richelsen1997elastic,ghosh2004computational,gudur2008neural,lenard2013primer, cawthorn2014comparison, minton2017mathematical}.  
However, we show that although the mesh resolution typically used in the scientific literature is high enough for most purposes, it is not high enough to properly resolve the oscillatory stress pattern of interest here. Using the highly careful FE model developed here, we will show below that the deforming and non-deforming zones inside the roll gap can exhibit a much more detailed pattern than previously reported (see, for example, \figref{fig:peeqintro} as opposed to \figref{fig:hartley}).

In Section~\ref{lit_review}, we provide a brief overview of existing analytical and FE cold-rolling models that consider through-thickness stress and strain variations, while an in-depth description of the ABAQUS/Standard and ABAQUS/Explicit FE formulations may be found in~\ref{app:abaqus_formulation_app}.
In Section \ref{sim_details} we provide details of the implicit and explicit FE cold rolling models.
In Section \ref{sec:model_refinement} we conduct a mesh study and compare implicit and explicit in terms of their ability to accurately predict through-thickness stress variations.
We show that roll force and torque are poor determinants of the optimal mesh size and of the steady state condition.
In Section~\ref{results} we present new physical phenomena predicted by our FE simulations, and show that the residual stress in the sheet is related to oscillatory mechanics inside the roll gap.  
In Section~\ref{results.2} we provide further validation by comparison with a new analytical model from~\citet{Mozhdeh_journ}.
In Section~\ref{conclusion} we draw conclusions and briefly discuss the future plans for our FE simulations.

\section{Literature review}
\label{lit_review}
Here we provide an overview of some existing analytical and FE rolling models that consider through-thickness variations in their analyses, and demonstrate their limitations.

FE simulations reported in this work were conducted using ABAQUS~2021 \citep{abaqus}.
Within the ABAQUS package there are two numerical solution types: implicit and explicit. 
For the cold rolling process it is not immediately clear which solution type is optimal as each type offers different advantages and disadvantages.
The explicit solver is often the preferred choice for modelling metal forming processes due to its computational speed and robust handling of complex contact conditions~\citep{exp_vs_imp, efficiency, mesh_sens, padhye2023mechanics}. 
A key disadvantage of choosing the explicit solver is the absence of equilibrium convergence enforcement which can lead to error accumulation~\citep{diaz2021residual}.
The potential for explicit solvers to accumulate error implies that implicit solvers provide stronger results, {since convergence is ensured at each time increment in implicit analyses}~\citep{jung1998step}.
However, implicit solvers typically consume more computational resources and can struggle to handle complex contact~\citep{padhye2023mechanics}.
In Section~\ref{sim_details} we will assess both implicit and explicit solvers for modelling cold rolling, with the specific aim of accurately capturing through-thickness variations of stress and strain quantities. 

Asymptotic analysis is useful for modelling physical processes where the smallness (or largeness) of a fundamental process parameter can be exploited \citep{murray2012asymptotic}.
This type of analysis is particularly useful for cold rolling, where the roll-gap aspect ratio is typically large \citep{cawthorn}, i.e., $1/\varepsilon = L/h_0 \gg1$, where $h_0$ is the initial half-thickness of the sheet and $L=\sqrt{2R \Delta h}$ is the approximate horizontal length of the roll gap (see \figref{fig:rolling_pic}).
\citet{domanti1995two} assume a small reduction in the sheet's thickness (i.e., $r =\Delta h/h_0 \ll 1$) along with the large-aspect-ratio assumption.
These assumptions enable an asymptotic analysis to be carried out to obtain accurate solutions in the limits as $\varepsilon \rightarrow 0$ and $r \rightarrow 0$.
However, these assumptions limit the range of validity of the analytical model.
\citet{Mozhdeh_journ} relax the assumption that the reduction of the sheet thickness is small and also resolve behaviour over two distinct length scales associated with $h_0$ and $L$.  
By contrast, \citet{domanti1995two} and other authors \citep[see, for example,][]{cawthorn} consider variations on a length scale of $L$ only.
The multiple-scales analysis of \citet{Mozhdeh_journ} recovers slab theory at leading order, and at higher orders predicts through-thickness oscillations in stress and velocity in the roll gap. 
These through-thickness variations have also been captured by the current authors in recent FE simulations \citep{flanagan2023new}.
The \citet{Mozhdeh_journ} model is limited to rigid-perfectly-plastic materials, thus elastic and strain-hardening effects are neglected.
FE models can integrate elastic-plastic material behaviour which captures elasticity and strain hardening.
Comparing FE outputs to analytical models such as the \citet{Mozhdeh_journ} model can give an understanding of what mathematical assumptions are valid or invalid.

Table~\ref{table:FE_settings} shows the common through-thickness variations from existing FE models.
As shown in Table~\ref{table:FE_settings}, comparing through-thickness variations from various works is difficult given the distinct differences in modelling criteria and assumptions (e.g., 2D or 3D simulation, number of elements through-thickness ($2N_e$), element type, etc.) made by authors.
The implication of this lack of comparability means that there is no clear mechanism for novices to recognise what constitutes a ``good" FE rolling simulation. 
While several authors report shear stress along the roller-sheet interface~\citep{exp_vs_imp,richelsen1996comparison,cawthorn}, Table~\ref{table:FE_settings} shows that few~\citep{yarita1985stress, cawthorn2014comparison, minton2017mathematical} provide the through-thickness shear stress variation.
Given the potential implications of oscillating shear stress on downstream quantities such as residual stress \citep{Mozhdeh_conf,flanagan2023new}, we argue that a ``good" FE model should be capable of accurately predicting the oscillatory shear stress (and other quantities) through the thickness of the sheet.

\citet{yarita1985stress} plotted the shear stress versus position in the rolling direction at four different heights from the sheet centre to the surface, since stresses were computed with $N_e=4$ reduced-integration quadrilateral elements through the half-thickness of the sheet.
We will see in subsection \ref{subsection:mesh_sens} that  this mesh is far too coarse to give accurate through-thickness predictions.
Oscillations are visible in the shear stress curves but were not discussed in the text, which implies a lack of investigation into these oscillations. 
\citet{cawthorn2014comparison} use $N_e=5$ reduced-integration elements in their FE model, which is implemented in ABAQUS/Explicit.
We show the limitations of ABAQUS/Explicit in subsection \ref{subsection:explicit_limitations}.
\citet{cawthorn2014comparison} present two-dimensional through-thickness vertical and shear stress components, but do not state the value for the roll-gap aspect ratio $1/\varepsilon$ or the half-thickness reduction $\Delta h$.
\citet{minton2017mathematical} considered a range of aspect ratios, with $\varepsilon$ varying from 0.05 to 1, and noted the emergence of an oscillatory pattern in shear, where the number of lobes in the oscillatory pattern is proportional to $1/\varepsilon$.  
In the limit as $\varepsilon \rightarrow 0$, many oscillatory lobes blur together so that none are visible.
For $\varepsilon=1$, there is a single sign change in shear, consistent with most analytical models.
\citet{minton2017mathematical} hypothesises that these small- and large-$\varepsilon$ results may explain why this oscillatory pattern has been missed in the literature. 
We note that the FE mesh used was still relatively coarse, with $N_e =9\text{--}15$, and that the means of determining a steady state is by analysing roll force and torque, which we will see (in subsection~\ref{sec:steady_state}) is a poor steady-state predictor.

\citet{olaogun2019heat} show a similar oscillatory pattern in their heat flux contour plots.
The maximum values of heat flux lie along a path that is similar to the path of the lines of maximum shear. 
Heat energy is dissipated during plastic flow \citep{johnson2013plane}, and plastic flow takes place in metals predominantly by shearing~\citep{howell2009applied}.
We therefore expect that a close relationship exists between the heat flux oscillations observed by~\citet{olaogun2019heat} and the shear stress oscillations that exist during rolling.
However, \citet{olaogun2019heat} do not present shear stress results since heat transfer during cold rolling is the main topic of this paper.

Finally, many other authors have illustrated through-thickness variation but only via quantities that are not significantly oscillatory, such as longitudinal stress, effective stress or equivalent strain 
\citep{thompson1982inclusion,mori1982simulation,lau1989explicit,yoshii1991analysis,richelsen1997elastic,ghosh2004computational,gudur2008neural,lenard2013primer}.
Others report through-thickness variations via mesh distortion \citep{mori1982simulation,liu1987finite,lau1989explicit}. 
\citet{tadic2023analysis} study the longitudinal residual stress induced by cold rolling through-thickness inhomogeneities via FE analysis, and state that it is ``imperative" to determine the through-thickness stress distributions during cold rolling to accurately predict and control the impact of longitudinal residual stress on the appearance and properties of cold-rolled strips. However, \citet{tadic2023analysis} select their mesh density ($N_e=8$) by satisfying convergence conditions concerning the mean value of the contact stress.
We will see in subsection \ref{subsection:mesh_sens} that accurate measurements of such surface quantities does not imply accurate through-thickness predictions, and that $N_e=8$ is far too coarse, especially when through-thickness predictions are desired.
\citet{li1982rigid} present contour plots of shear strain rate at a low resolution inhibiting any oscillatory pattern observation.
\citet{liu1988analysis} presented cross-sectional profiles of strain components from three-dimensional simulations which contain hints of oscillations but these patterns are not commented on.
The results of \citet{valjanju2016deformations} for displacement and velocity also hint at a possibility of oscillations, but are only presented at the sheet centre and surface.

It is clear that a reliable FE model, one capable of accurately predicting through-thickness stress and strain variations, is absent from existing literature. 
We aim to provide such a model, which can be used to guide and compare with faster analytical models.

\begin{landscape}
\global\pdfpageattr\expandafter{\the\pdfpageattr/Rotate 90}
\centering%
\captionsetup{type=table}\captionof{table}{Summary of FE models in the literature that consider through-thickness variations. Note that $2N_e$ represents the number of elements through the full thickness of the sheet, PEEQ is plastic equivalent strain and the material models are elastic-plastic (EP), rigid-plastic (RP), elasto viscoplastic (EVP) and elastoplastic (EoP). The * indicates that oscillations are visible in certain FE results, but are not directly discussed in the paper.}%
\label{table:FE_settings}%
\renewcommand*{\arraystretch}{1.4}\begin{longtable}{|>{\centering\arraybackslash}p{3.5cm}|X|>{\centering\arraybackslash}p{0.75cm}|>
{\centering\arraybackslash}p{1cm}|>
{\raggedright\arraybackslash}p{3.2cm}|>
{\centering\arraybackslash}p{1.5cm}
|p{4.5cm}|>{\centering\arraybackslash}p{1.9cm}|>{\centering\arraybackslash}p{1.9cm}|}
\toprule
\textbf{Reference} 
& \textbf{2D or 3D}
& \textbf{$2N_e$} 
& \multicolumn{1}{p{3.2cm}|}{\centering\textbf{Element type}}
& \textbf{Material model}
& \multicolumn{1}{p{4.5cm}|}{\centering\textbf{Through-thickness variations}}
& \textbf{Symmetric system}
& \textbf{Oscillatory pattern}\\
\midrule
\hline
Current work  
& 2D
& 10--80 
& Plane-strain 4-node, reduced-integration
& EP
& Shear, von Mises, PEEQ, PEEQ rate
& \cmark
& \cmark\\
\hhline{|=|=|=|=|=|=|=|=|}
\citet{li1982rigid} 
& 2D  
& 10
& 
& RP
& Mesh distortion, relative velocities, normal/shear strain rates, PEEQ
& \cmark
&  \\
\hline
\citet{thompson1982inclusion} 
& 2D
& 12
& Six-node triangular
& EVP
& Effective stress
& \cmark
&  \\
\hline
\citet{mori1982simulation} 
& 2D
& 10 
& Isoparametric quad full integration
& RP
& Mesh distortion, PEEQ rate, normal stress
& \cmark
&  \\
\hline
\citet{liu1985elastic} 
& 3D
& 8--10 
& 8-node brick
& EP
& Velocity flow patterns
& \cmark
&  \\
\hline
\citet{yarita1985stress} 
& 2D
& 8 
& Quad, reduced-integration
& EP
& Stress components
& \cmark
& \cmark$^*$ \\
\hline
\citet{liu1987finite} 
& 3D
& 6
&
& EP
& Mesh distortion
& \cmark
&  \\
\hline
\citet{mori1987analysis} 
& 2D 
& 14 
& Quad
& RP
& PEEQ rate, pressure
& \cmark
&  \\
\hline
\citet{liu1988analysis} 
& 3D
& 3 
& 8-node brick
& EP
& Incremental strain components
& \cmark
& \cmark$^*$ \\
\hline
\citet{lau1989explicit} 
& 2D 
& 20 
& Quad
& EP
& Mesh distortion, PEEQ
& \cmark
&  \\
\hline
\citet{yoshii1991analysis} 
& 2D 
& 8
&
& RP
& PEEQ, PEEQ rate
& Asymmetric  
& \\
\hline
\citet{MALINOWSKI1992273} 
& 2D 
& 18 
& 4-node
& EoP
& Stress (centre \& surface only)
& \cmark
&  \\
\hline
\citet{richelsen1997elastic} 
& 2D 
& 18 
& Quad
& EVP
& Mesh distortion, PEEQ, horizontal stress
& Asymmetric  
&  \\
\hline
\citet{ghosh2004computational} 
& 3D
& 10 
& 8-node brick, reduced-integration
& EP
& Effective stress, PEEQ
& \cmark
&  \\
\hline
\citet{gudur2008neural} 
& 2D 
& 8 
& Rectangular
& RP
& PEEQ
& \cmark
&  \\
\hline
\citet{shahani2009prediction} 
& 2D 
& 16--18 
& Plane-strain 4-node
& VP
& Temperature, strain, strain rate, effective stress
& \cmark
&  \\
\hline
\citet{lenard2013primer} 
& 2D 
&
& Isoparametric quad elements
& EP
& PEEQ
& \cmark
&  \\
\hline
\citet{cawthorn2014comparison} 
& 3D
& 10 
& 8-node brick reduced-integration
& EP
& Vertical, shear stress components
& \cmark
&  \\
\hline
\citet{zhou2016constitutive} 
& 3D
& 12 
& 8-node brick reduced-integration temperature-coupled
& EP
& Normal stress, pressure, displacements (edge only)
& \cmark
&  \\
\hline
\citet{valjanju2016deformations} 
& 2D 
& 16 
& Square isoparametric
& RP
& PEEQ, displacements, velocities
& \cmark
& \cmark$^*$\\
\hline
\citet{koohbor2017finite} 
& 2D 
& 36 
& 3-noded triangular
& RP
& PEEQ, temperature
& \cmark
&  \\
\hline
\citet{minton2017mathematical} 
& Both
& 2D: 18-30, 3D: 11
& 2D: Plane-strain 4-node reduced-integration, 3D: 8-node reduced-integration
& EP
& Stress, velocity components
& Symmetric and asymmetric 
& \cmark\\
\hline
\citet{olaogun2019heat} 
& 2D 
& 20 
& 4-node isoparametric
& EP
& Temperature, heat flux
& \cmark
& \cmark\\
\hline
\citet{tadic2023analysis} 
& 2D
& 16
& 4-node square elements
& EP
& Longitudinal, shear and effective strain, longitudinal and von Mises stress
& \cmark
& 
\\
\hline
\citet{flanagan2023new} 
& 2D
& 12--100 
& Plane-strain 4-node reduced-integration
& EP
& Vertical velocity, von Mises
& \cmark
& \cmark\\
\hline
\bottomrule
\end{longtable}
\end{landscape}
\global\pdfpageattr\expandafter{\the\pdfpageattr/Rotate 0}

\section{Simulation details}
\label{sim_details}
In this section we give a complete description of the FE models used in this work.
We include key information about the geometry, material, dynamics, mesh, contact and computer precision settings.

\subsection{Model geometry}\label{sec:model-geometry}
Since the metal sheet's width--thickness ratio is, in general, large in cold rolling (for example, \citet{hacquin1996steady,yao2019real, yao2020finite} and \citet{li2020numerical} all employ a width--thickness ratio that is at least 100), we consider an idealised two-dimensional geometry by assuming plane-strain conditions apply~\citep{richelsen1996comparison, jacobs2023quantification, lenard2013primer}.
Readers are referred to Figure~\ref{fig:rolling_pic} for a diagram of the model geometry.
Only the top roller and the top half of the sheet are modelled in this work due to the expected symmetry about the symmetry axis at $z=0$.
We consider a rigid roller of radius $R=257.45\,\si{mm}$, rolling a sheet of initial half-thickness $h_0 = 2\,\si{mm}$, and impose an approximate half-thickness reduction of $\Delta h=0.5\,\si{mm}$.
This corresponds to $\varepsilon = h_0/L = 0.125$, where $L = 16.05\,\si{mm}$ is the approximate horizontal length of the roll gap.

\subsection{Material properties}
We consider a sheet of elastic-plastic mild steel (grade DC04), a common material used in cold metal forming~\citep{spittel2009steel}.
The DC04 material deforms elastically up to an initial yield stress of $Y = 477.2\,\si{MPa}$, and the yield stress increases to $650.25\,\si{MPa}$, at a true plastic strain of 1.1, as a result of strain hardening.
The Young's modulus and Poisson's ratio are taken to be $E = 206.3\,\si{GPa}$ and $\nu=0.3$ respectively.
For all explicit analyses, the density of the material is defined as $\rho = 7831.3\, \si{kg.m^{-3}}$.

\subsection{General FE model conditions}
\label{subsec:general_fe_conditions}
The model consists of two steps: the \textit{bite} step where contact between the roller and the sheet is initialised, and the \textit{rolling} step, where the sheet is horizontally displaced due to rotation of the roller.
To ensure contact convergence between the sheet and the roller the \textit{bite} step occurs over a time increment of $1\,\si{s}$.
The \textit{rolling} step occurs over a time increment of $0.1\,\si{s}$.
This is enough time for approximately 8 roll-gap lengths to be rolled when $R=257.45\,\si{mm}$.
The step type is dependent on the solver implemented (i.e. implicit or explicit), and readers are referred to Table~\ref{table:FE model specifics} for the specific step types used in simulations discussed here.

The sheet's horizontal centre line (i.e., the symmetry axis in \figref{fig:rolling_pic}) is constrained vertically to enforce the expected symmetry.
This modelling choice reduces the computational intensity of the problem, and in turn facilitates the use of a finer mesh.
The centre of the roller is defined as a reference point and boundary conditions, depicting the movement of the roller, are applied to this location.
Initially, the roller and the sheet are not in contact, but are set up such that the roller is positioned above the sheet. 
In the \textit{bite} step the roller is slowly translated vertically (via a velocity boundary condition) to indent the sheet, but no rotation of the roller occurs in the \textit{bite} step.
In the \textit{rolling} step, the roller reference point is fixed so its only degree of freedom is rotation in the rolling direction, and the roller is given a tangential velocity of $R \Omega = 1287.25\,\si{mm\,s^{-1}}$.
The \textit{bite} step is used here, instead of starting the strip in front of the roll gap and feeding it in in some way, as the bite-then-roll simulations showed fewer initial contact issues and converged more quickly to steady state (as also described by \citet{minton2017mathematical}). 
Note that the above boundary conditions are used for all analyses below unless otherwise stated. 

As the formulation differs for implicit and explicit simulations the model specifications understandably diverge.
Table \ref{table:FE model specifics} provides model settings unique to implicit and explicit simulations. 
The amplitude curve information given in Table~\ref{table:FE model specifics} relates to the application of loading on the roller during the \textit{rolling} step. 
Implicit simulations detailed here apply an instantaneous $\Omega=5~\si{rad\,s^{-1}}$ while explicit simulations smoothly ramp loading to $\Omega=5~\si{rad\,s^{-1}}$ over a fraction of the \textit{rolling} step.

In the analysis below we compare results from the four-node, plane-strain elements. 
As detailed in Table~\ref{table:FE model specifics}, implicit analyses include comparisons of full-integration (CPE4), incompatible (CPE4I) and reduced-integration (CPE4R) element types. 
Since full-integration (CPE4) and incompatible type (CPE4I) elements are not offered by ABAQUS/Explicit only reduced-integration (CPE4R) elements are considered.
Meshing is applied using a global seed formulation with a structured meshing technique.
Varying levels of mesh density are considered and the optimal mesh is discussed in detail in Section~\ref{sec:model_refinement}.
\begin{table*}
\caption{Unique FE model settings for the ABAQUS/Standard and ABAQUS/Explicit rolling models used in this work.}
\centering
\vspace{5pt}
\begin{tabular}{ |>{\centering\arraybackslash}p{5.5cm}||>{\centering\arraybackslash}p{4.5cm}|>
{\centering\arraybackslash}p{4.5cm}|}
 \hline
\textbf{Feature} & \textbf{ABAQUS/Standard} &  \textbf{ABAQUS/Explicit}\\
\hline
\hline
Density & N/a & $7831.3\, \si{kg.m^{-3}}$ \\ \hline
Step type & Static, general & Dynamic, explicit \\ \hline
Amplitude curve & Instantaneous & Smooth step \\ \hline
Contact tracking & State-based & N/a \\ \hline
Contact discretisation & Surface-to-surface & N/a \\ \hline
Mechanical constraint formulation & N/a & Kinematic contact\\ \hline
Element types considered & CPE4/CPE4I/CPE4R & CPE4R\\ \hline
ABAQUS/Explicit precision & N/a & Double, analysis \& packager\\
 \hline
\end{tabular}
\label{table:FE model specifics}
\end{table*}

\subsection{Contact conditions}
The roller, as an analytically rigid geometry, is defined as the master surface for contact between the roller and the sheet.
The surface of the sheet in contact with the roller is defined as the slave (i.e. deformable) surface.
Hard pressure-overclosure contact is the most common contact pressure-overclosure relationship and is defined between the roller and the sheet~\citep{abaqus}. 
In theory, by using this hard contact relationship, contact pressure is zero until contact actually occurs, and immediately increases to infinity upon contact. 
This infinite contact force prevents penetration of the slave surface into the master surface. 
In practice, however, this infinite spike in contact pressure is numerically challenging. 
The zero-penetration condition may or may not be strictly enforced depending on the constraint enforcement method used. 
The default penalty constraint enforcement method is implemented here, which approximates hard pressure-overclosure behavior, so some degree of penetration will occur.
For cold-rolling processes we expect significant relative motion between the rigid roller and the deformable sheet.
To account for relative motion between the roller and sheet surfaces, a finite sliding formulation is adopted.
Friction between the roller and the sheet is defined through the isotropic Coulomb friction model. 
Coulomb friction is regularly used in FE simulations of rolling (e.g., \citep{li1982rigid,mori1982simulation,yoshii1991analysis,gudur2008neural,minton2017mathematical}).
For industrial cold rolling mills the friction coefficient typically lies in the range of $\mu= 0.02 - 0.15$~\citep{rolling_params2}.
FE models in this work enforce a friction coefficient of $\mu=0.1$.

\subsubsection{Implicit analyses}
\label{h:contact_implicit}
Implicit simulations employ multiple types of contact-tracking algorithms. 
For finite-sliding contact interactions between analytically rigid and deformable surfaces the state-based algorithm is recommended~\citep{abaqus}.
The state-based contact algorithm evaluates contact interactions based on the current state of the simulation, together with geometric information.
Contact is discretised using the surface-to-surface method. 

\subsubsection{Explicit analyses}
Mechanical constraints in explicit analyses are described using the default kinematic contact method.

\subsection{Transient effects and job precision}
Our focus in this paper is on steady-state behaviour, so we aim to minimise transient effects during the dynamic explicit analysis as outlined in subsection \ref{sec:explicittransient}.

\subsubsection{Implicit analyses}
To ensure computational precision, full nodal output was requested for all simulations in this work. 
This is equivalent to double precision in explicit simulations.

\subsubsection{Explicit analyses}\label{sec:explicittransient}
To minimise transient effects loading in explicit simulations was applied using the smooth amplitude method.
Similarly mass scaling and/or increased loading rates were not applied.
To further increase computational precision explicit simulations were conducted using double precision for both the packager and analyses.
This is equivalent to the full nodal precision requested in implicit analyses. 

\section{FE model refinement}
\label{sec:model_refinement}
Here we investigate the effect of element type, solver types (i.e., implicit and explicit) and mesh size on through-thickness stress and strain predictions. 
Through careful consideration of these variables we obtain a reliable FE model for the rolling process.
We then describe the process of ensuring that the simulated rolling process is in a steady state.

\subsection{Element type}
\label{subsection:element_type}
In this work, first-order plane-strain elements are employed to discretise the sheet. 
These elements provide sufficient accuracy and are less computationally-expensive compared to second-order elements~\citep{patel2016mechanisms}. 
Others~\citep{exp_vs_imp, MALINOWSKI1992273, shangwu1999simulation, shahani2009prediction, olaogun2019heat} similarly used first-order elements to model rolling.
As previously discussed (see subsection~\ref{subsec:general_fe_conditions}), ABAQUS limits the element type depending on solver implementation. 
The element types applied in this work are outlined in Table~\ref{table:FE model specifics} which differentiates by solver implementation.
In the following sections we discuss the benefits (and potential drawbacks) of each element type to demonstrate the effect of element selection on FE model output. 

\subsubsection{Implicit analyses}

The material response is evaluated by Gaussian quadrature at each element integration point~\citep{abaqus}. 
Fully-integrated plane-strain four-node (CPE4) elements contain a total of four integration points.
The number of nodes and/or number of integration points correlates to the degree of freedom any given element can exhibit.
Reduced-integration elements, such as CPE4R elements, use just a single integration point in the centre of the element (compared to the four integration points in a fully-integrated CPE4 element).
Therefore, in principle, CPE4 elements are more accurate than the CPE4R elements.
However, in the presence of bending moments the inherent linear geometry of CPE4 elements can give rise to large, artificial shear strains. 
This problematic behaviour is often termed ``shear locking"~\citep{safaei2011towards,qiao2016modeling}.
Shear locking is a purely computational behaviour and is not representative of physical processes.
FE models exhibiting shear locking provide erroneous displacement and stress values~\citep{safaei2011towards}.

\begin{figure}
    \centering
    \includegraphics[width=\linewidth]{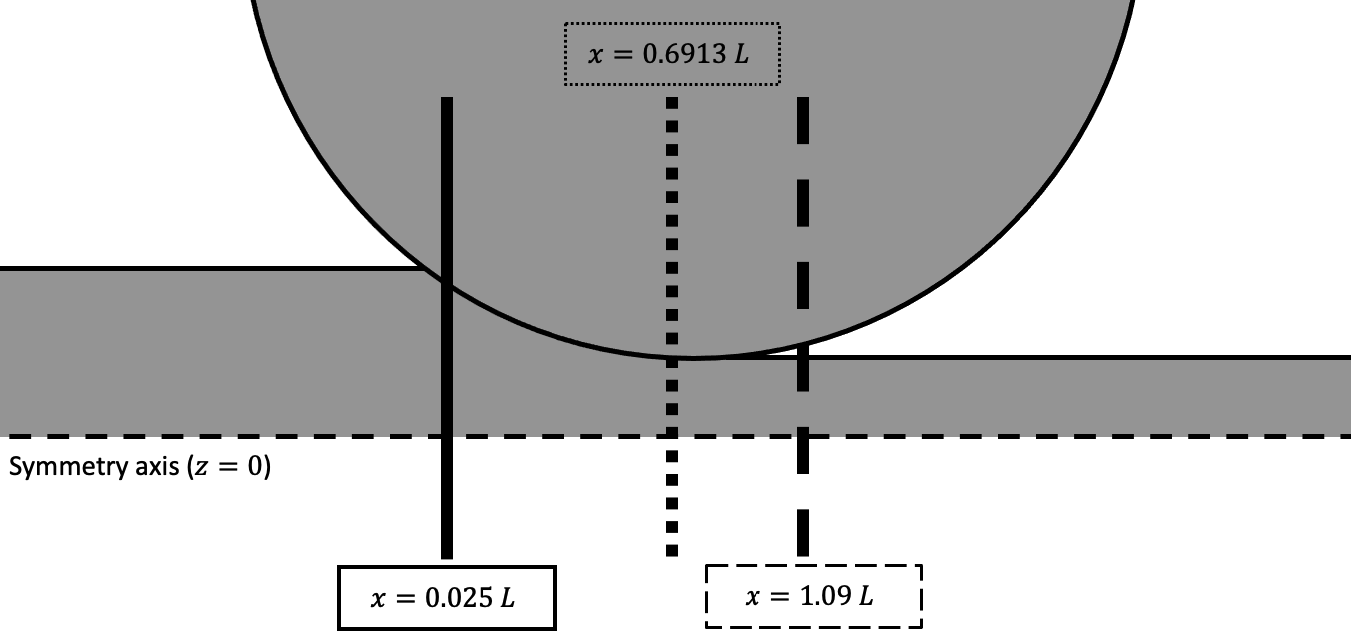}
   \caption{Interpolation is carried out in Figures \ref{fig:CPE4_vs_CPE4R}, \ref{fig:mesh_sens_CPE4R} and \ref{fig:steady_state_CPE4R} in the $z$-direction to facilitate comparison between different FE results. The solid, dotted and dashed curves in those figures depict results in the sheet at $x=0.025\,L$, $x=0.6913\,L$ and $x=1.09\,L$, respectively, as shown here. These are the three $x$-positions of the maximum relative errors (see \ref{relative_error_appendix}), which are calculated with respect to the simulation with $N_e=40$.}
   \label{fig:interp_positions} 
\end{figure}%

\begin{figure}
   \centering
    \includegraphics[width=\linewidth]{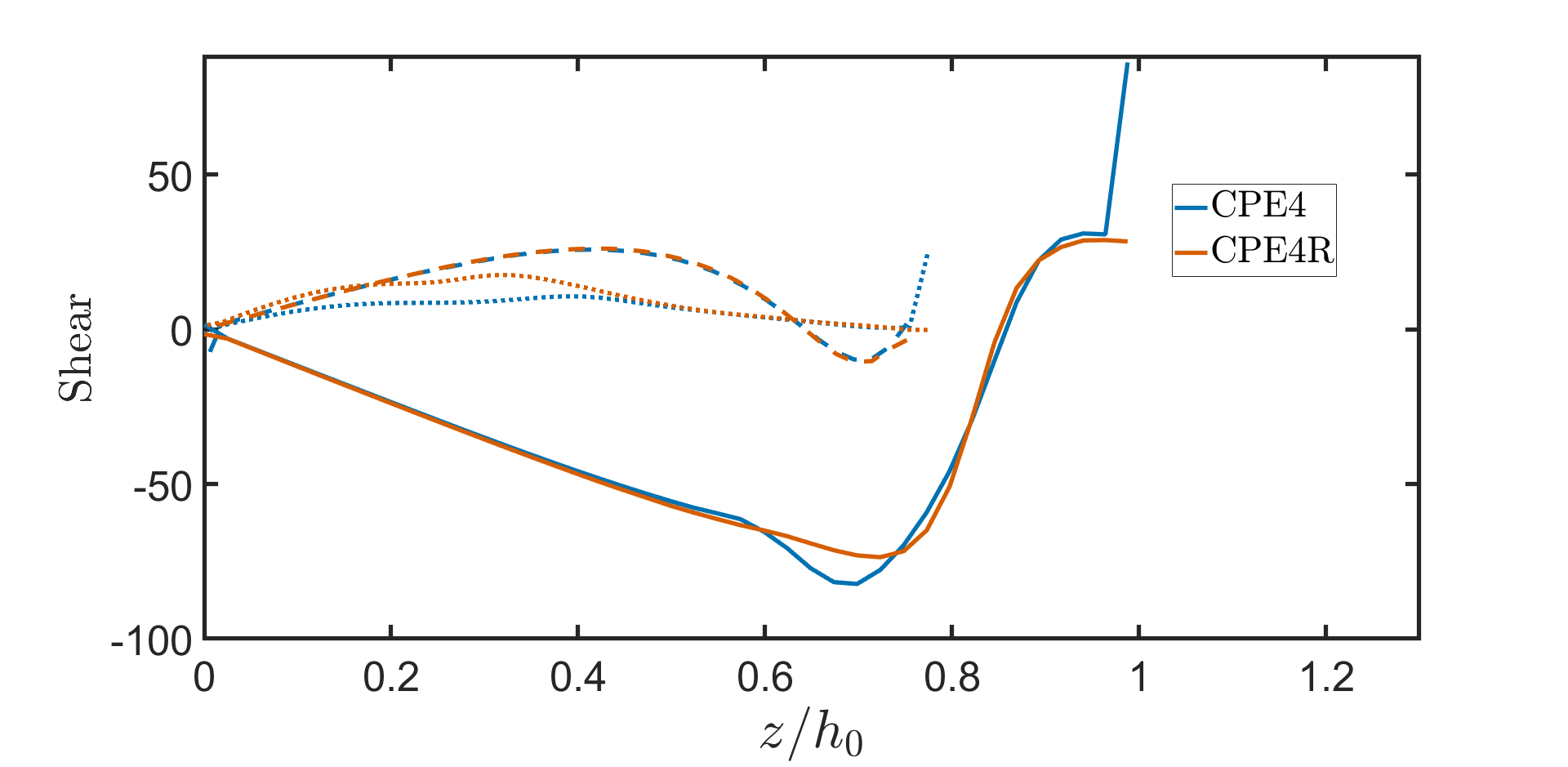}
    \caption{Interpolated shear results from two separate simulations that employ $N_e=40$ CPE4 and CPE4R elements through the half-thickness respectively. Readers should refer to \figref{fig:interp_positions} which identifies the horizontal position for the solid, dotted and dashed lines, respectively. The horizontal centre (symmetry axis in \figref{fig:rolling_pic}) of the sheet corresponds to $z/h_0=0$.
    Both simulations employ a radius $R=257.45\,\si{mm}$ and $\varepsilon=0.125$.}
    \label{fig:CPE4_vs_CPE4R}
\end{figure}%
Figure~\ref{fig:CPE4_shear_locking} shows a checkerboard pattern in the shear stress which is evaluated at the integration points of CPE4 elements during a rolling simulation with $N_e=40$ elements through the half-thickness of the sheet.
This checkerboard pattern is typical of shear locking behaviour~\citep{abaqus}.
Figure \ref{fig:CPE4_vs_CPE4R} shows a quantitative comparison between the aforementioned CPE4 shear results and shear outputs from a simulation that employs $N_e=40$ CPE4R elements through the half-thickness. 
Interpolation is carried out in the $z$-direction to facilitate comparison between different mesh types.
The solid, dotted and dashed curves depict results in the sheet at $x=0.025\,L$, $x=0.6913\,L$ and $x=1.09\,L$, respectively (see \figref{fig:interp_positions}).
These three $x$-positions represent the positions where maximum relative errors were observed in our convergence analysis (see~\ref{relative_error_appendix} for more details).
Figure~\ref{fig:CPE4_vs_CPE4R} shows sharp changes in the nodal CPE4 shear results, particularly towards the surface of the sheet (i.e., higher $z/h_0$ values) at $x=0.025\,L$ and $x=0.6913\,L$ (the solid and dotted curves respectively), highlighting the non-smooth nature of the shear results that is also visible in Figure~\ref{fig:CPE4_shear_locking}. 
In contrast, the CPE4R results are quite smooth everywhere, confirming that shear locking is not an issue for CPE4R elements.
The checkerboard pattern (Figure~\ref{fig:CPE4_shear_locking}) and the interpolated shear results (Figure~\ref{fig:CPE4_vs_CPE4R}) imply that CPE4 elements are not suitable for this test case and
hence this element type is not considered for the remainder of this work.

Unlike CPE4 elements, plane-strain reduced-integration (CPE4R) and incompatible (CPE4I) elements do not suffer from shear-locking behaviour.
Low-order elements lack the necessary shape functions to describe bending. 
Incompatible elements such as CPE4I elements artificially add the required shape functions to adequately describe this behaviour~\citep{sun2006shear, souto2022determination}. 
Simulations conducted using CPE4R and CPE4I elements do not exhibit the typical shear-locking checkerboard pattern observed in Figure~\ref{fig:CPE4_shear_locking}. 
This indicates that employing either reduced-integration or incompatible elements would be sufficient for removing shear-locking behaviour.
However, incompatible elements are slow to converge in simulations with large compressive strains~\citep{abaqus}.
For example, a simulation with $N_e=15$ CPE4I elements through the half-thickness took over 5.5 times as long as a simulation with $N_e=15$ CPE4R elements, (see~\ref{CPE4I_app}, where the performance of CPE4I and CPE4R elements are compared in more detail).
The large computational time severely restricts the mesh size that can be employed with the CPE4I element type, and hence CPE4I elements are not considered for the remainder of this work.

\subsubsection{Implicit and explicit analyses}
While reduced-integration elements are often recommended as a means of mitigating shear locking~\citep{sun2006shear,hadadian2019investigation,zhang2022damage}, they can suffer from spurious zero-energy element deformations.
These zero-energy deformations are often known as “hourglass” modes~\citep{safaei2011towards,yang2019numerical}.
ABAQUS offers controls to mitigate against this unrealistic zero-energy behaviour. 
The default total stiffness hourglass controls are applied to simulations in this work.
Default hourglass controls apply additional stiffness to elements to counteract hourglass modes.
To obtain physically reliable results, the artificial energy used to control hourglassing must be small (within $5$--$10\%$) compared to the internal energy in the model~\citep{yang2019numerical,abaqus,matsubara2024computational,gao2024effect}.
In this work the artificial energy is less than $0.5\%$ of the internal energy for all simulations.
Given the small energy ratio and considering that hourglass behaviour is not visible in the deformed mesh of any simulation, we conclude that hourglassing does not occur in simulations reported here.
Therefore CPE4R elements are the most appropriate element for this research.

\subsection{Limitations of ABAQUS/Explicit}
\label{subsection:explicit_limitations}
We now consider results from ABAQUS/Explicit simulations.
Mass scaling (or increased loading rate) was not used in any explicit simulation in this paper.
Given that the kinetic-to-internal energy ratio remains in the desired range (5--10\%)~\citep{web,abaqus}, we consider inertia effects to be insignificant.

\begin{figure}
\centering
\begin{subfigure}[b]{\linewidth}
   \centering
   \includegraphics[width=\linewidth]{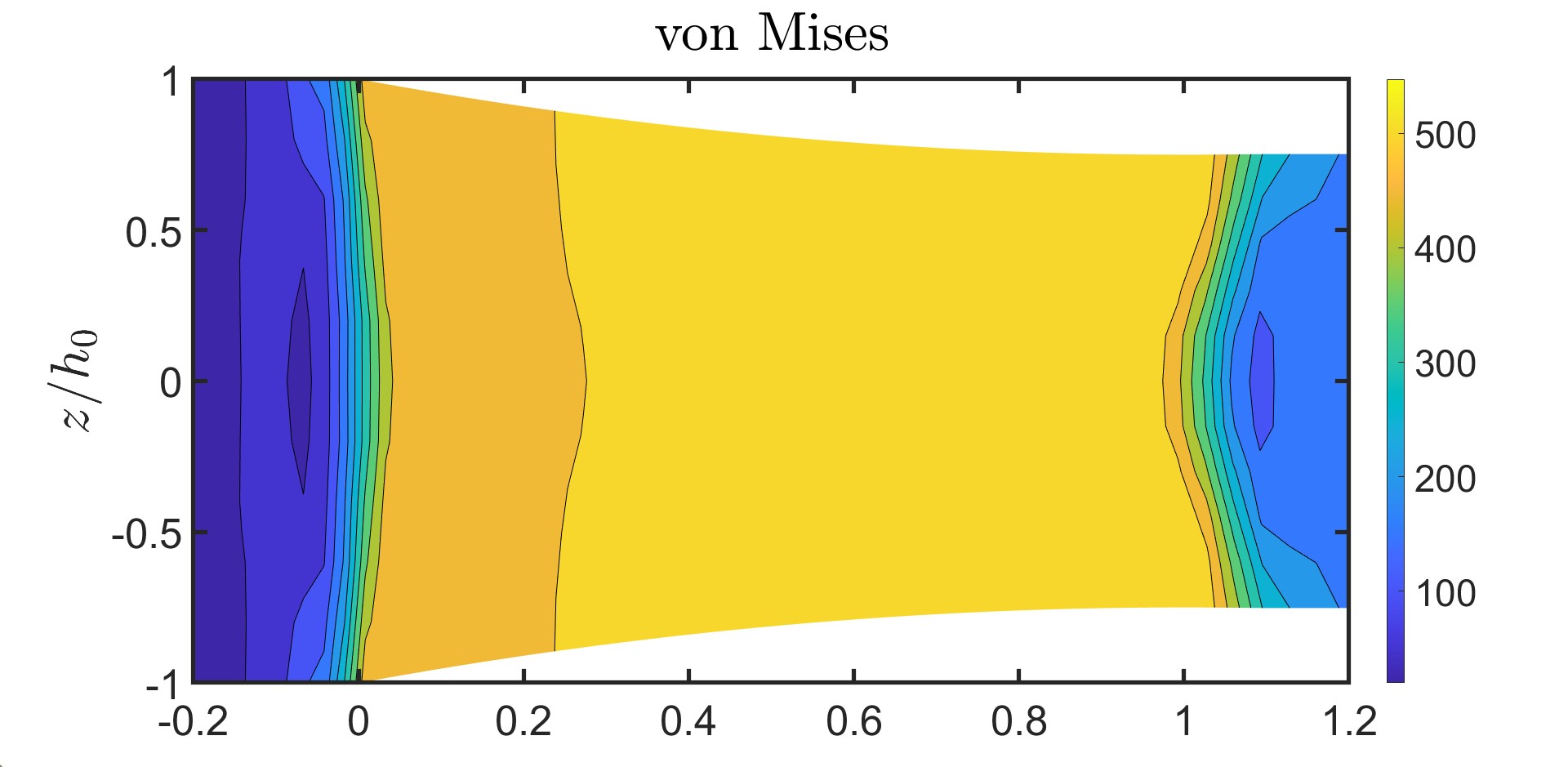}
   \caption{}
   \label{fig:exp_5tt} 
\end{subfigure}

\begin{subfigure}[b]{\linewidth}
   \centering
   \includegraphics[width=\linewidth]{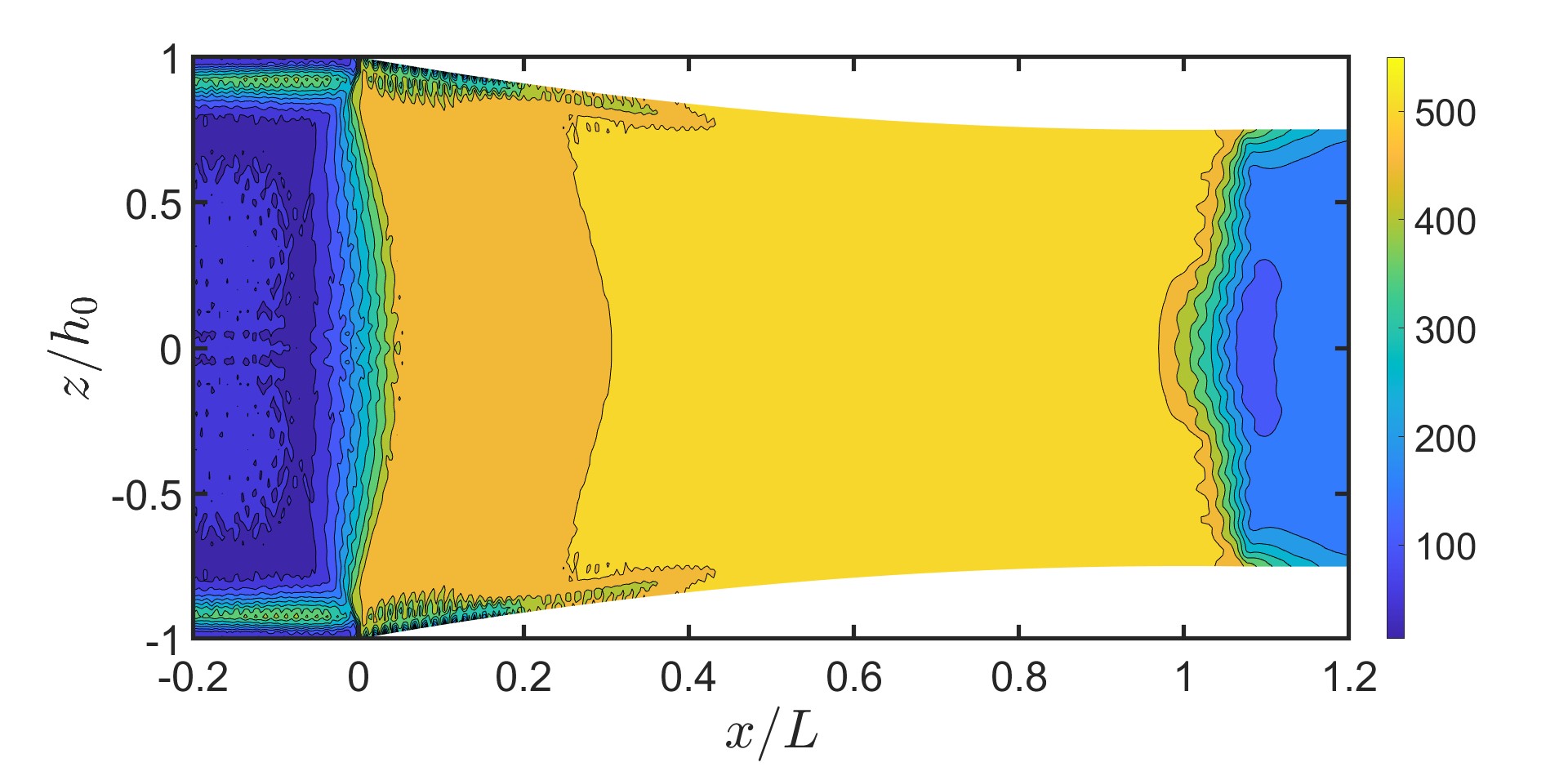}
   \caption{}
   \label{fig:exp_40tt}
\end{subfigure}
\caption{Contour plots showing von Mises stress ($\si{MPa}$) from explicit simulations with (a) $N_e=5$ CPE4R elements and (b) $N_e=40$ CPE4R elements, where the symmetry axis is located at $z/h_0=0$ (see Figure~\ref{fig:rolling_pic}). Both simulations have roller radius $R=257.45\,\si{mm}$ and $\varepsilon=0.125$.} \label{fig:explicit_vM}
\end{figure}%
Figures \ref{fig:exp_5tt} and \ref{fig:exp_40tt} compare the von Mises stress for two explicit simulations with mesh densities of $N_e=5$ and $N_e=40$ elements through the half-thickness, respectively. 
The most notable difference between the plots is the smoothness of the solution. 
As mesh density increases, it is expected that the calculation accuracy should, in general, improve~\citep{qian2010research}.
However, \figref{fig:explicit_vM} shows that increasing the mesh density increases noise, indicating that the solution diverges as mesh density increases.
In particular, we note the numerical noise near the surface of the sheet towards the entrance to the roll gap, $x/L=0$.
This noise is likely an artefact of the lack of convergence checks performed by ABAQUS/Explicit at each time increment, which allows the solution to diverge without restriction.
\citet{flanagan2023new} successfully removed such noise by implementing additional stiffness-proportional damping.
However, the damping approach increases the required computational time (700 CPU hours compared to 10 CPU hours without damping for a particular rolling simulation conducted by \citet{flanagan2023new}).
Although the explicit simulations are typically quicker to execute than their implicit counterparts (for example, with $N_e=40$ CPE4R elements through the half-thickness, the explicit simulation's CPU time is less than 37\% than that of the equivalent implicit simulation), the divergent nature of the explicit results and the lack of equilibrium checks lead us to proposing that implicit simulations outperform explicit simulations for this case study. Therefore the remainder of this work will focus on simulation results from implicit simulations only.

\subsection{Mesh sensitivity study}
\label{subsection:mesh_sens}
In subsection~\ref{subsection:element_type} we demonstrated that CPE4R elements are the preferred element type for this simulation case study.
We also demonstrated (subsection~\ref{subsection:explicit_limitations}) that explicit analyses are subject to increasing noise with increasing mesh density.
In this section we perform a mesh convergence study for implicit simulations with CPE4R elements that results in the optimum trade-off between accuracy and computational intensity.

\begin{figure}[tb]
\centering
\begin{subfigure}[b]{\linewidth}
   \centering
   \includegraphics[width=\linewidth]{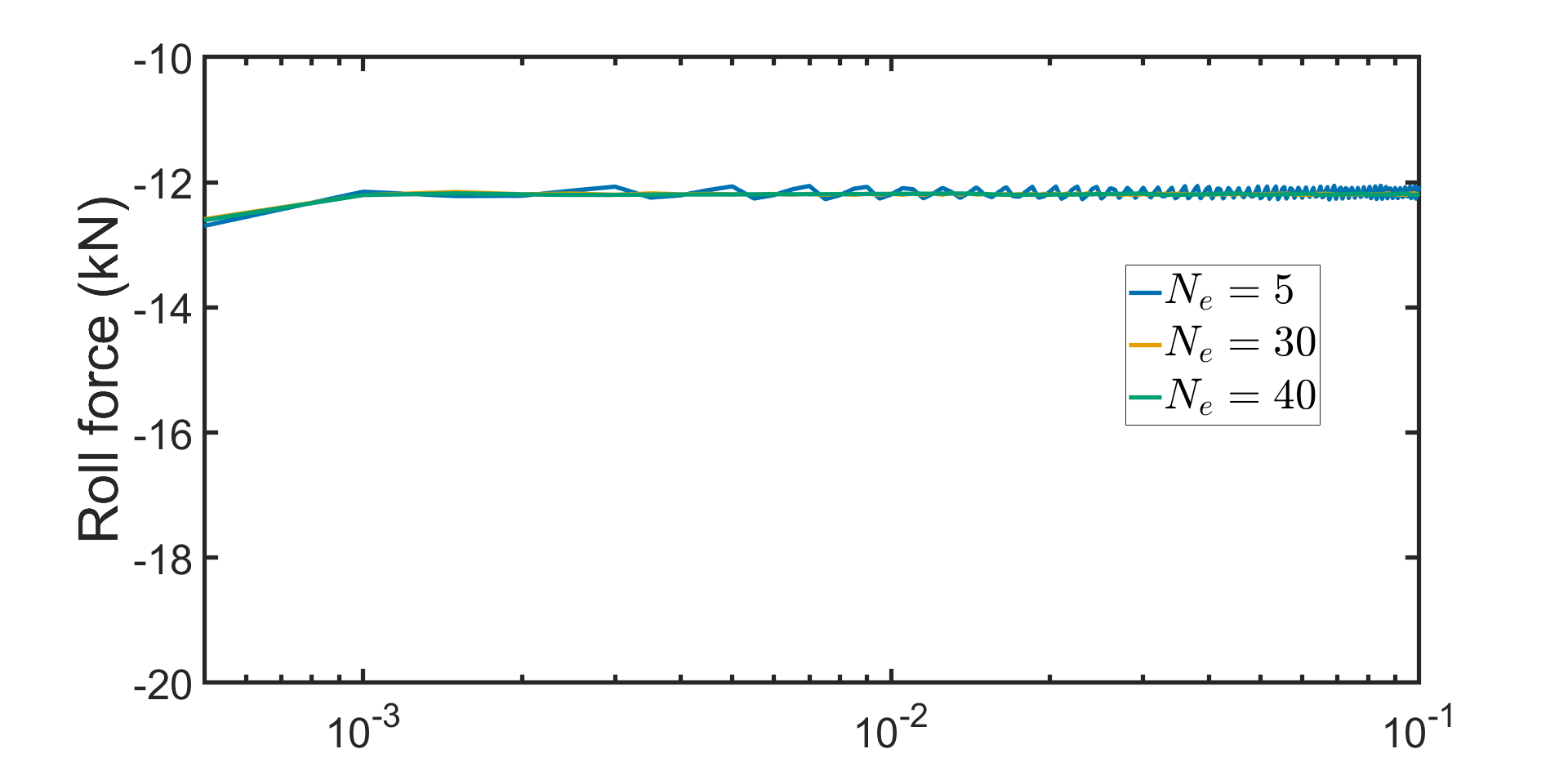}
   \caption{}
   \label{fig:roll_force} 
\end{subfigure}

\begin{subfigure}[b]{\linewidth}
   \centering
   \includegraphics[width=\linewidth]{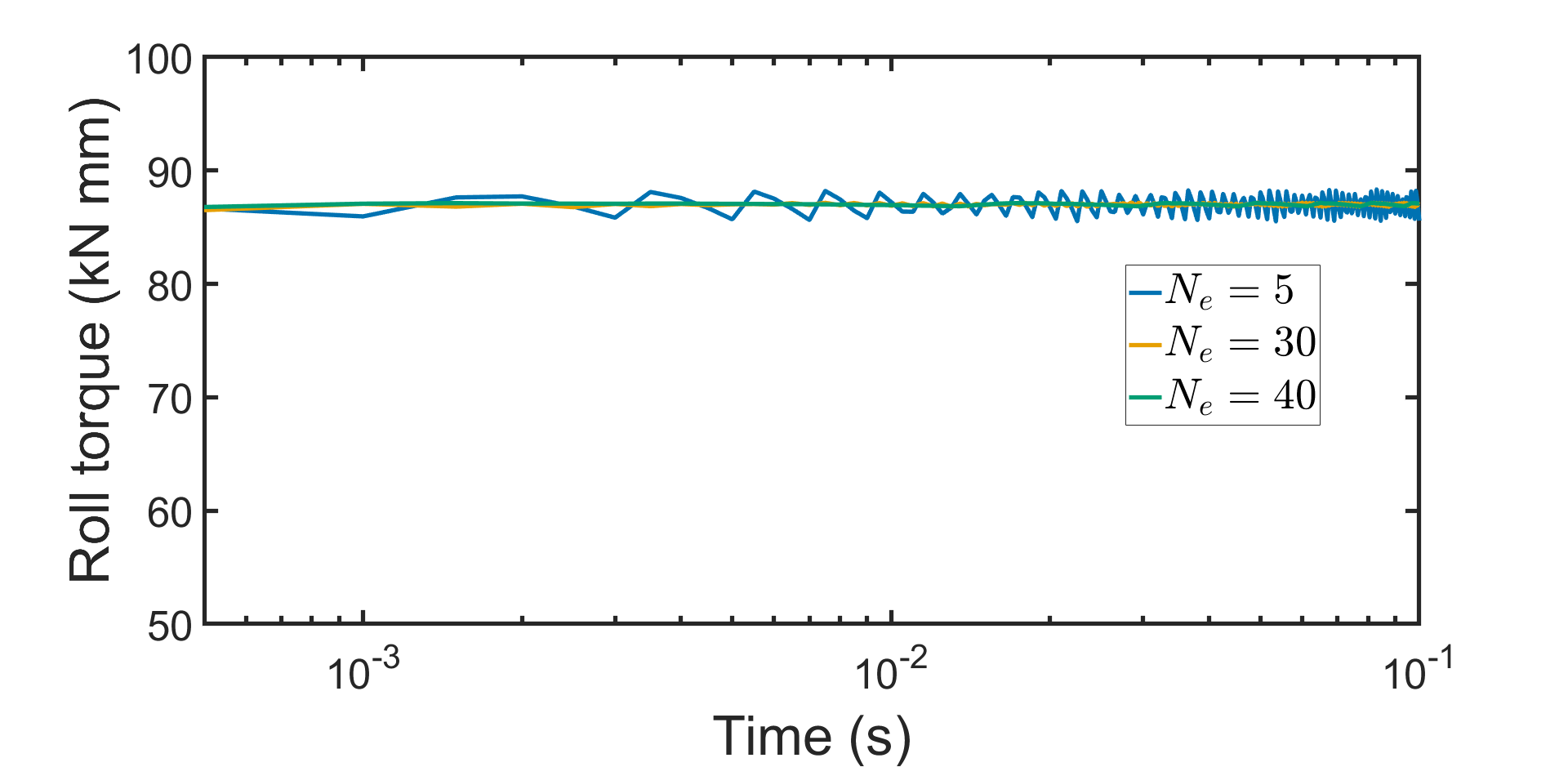}
   \caption{}
   \label{fig:roll_torque}
\end{subfigure}
\caption{Time-history plots of (a) roll force and (b) torque per unit width during the \textit{rolling} analysis step for implicit simulations with $N_e=5$, $N_e=30$ and $N_e=40$ CPE4R elements. Relatively steady average force and torque values are observed after $0.001\,\si{s}$ for each simulation. All simulations have radius $R=257.45\,\si{mm}$ and $\varepsilon=0.125$.} \label{fig:roll_force_torque}
\end{figure}
\begin{figure}
\centering
\begin{subfigure}[b]{\linewidth}
   \centering
   \includegraphics[width=\linewidth,height=0.25\textheight,keepaspectratio]{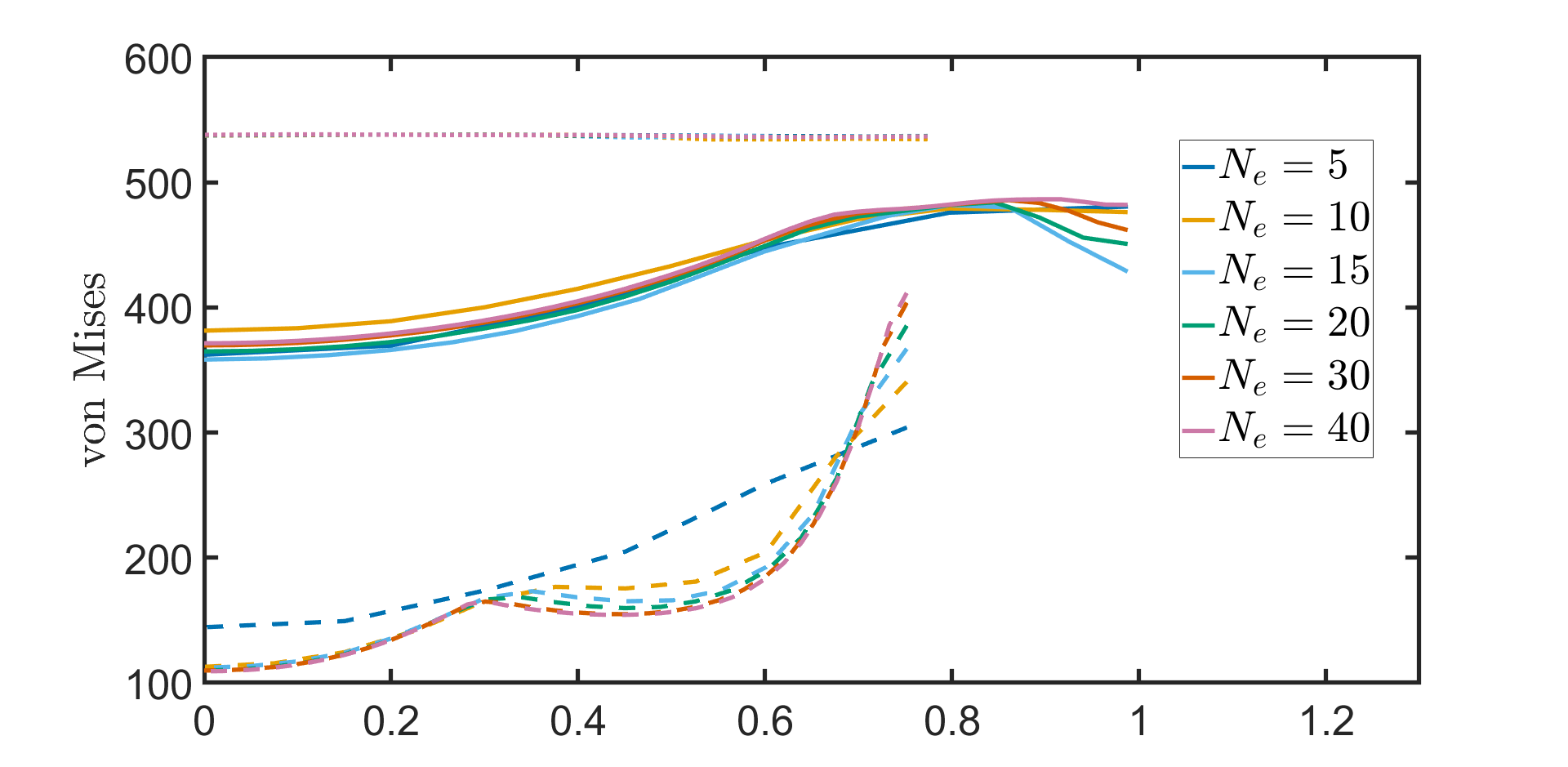}
   \caption{}
   \label{fig:vM_vertical} 
\end{subfigure}

\begin{subfigure}[b]{\linewidth}
   \centering
   \includegraphics[width=\linewidth,height=0.25\textheight,keepaspectratio]{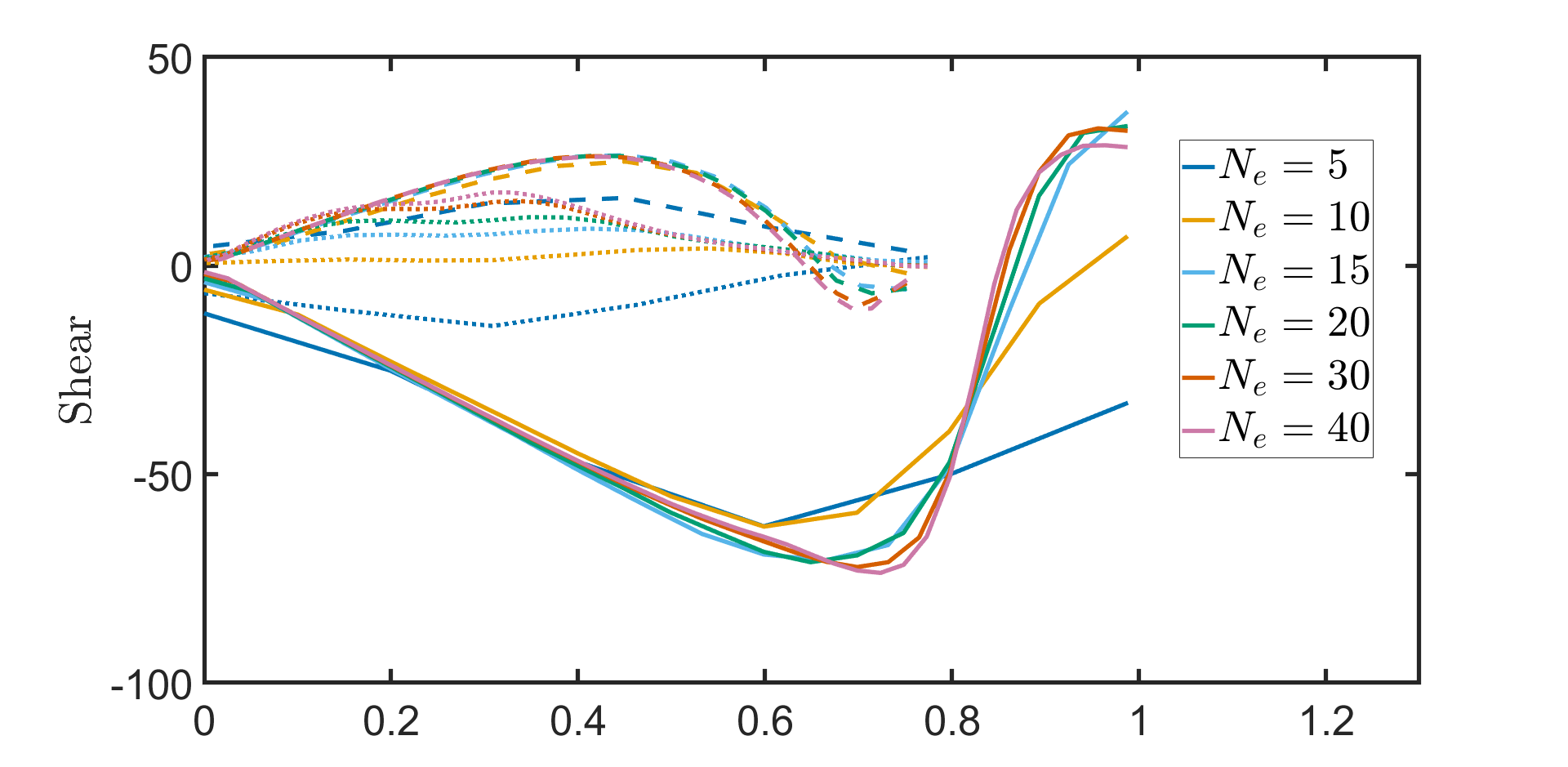}
   \caption{}
   \label{fig:shear_vertical}
\end{subfigure}

\begin{subfigure}[b]{\linewidth}
   \centering
   \includegraphics[width=\linewidth,height=0.25\textheight,keepaspectratio]{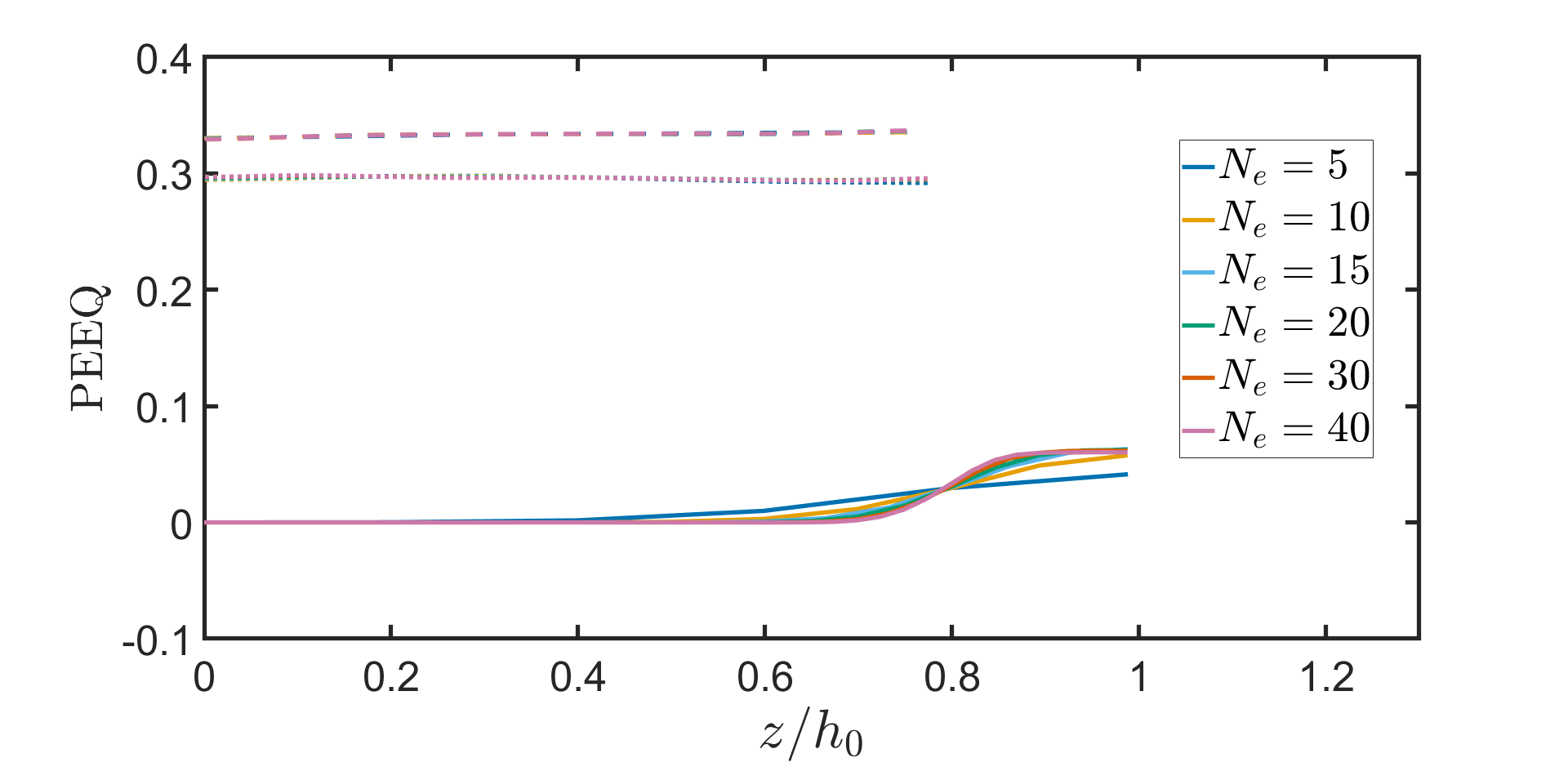}
   \caption{}
   \label{fig:PEEQ_vertical}
\end{subfigure}
\caption{Comparing the number of elements through half-thickness $\left(N_e\right)$ for simulations with radius $R=257.45\,\si{mm}$ and $\varepsilon=0.125$. 
Plotting interpolated (a) von Mises stress, (b) shear stress and (c) PEEQ (plastic equivalent strain). 
Readers should refer to \figref{fig:interp_positions} which identifies the horizontal position for the solid, dotted and dashed lines, respectively.}
\label{fig:mesh_sens_CPE4R}
\end{figure}

\figref{fig:roll_force_torque} plots (a) roll force and (b) torque per unit width as a function of simulation time during the \textit{rolling} step for various mesh densities.
We note that all force and torque values take on a relatively constant value after 0.001$\si{s}$, and that the $N_e=5$ simulation gives average roll force and torque values similar to the values outputted from the simulations with $N_e=30$ and $N_e=40$.
Some authors use roll force and torque in their mesh convergence studies by identifying the mesh size at which no further significant changes in force or torque occur \citep{rao1977finite,mesh_sens}.
By extension, we could take the $N_e=5$ results as sufficient here.
However, as we will see in \figref{fig:mesh_sens_CPE4R}, the simulation with $N_e=5$ performs poorly in predicting the von Mises stress, shear stress and plastic equivalent strain (PEEQ) values through the thickness of the sheet, thus proving that roll force and torque are poor measures of accuracy when through-thickness predictions are desired.

\figref{fig:mesh_sens_CPE4R} shows interpolated von Mises stress, shear stress and PEEQ results for mesh densities in the range $N_e=5\text{--}40$.
Interpolation is carried out in the $z$-direction to facilitate comparison between different mesh densities.
The results are extracted at $x=0.025\,L$, $x=0.6913\,L$ and $x=1.09\,L$ (see Figure~\ref{fig:interp_positions} for details on the three $x$-positions chosen).
The reader is reminded that each set of curves (i.e., solid, dashed and dotted) represent the interpolated results at a particular $x$-position, and so if the mesh is sufficiently refined, the curves at each $x$-position should be indistinguishable for subsequent mesh sizes.
The simulation with $N_e=40$ has a CPU time of the order of days (see Table~\ref{table:simulations}) and is therefore not suitable for practical use but is useful in benchmarking the error of lower-mesh-density simulations.
Errors decrease as the mesh density and CPU time increase, as shown in Table~\ref{table:simulations} (and in~\ref{relative_error_appendix}).

Figure~\ref{fig:vM_vertical} shows von Mises stress as a function of through-thickness position $\left(z/h_0\right)$ for simulations with mesh densities in the range $N_e=5\text{--}40$.
The dotted curves (representing results from $x = 0.6913\,L$) are almost indistinguishable from mesh size to mesh size.
The largest numerical errors for von Mises occur at $x = 0.025\,L$ and $x = 1.09\,L$, (see the solid and dashed curves in \figref{fig:vM_vertical}, respectively). 
This makes sense because the von Mises stress satisfies the yield condition uniformly in the interior of the contact region, except very close to the start and end where the material may be behaving elastically.
For the shear stress, Figure~\ref{fig:shear_vertical} shows significant differences from mesh size to mesh size at all three $x$-positions.
The largest shear stress errors occur at $x=0.025\,L$ (see the solid curves in \figref{fig:shear_vertical}).
The maximum relative difference in shear between the simulations with $N_e=30$ and $N_e=40$ is 4.9\% at this position.
Finally, the deviations in PEEQ between different mesh densities are largest near $x=1.09\,L$ (see the dashed curves in \figref{fig:PEEQ_vertical}), but the PEEQ errors are much smaller than for von Mises stress and shear stress (see Table~\ref{table:simulations}).

We note that the simulation with $N_e=5$ does not capture trends such as non-monotonic von Mises stress at $x=1.09\,L$ (\figref{fig:vM_vertical}) or the curvature of the shear stress (\figref{fig:shear_vertical}).
\citet{yarita1985stress} and \citet{cawthorn2014comparison} used just $N_e=4$ and $N_e=5$ reduced-integration elements respectively through the half-thickness of the sheet, which our results suggest is far too coarse a mesh to accurately capture the through-thickness stress and strain distributions in the sheet.
We use $N_e=30$ CPE4R elements through the half-thickness of the sheet in the remainder of this work since the maximum difference from the simulation with $N_e=40$ is 4.9\%, and the computation time is 4.83 times faster compared to the $N_e=40$ simulation.

\begin{table*}
\caption{CPU time and maximum percentage error in von Mises stress, PEEQ (plastic equivalent strain) and shear stress for simulations with $N_e=$ 5--30 benchmarked against interpolated results from the simulation with $N_e=40$. Errors are calculated relative to the maximum absolute roll-gap value from the simulation with $N_e=40$. All simulations in this table are implicit simulations with CPE4R elements.}
\centering
\vspace{5pt}
\begin{tabular}{ |>{\centering\arraybackslash}p{3cm}||>{\centering\arraybackslash}p{2.2cm}|>
{\centering\arraybackslash}p{1.9cm}|>
{\centering\arraybackslash}p{1.9cm}|>
{\centering\arraybackslash}p{1.9cm}|  }
 \hline
\textbf{Elements through half-thickness} & \textbf{CPU time (hours)} &  \textbf{Maximum von Mises \% error } &  \textbf{Maximum shear \% error }& \textbf{Maximum PEEQ \% error }\\
\hline
\hline
5 & 0.25 & 20.55 & 45.48 & 7.24\\
\hline
10 & 1.16 & 12.97 & 24.3 & 4.33\\
\hline
15 & 3.2 & 9.74 & 14.51 & 2.93\\
\hline
20 & 7.09 & 5.71 & 9.24 & 2.02\\
\hline
30 & 15.89 & 3.68 & 4.9 & 1.09\\
\hline
40 & 76.75 & - & - & -\\
 \hline
\end{tabular}
\label{table:simulations}
\end{table*}

\subsection{Steady-state determination}
\label{sec:steady_state}
\begin{figure}
\centering
\begin{subfigure}[b]{\linewidth}
   \centering
   \includegraphics[width=\linewidth,height=0.25\textheight,keepaspectratio]{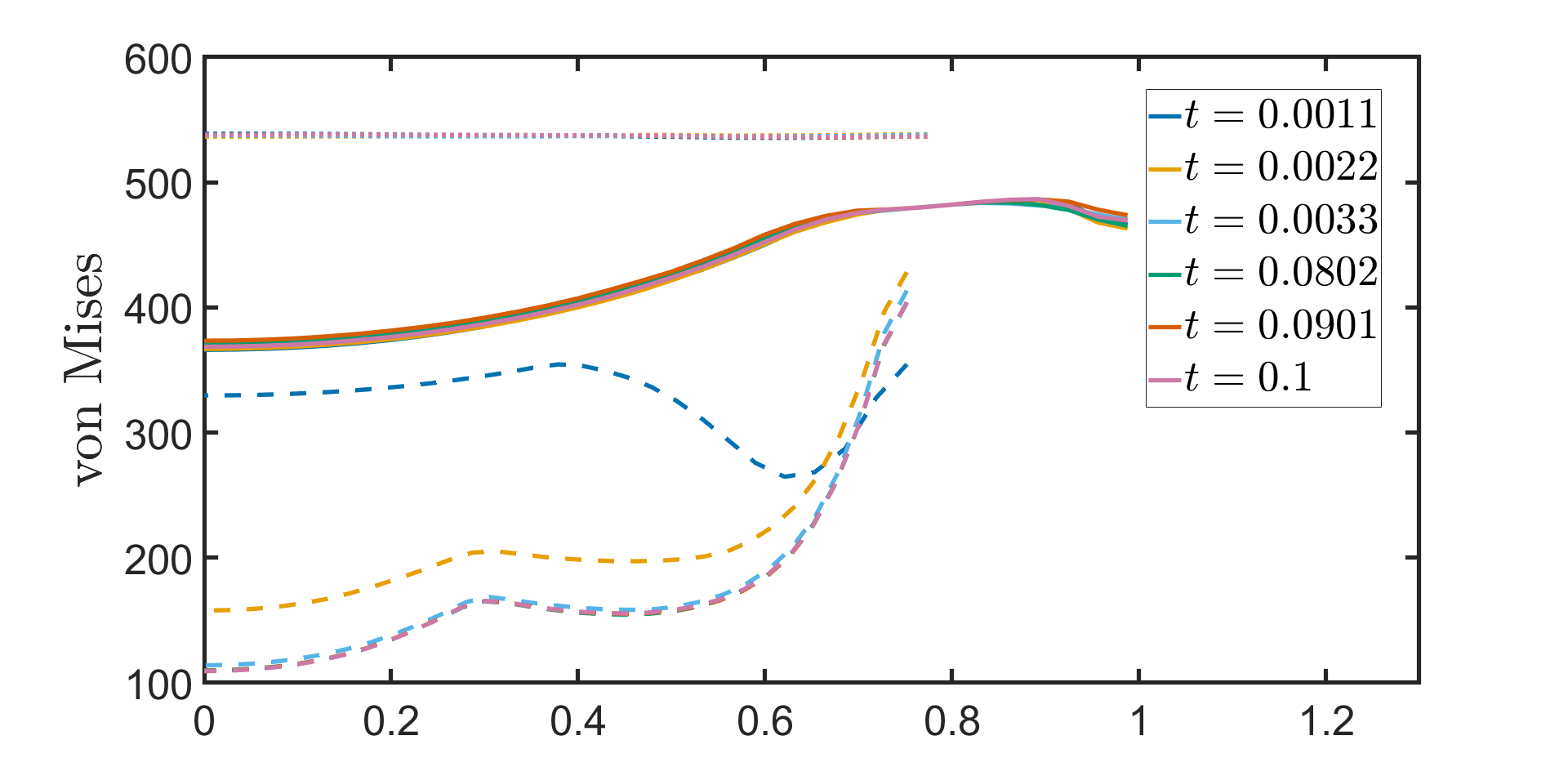}
   \caption{}
   \label{fig:vM_vertical_steady} 
\end{subfigure}
\begin{subfigure}[b]{\linewidth}
   \centering
   \includegraphics[width=\linewidth,height=0.25\textheight,keepaspectratio]{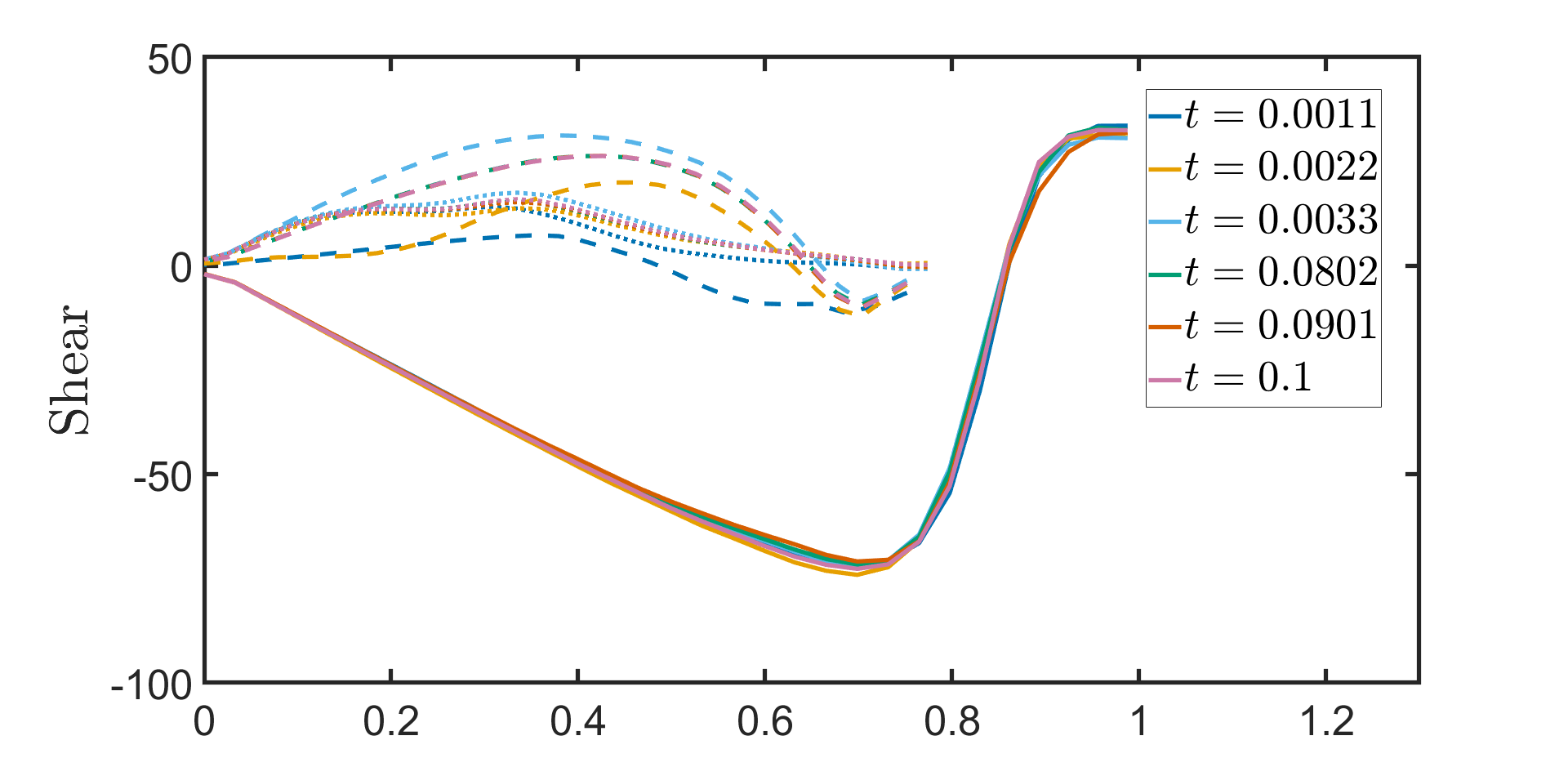}
   \caption{}
   \label{fig:shear_vertical_steady}
\end{subfigure}
\begin{subfigure}[b]{\linewidth}
   \centering
   \includegraphics[width=\linewidth,height=0.25\textheight,keepaspectratio]{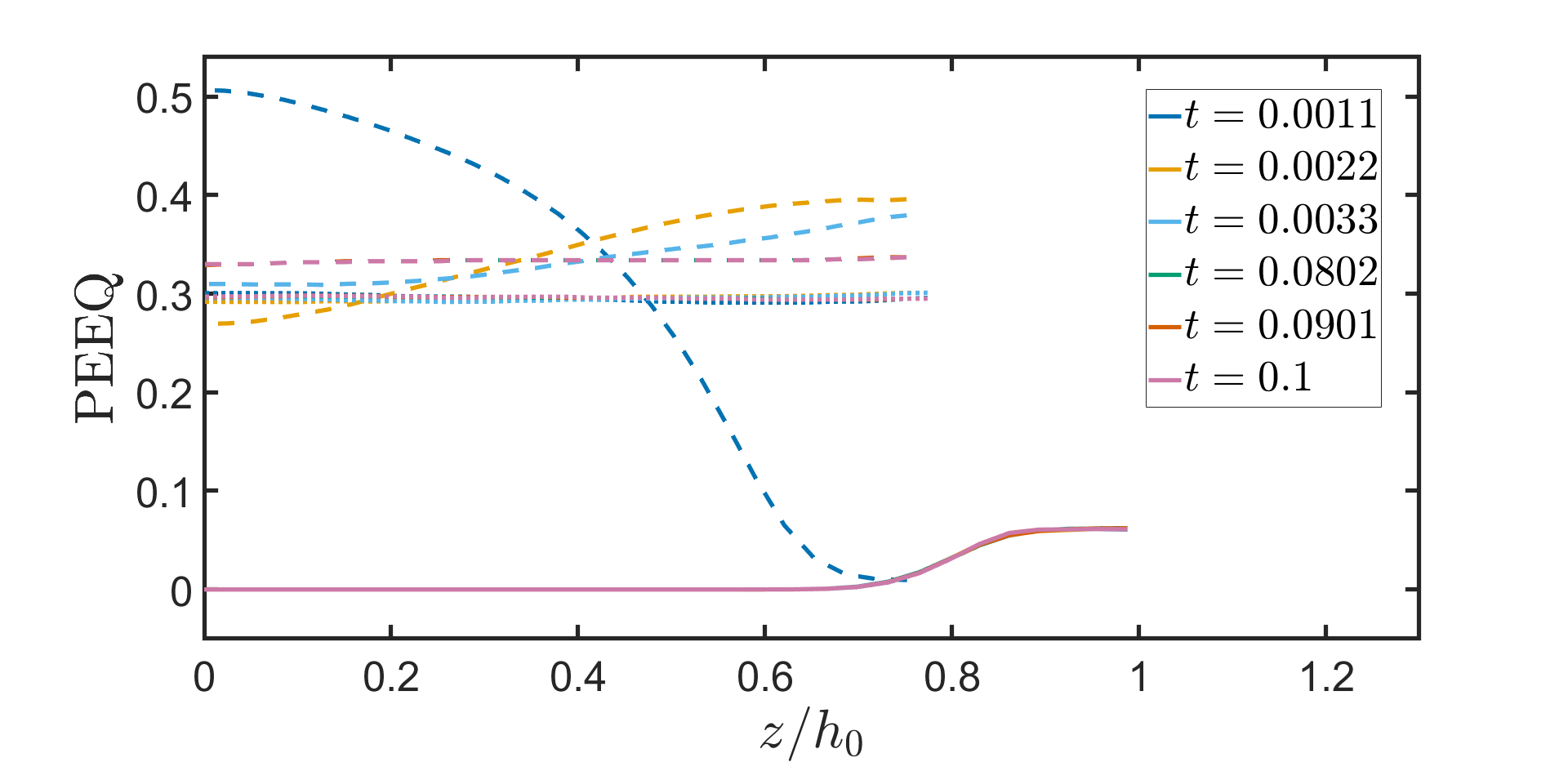}
   \caption{}
   \label{fig:PEEQ_vertical_steady}
\end{subfigure}
\caption{Comparing results from different time points of the implicit simulation with $N_e=30$ CPE4R elements, radius $R=257.45\,\si{mm}$ and $\varepsilon=0.125$, by plotting interpolated (a) von Mises stress, (b) shear stress and (c) PEEQ (plastic equivalent strain).
Readers should refer to \figref{fig:interp_positions} which identifies the horizontal position for the solid, dotted and dashed lines, respectively.} \label{fig:steady_state_CPE4R}
\end{figure}

In Section~\ref{lit_review} we noted that steady-state conditions are often defined in FE simulations through roll-force and/or roll-torque results in the literature. 
\figref{fig:roll_force_torque} suggests that for all mesh sizes, both roll force and torque reach a relatively steady value by time $t \approx 0.001\,\si{s}$.
With the help of \figref{fig:steady_state_CPE4R} however we demonstrate here how roll force and torque are insufficient at truly assessing whether steady-state conditions have been achieved, and show how to accurately determine if a true steady state has been reached.

\figref{fig:steady_state_CPE4R} shows interpolated von Mises stress, shear stress and PEEQ results from six different time points throughout the $N_e=30$ simulation. 
Interpolation is carried out as outlined in subsection~\ref{subsection:mesh_sens}.
The purpose of \figref{fig:steady_state_CPE4R} is to assess the time at which a true through-thickness steady state has been achieved. 
Since each set of curves (i.e., solid, dotted and dashed) represent the interpolated results at a particular $x$-position, we expect that under steady-state conditions the shape of each set of through-thickness curves should remain stagnant as a function of time.
That is to say that, in a steady state, the profile of the curve should not change from time-to-time because significant differences from consecutive time increments indicates that a steady state has not been achieved.

Firstly we draw the readers attention to Figure~\ref{fig:vM_vertical_steady} which details the von Mises stress as a function of through-thickness position ($z/h_0$) for the simulation with $N_e=30$ CPE4R elements through the half-thickness. 
We previously showed this element size and type to provide the optimal trade off between computational expense and accuracy. 
In Figure~\ref{fig:vM_vertical_steady} we plot results from six time points throughout the simulation, at identical horizontal positions that were analysed previously (see Figure~\ref{fig:interp_positions}).
The solid and dotted curves in Figure~\ref{fig:vM_vertical_steady} all have very similar profiles to their respective curves for all time points throughout the simulation.
However, the dashed curves (representing results from $x=1.09\,L$) reveal unsteady behaviour at times $t=0.0011\,\si{s}$ and $t=0.0022\,\si{s}$, where these curves are noticeably different to the $t\geq0.0033\,\si{s}$ curves at the same $x$-position.
Figure~\ref{fig:shear_vertical_steady} depicts the interpolated shear results at the same three $x$-positions. 
Once again, the solid curves all lie on top of each other, showing no unsteady behaviour at $x=0.025\,L$ in terms of shear stress.
However, the dotted and dashed curves vary significantly from time-to-time until $t>0.0033\,\si{s}$.
Similarly, the PEEQ (Figure~\ref{fig:PEEQ_vertical_steady}) results at $x=1.09\,L$ (dashed curves) do not indicate steady-state conditions until time $t>0.0033\,\si{s}$.
It is therefore clear that the through-thickness stress and strain states do not reach a steady state at all $x$-positions until $t>0.0033\,\si{s}$.
However, as pointed out earlier, \figref{fig:roll_force_torque} suggests that roll force and torque reach a steady value by $t = 0.001\,\si{s}$, highlighting the inadequacy of roll force and torque in terms of predicting the steady state of the rolling process when through-thickness accuracy is desired since stress and strain components are still significantly changing at $t = 0.001\,\si{s}$, as shown here.
Despite this inadequate steady-state determination, roll force and torque are used by various authors \citep[see, for example,][]{exp_vs_imp,minton2017mathematical} for determining if a steady state has been reached.

The interpolated results from the last three time points in Figure~\ref{fig:steady_state_CPE4R} are indistinguishable for all stress and strain quantities at all three $x$-positions, and so we can say with confidence that a steady state is reached by $t=0.0802\,\si{s}$.
Hence, we extract the data used for the remainder of this paper at $t=0.0802\,\si{s}$ to both ensure steady results and to facilitate the velocity calculation in the postprocessing (see~\ref{vel_app}).

\section{Numerical results}
\label{results}
In this section, we present some notable outputs from our FE simulations, highlighting the emergence of a deterministic oscillatory pattern in the shear, PEEQ rate and velocity profiles of metal sheets during cold rolling.
We also compute the residual stress in the sheet after rolling and investigate its connection to the oscillatory pattern in stress and strain quantities in the roll gap.
This connection is likely to have a strong relationship with curvature in asymmetric rolling, a phenomenon that is often incorrectly predicted \citep{minton2017meta}.

\subsection{Oscillations}\label{results.new}
\subsubsection{Shear lobes}

\begin{figure*}
    \vspace{-2.95cm}
    \centering
    \begin{minipage}{0.44\textwidth}
        \centering
        \begin{subfigure}{\textwidth} 
            \centering
            \includegraphics[width=\linewidth]{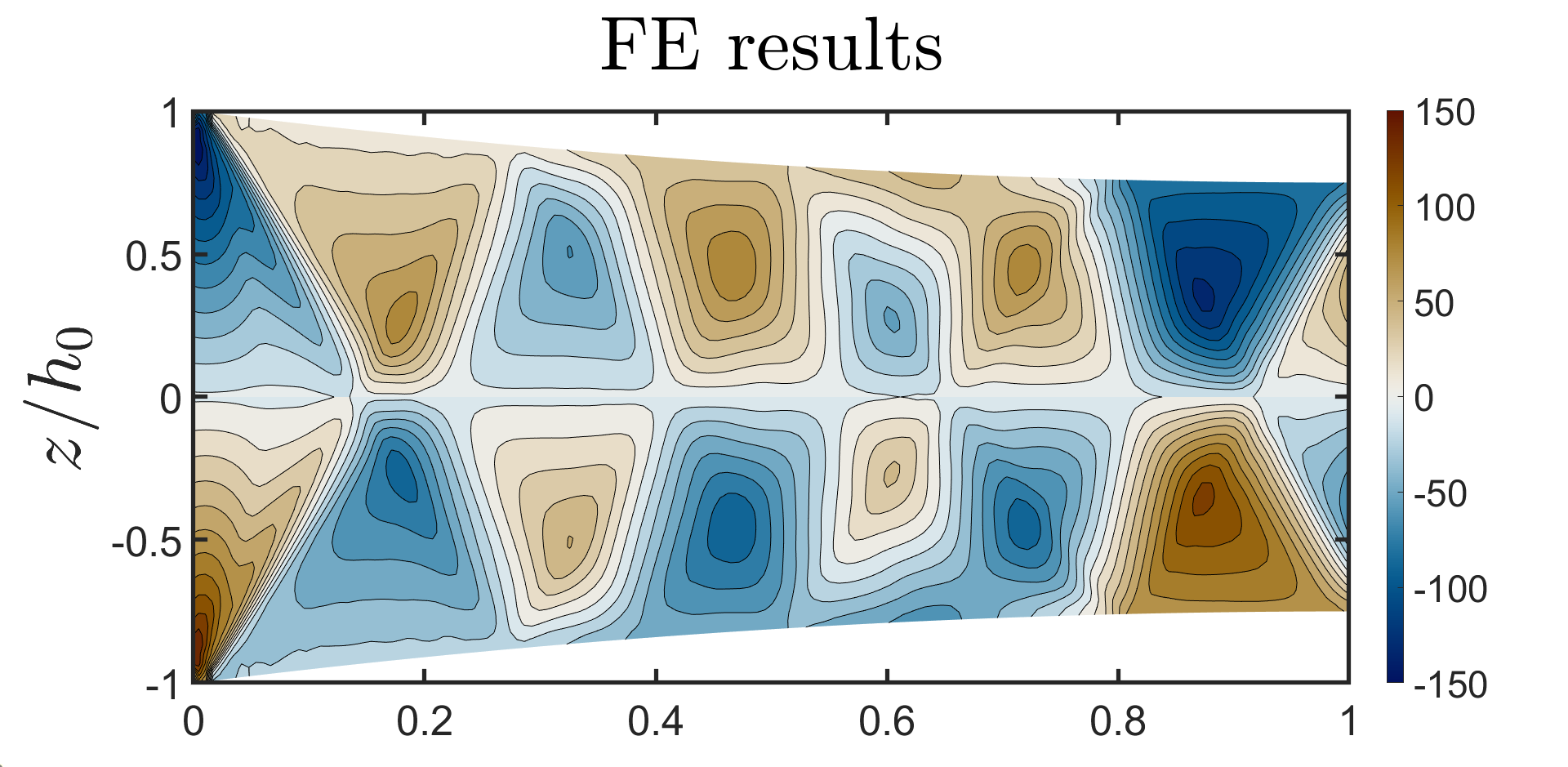}
            \caption{}
            \label{fig:7a}
        \end{subfigure}\hfill
        \begin{subfigure}{\textwidth} 
            \centering
            \includegraphics[width=\linewidth]{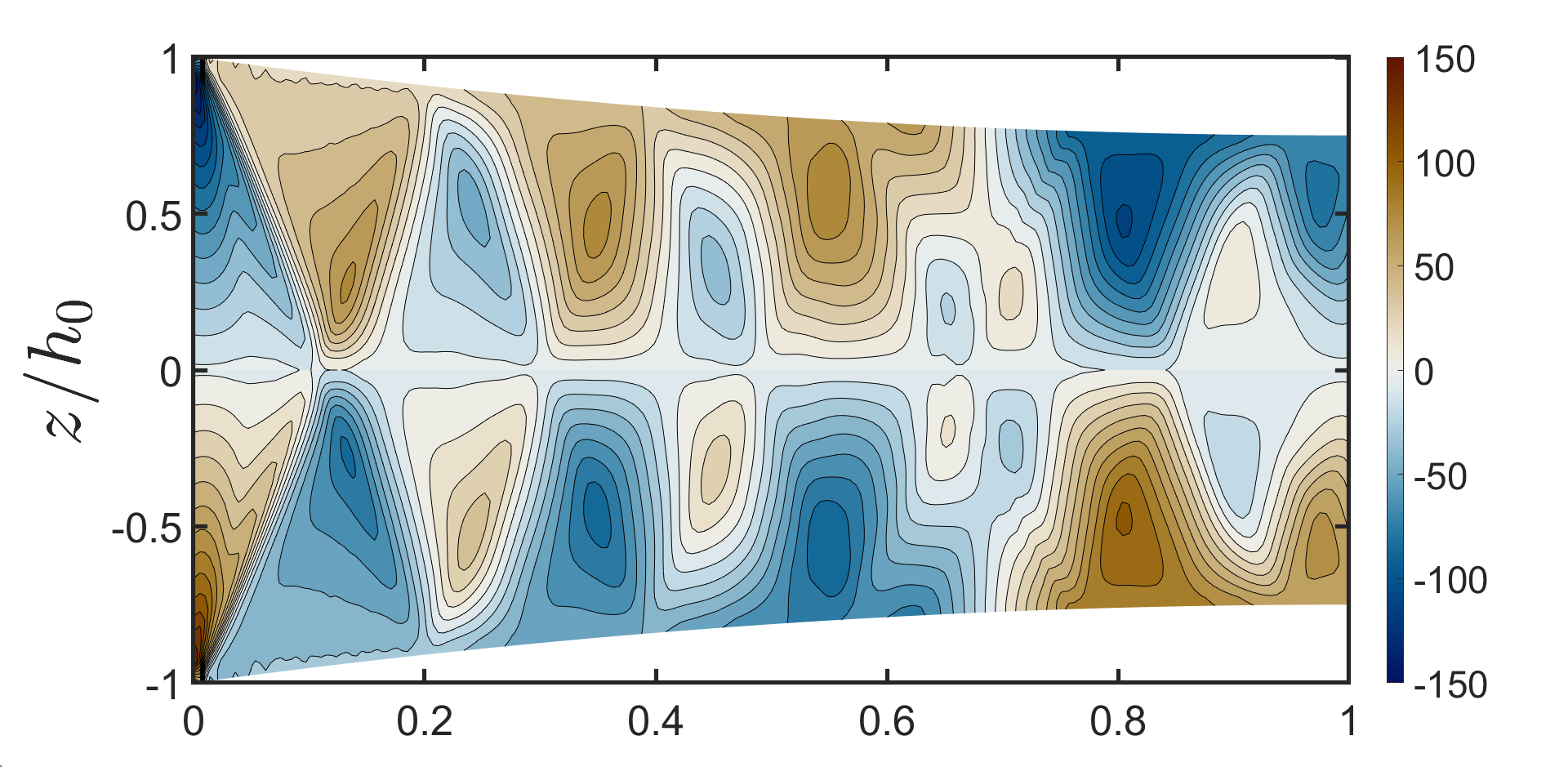}
            \caption{}
            \label{fig:7b}
        \end{subfigure}\vspace{0.05cm}
        \begin{subfigure}{\textwidth} 
            \centering
            \includegraphics[width=\linewidth]{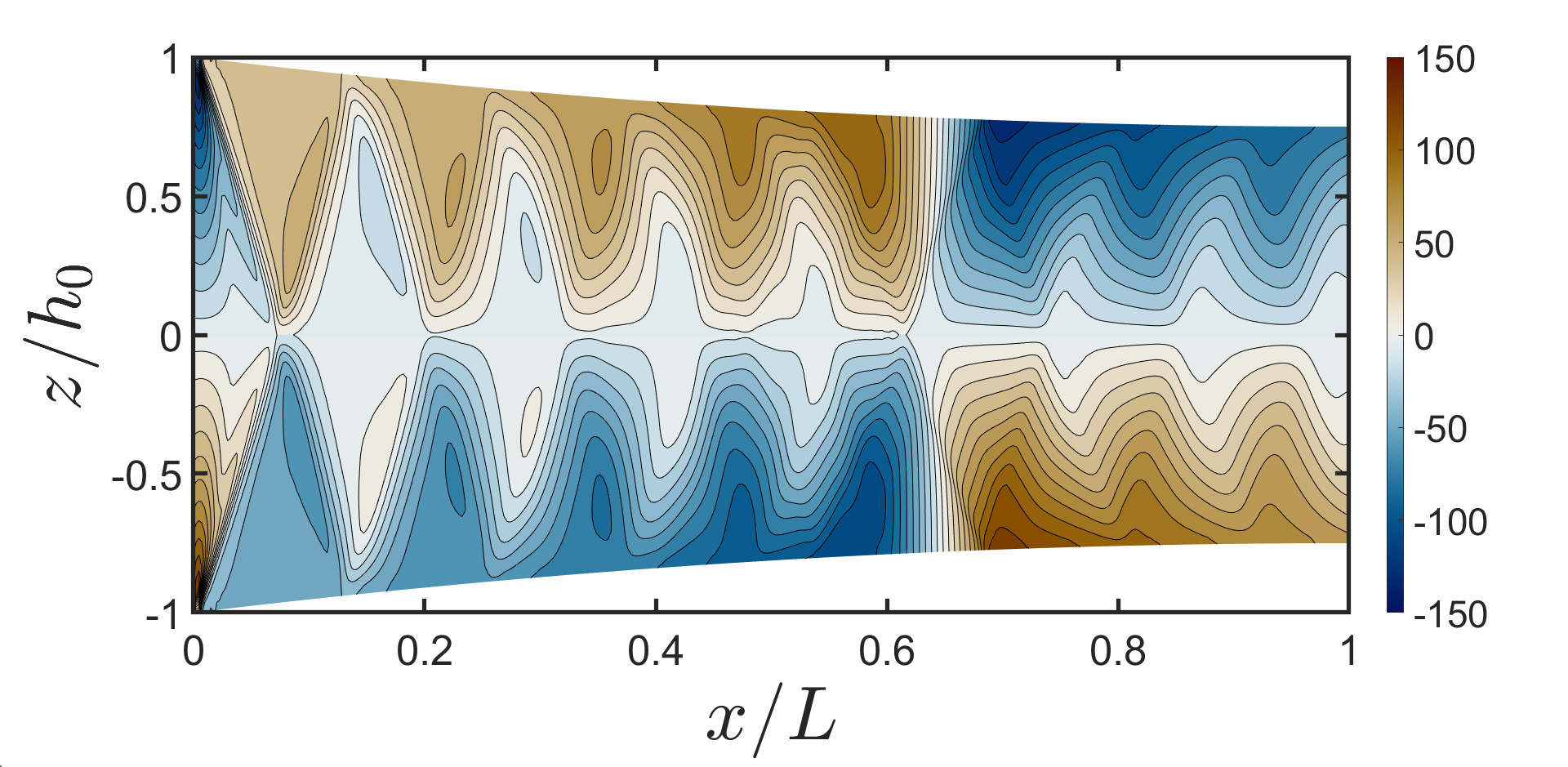}
            \caption{}
            \label{fig:7c}
        \end{subfigure}
    \end{minipage}\hfill
    \begin{minipage}{0.12\textwidth}
        \vspace{4.8cm} 
        \parbox{6.2cm}{
            \scriptsize
            $R=130.6\,\si{mm}$, \\$\varepsilon = 0.175$ \\[2.9cm] 
            $R=257.45\,\si{mm}$, \\$\varepsilon = 0.125$ \\[2.9cm] 
            $R=710\,\si{mm}$, \\$\varepsilon = 0.075$
        }
        \rule{0pt}{5.4cm}
    \end{minipage}\hfill
    \begin{minipage}{0.44\textwidth}
        \centering
        \begin{subfigure}{\textwidth} 
            \centering
            \includegraphics[width=\linewidth]{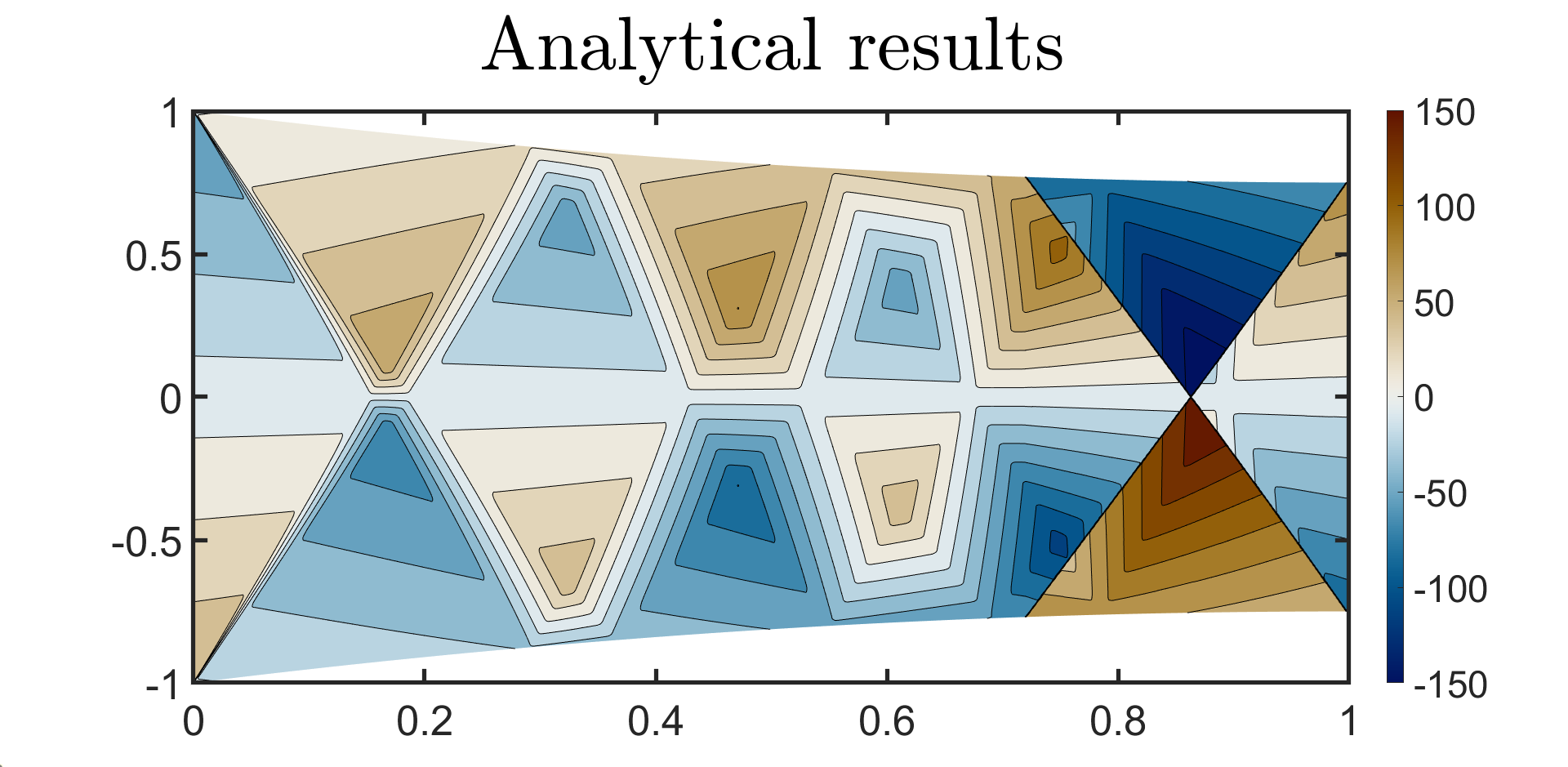}
            \caption{}
            \label{fig:7d}
        \end{subfigure}\hfill
        \begin{subfigure}{\textwidth} 
            \centering
            \includegraphics[width=\linewidth]{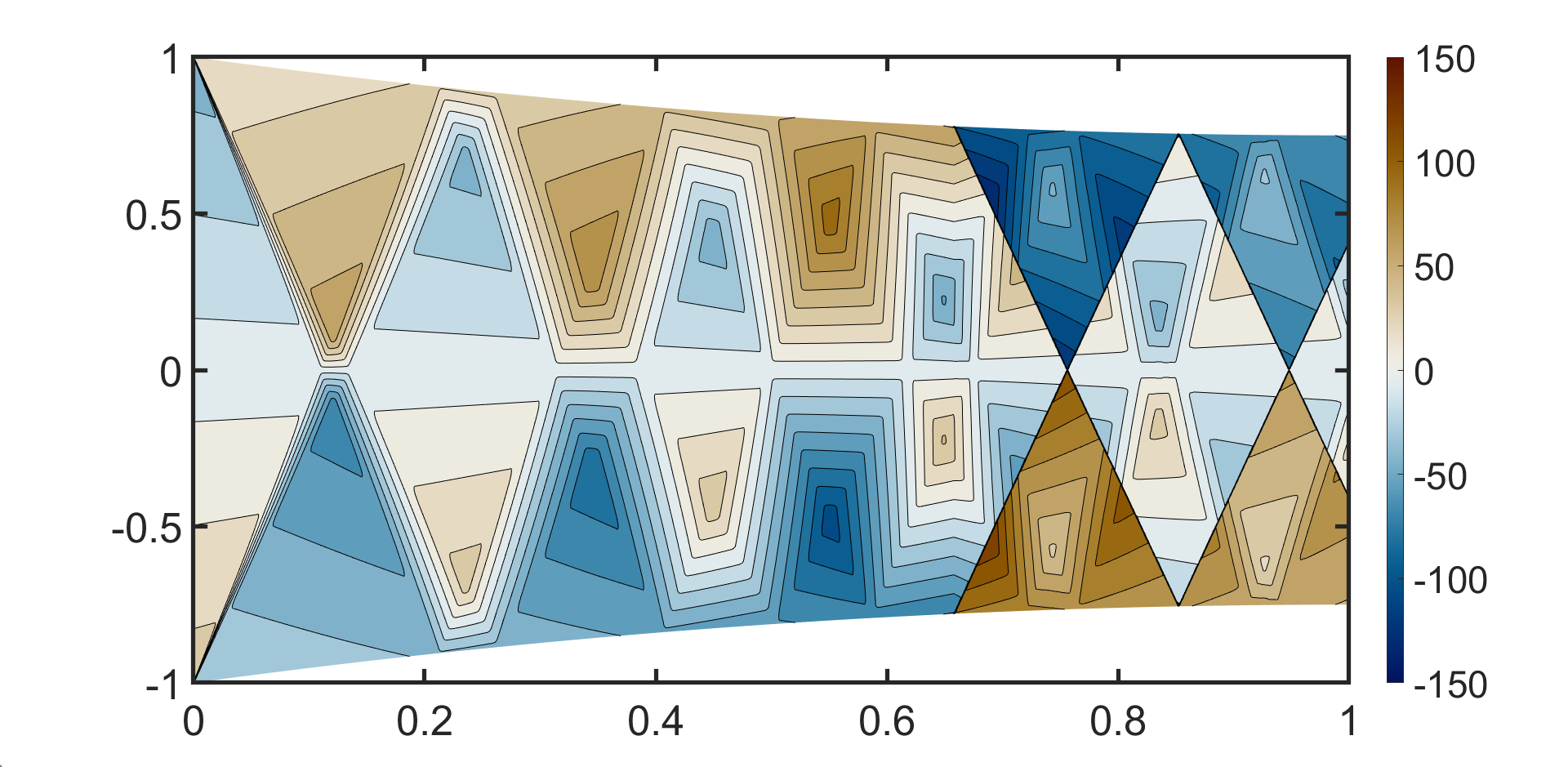}
            \caption{}
            \label{fig:7e}
        \end{subfigure}\vspace{0.05cm}
        \begin{subfigure}{\textwidth}
            \centering
            \includegraphics[width=\linewidth]{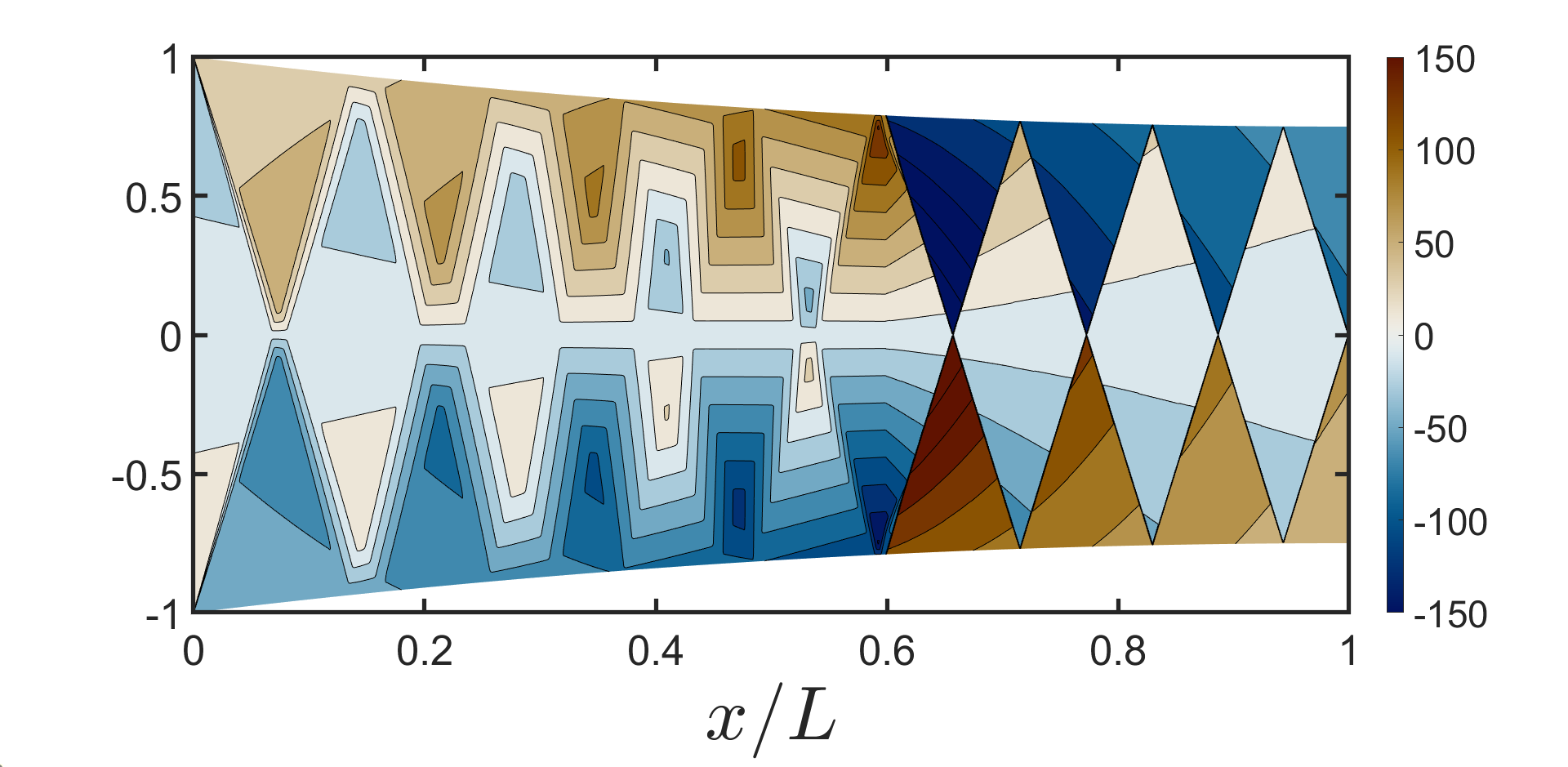}
            \caption{}
            \label{fig:7f}
        \end{subfigure}
    \end{minipage}
    \vspace{-2.7cm}
    \caption{Contour plots of shear stress results from simulations with $N_e=30$ elements, where the symmetry axis is located at $z/h_0=0$ (see Figure~\ref{fig:rolling_pic}). The left panel (a-c) shows FE simulations and the right panel (d-f) shows results from the analytical model of \citet{Mozhdeh_journ}, for three different $\varepsilon$ values: $\varepsilon=0.175$ (top row, $R=130.6\,\si{mm}$), $\varepsilon=0.125$ (middle row, $R=257.45\,\si{mm}$) and $\varepsilon=0.075$ (bottom row, $R=710\,\si{mm}$). Data is presented for the roll gap only, between $x/L=0$ and $x/L=1$, in all cases since \citet{Mozhdeh_journ} do not consider deformations outside the roll gap. The neutral point is the horizontal $x/L$ position at which the surface shear stress changes sign.}
    \label{fig:shear_FE_analytical}
\end{figure*}%
The left column of \figref{fig:shear_FE_analytical} shows our FE prediction of shear stress in sheets undergoing rolling in three different roll gaps corresponding to aspect ratios $1/\varepsilon = 1/0.175$, $1/0.125$ and $1/0.075$.  In each of the three cases, we observe a clear oscillatory pattern, with the shear stress varying from $-150\si{MPa}$ to $+150\si{MPa}$, and with lobes of positive and negative shear stress appearing with regular spacing on both sides of the sheet mid-surface.  The number of lobes increases as $1/\epsilon$ increases, which is consistent with preliminary findings from~\cite{minton2017mathematical}.
These lobes are not typically reported in FE modelling of cold rolling (c.f.\ Table~\ref{table:FE_settings}) and could be mistaken for numerical error, but they have also been predicted by recent analytical work (c.f.\ the right column of Figure~\ref{fig:shear_FE_analytical} and Section~\ref{results.2}).   

\subsubsection{Deforming and non-deforming zones}

\begin{figure}
    \centering
        \centering
        \begin{subfigure}{\linewidth}
            \centering
            \includegraphics[width=\linewidth,height=0.25\textheight,keepaspectratio]{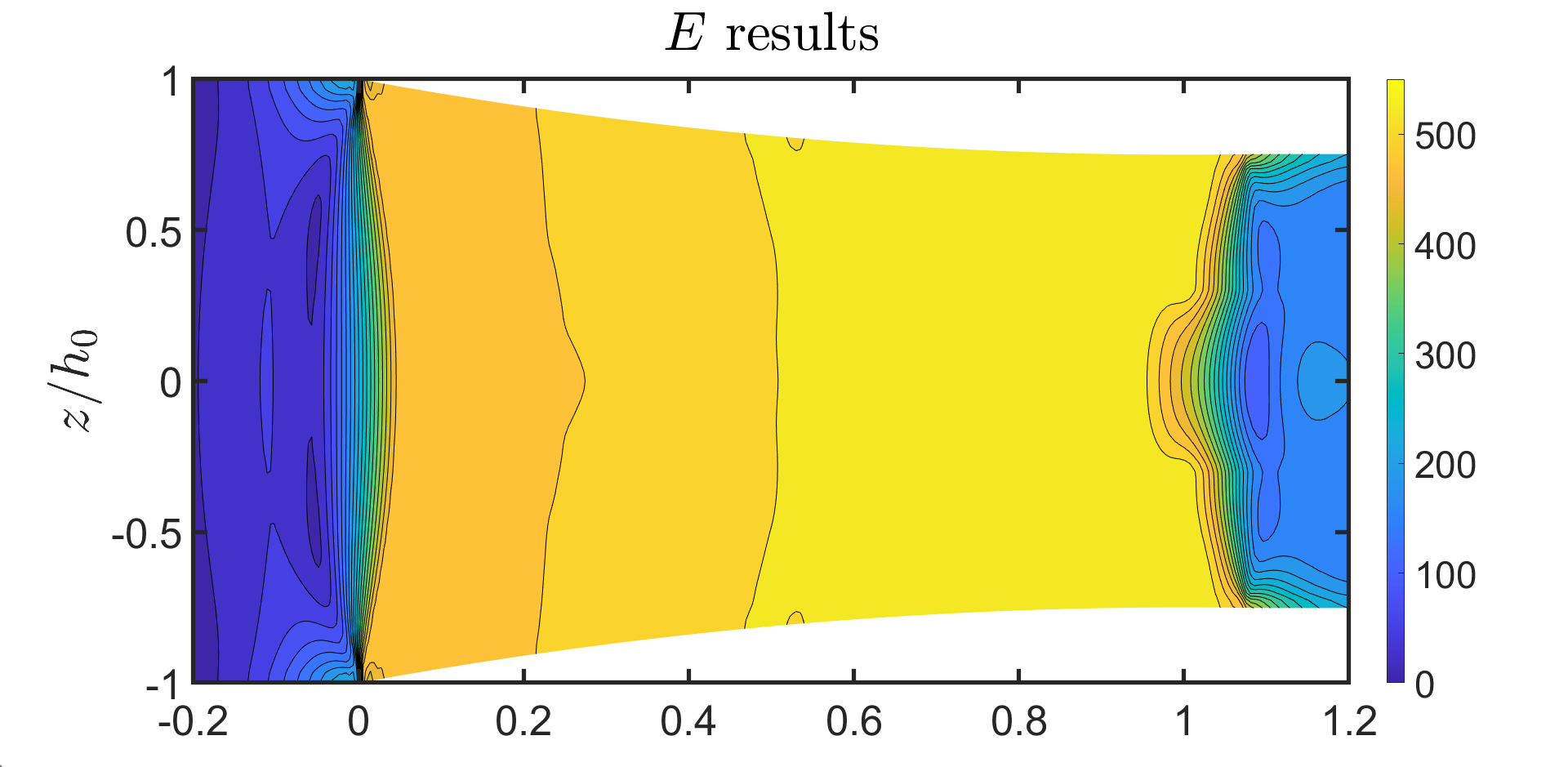}
            \caption{}
            \label{fig:vM_E}
        \end{subfigure}\hfill
        \begin{subfigure}{\linewidth}
            \centering
            \includegraphics[width=\linewidth,height=0.25\textheight,keepaspectratio]{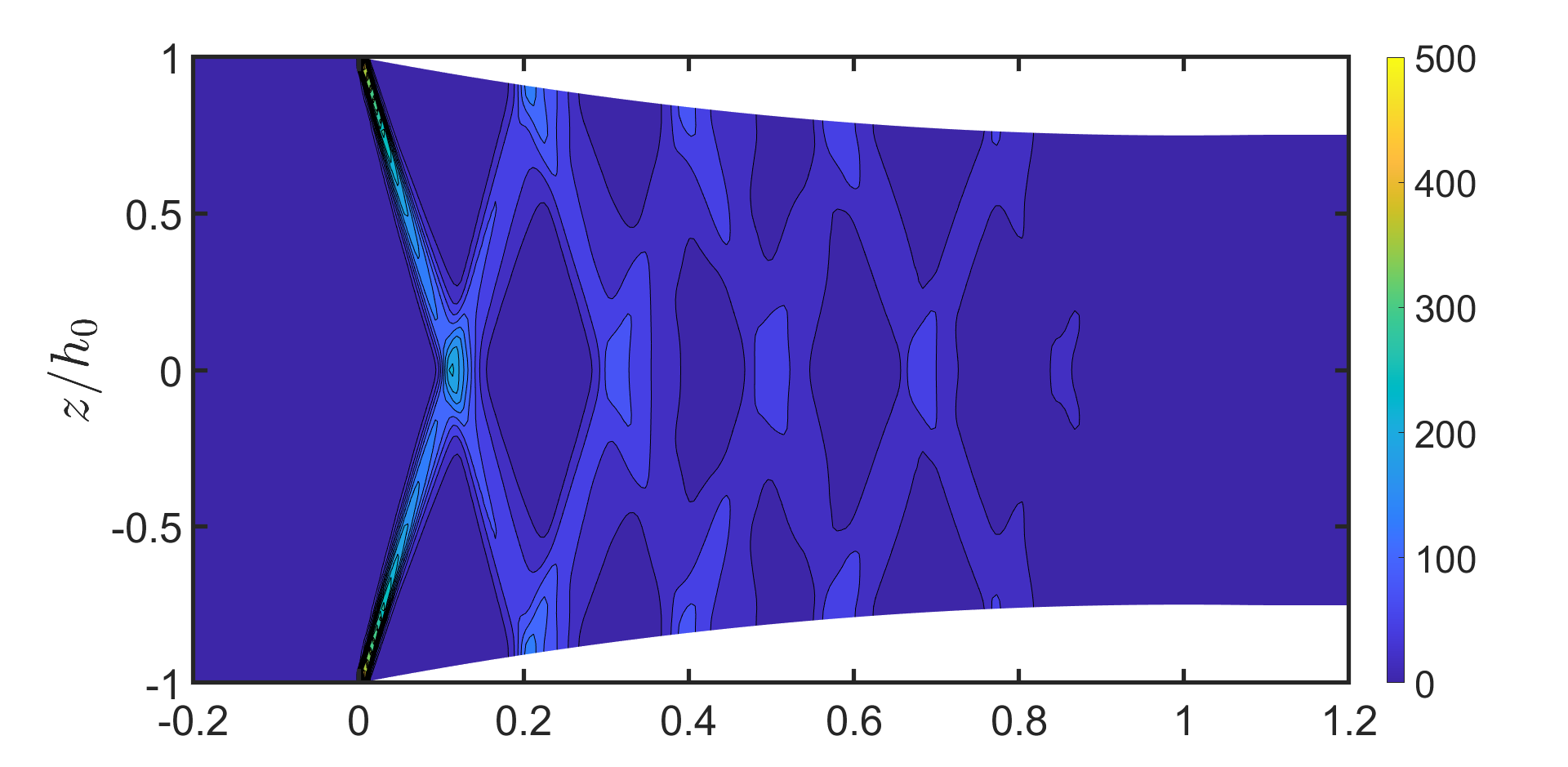}
            \caption{}
            \label{fig:PEEQ_rate_E}
        \end{subfigure}\hfill
        \begin{subfigure}{\linewidth}
            \centering
            \includegraphics[width=\linewidth,height=0.25\textheight,keepaspectratio]{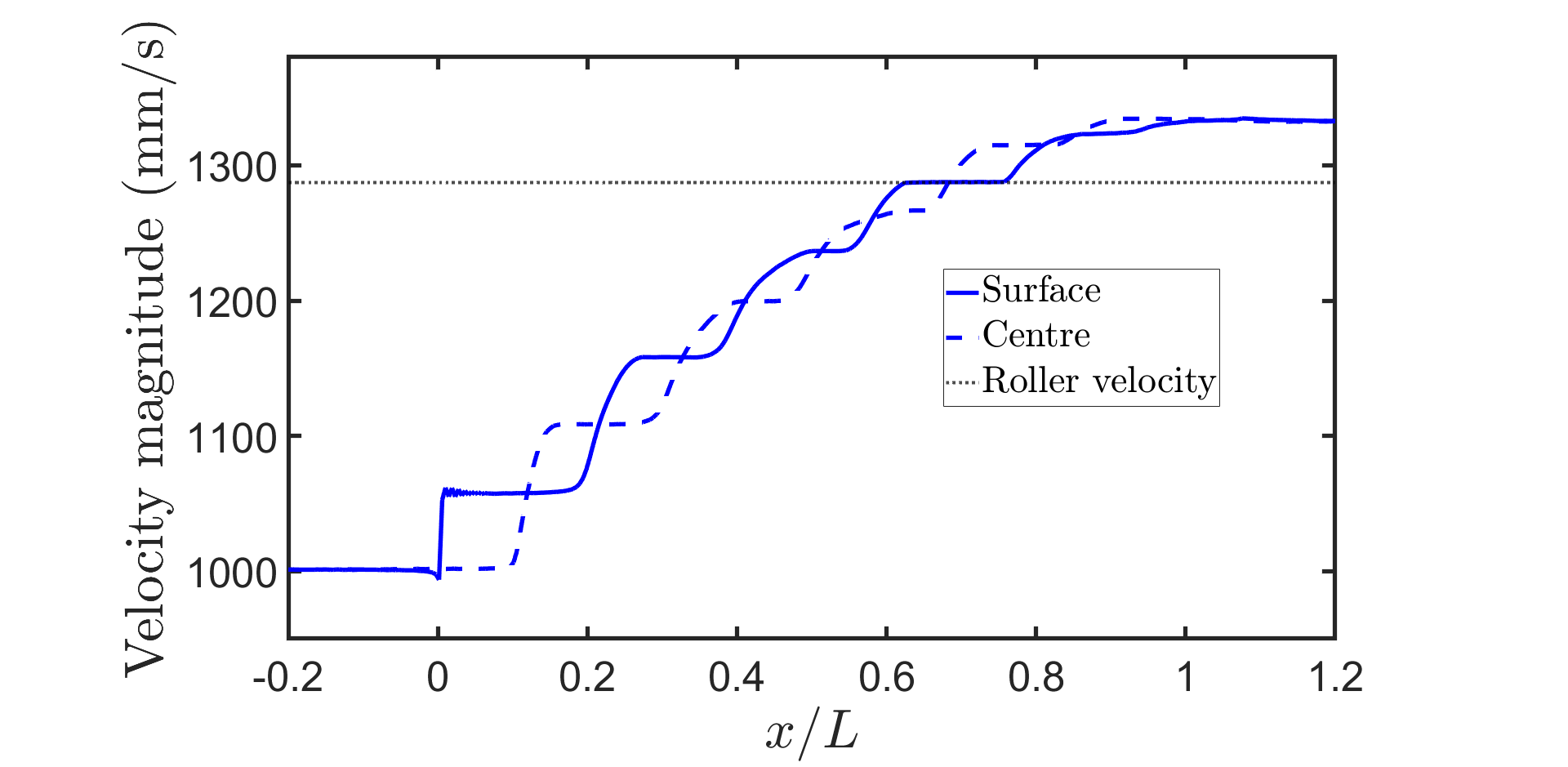}
            \caption{}
            \label{fig:vel_E}
        \end{subfigure}
    \caption{(a) Von Mises stress, (b) PEEQ rate and (c) velocity profiles for the simulation presented in \figref{fig:7b} ($R=257.45\,\si{mm}$, $\varepsilon = 0.125$). In the contour plots, the symmetry axis is located at $z/h_0=0$ (see Figure~\ref{fig:rolling_pic}).
    The PEEQ rate and velocity values are obtained through postprocessing of the PEEQ values and displacements respectively (see~\ref{vel_app}).}
    \label{fig:yield_plots_part1}
\end{figure}%
\figref{fig:yield_plots_part1} shows von Mises stress and PEEQ rate contour plots, along with the sheet's surface and central velocity magnitudes, for the simulation presented in \figref{fig:7b} ($R=257.45\,\si{mm}$, $\varepsilon = 0.125$).
The PEEQ rate results were calculated in a similar manner to the velocities (see~\ref{vel_app}).
From \figref{fig:vM_E}, we see that the von Mises stress does not drop below yield in the interior of the roll gap, as expected.
However, the PEEQ rate plot in \figref{fig:PEEQ_rate_E} reveals a far richer and more complex layer of dynamics.  A clear wave-like structure can be seen, mirroring the oscillatory pattern in shear in \figref{fig:shear_FE_analytical}.  The plastic deformation is highly localized in very thin bands, oriented diagonally and oscillating back and forth across the sheet.
In the regions between these bands, where the PEEQ rate is zero, the material appears to translate as a solid body, with no plastic flow, even though it is at yield.
This is unexpected and may have major implications for hardening and the development of microstructure in the rolled sheet.

We investigate the emergence of rapidly-deforming and non-deforming zones further by plotting the velocity magnitude at the surface and centre of the sheet in \figref{fig:vel_E} (solid and dashed curves, respectively).
The speed of the roller (black dotted line) is also shown for comparison.
According to these curves, the sheet surface initially accelerates when it enters the roll gap and comes into contact with the roller, as expected from Coulomb friction, while the sheet centreline remains at its initial speed.
{As shown in Figure~\ref{fig:vel_E} the velocity at the sheet centre does not increase until $x/L \approx 0.1$ while the sheet surface velocity begins to increase at $x/L \approx 0$.
From Figure~\ref{fig:shear_FE_analytical} we know that the shear stress has a non-linear distribution through the sheet thickness, and this delayed response to velocity changes is therefore expected.}
{The velocity of the material at the sheet's horizontal centre does not change until the information from the surface reaches the sheet's centre.
The acceleration is transmitted to the centre of the sheet via the shear, as shown in \figref{fig:shear_FE_analytical}, so it begins to accelerate later.} 
However, the surface then stops accelerating, in line with the translation-only (i.e., zero PEEQ rate) region observed in the PEEQ rate results.
This means that the centreline temporarily accelerates to a faster velocity than the surface.
This behaviour is, again, somewhat unexpected, since one might intuitively expect the material velocity to be highest near the sheet surface, where the forcing is imposed.
The pattern then repeats, and both regions move through consecutive intervals of acceleration and rigid translation, with the respective velocities leap-frogging each other repeatedly.   Furthermore, we observe that one of the velocity plateaus on the surface coincides with the roll velocity, so a ``stick'' region emerges naturally around the neutral point\footnote{
The neutral point is defined as the position at which friction changes direction on the surface (see Figure~\ref{fig:shear_FE_analytical}). }.
This allows the shear stress to smoothly change from positive to negative on the surface as the sheet sticks to the rollers for a small portion of the roll gap, as was observed at the sheet's surface near the neutral point in the FE results in Figures \ref{fig:7a}--\ref{fig:7c}.

\subsection{Residual stress}
\label{residual_stress}

\begin{figure}
\centering
\begin{subfigure}[b]{\linewidth}
   \centering
   \includegraphics[width=\linewidth]{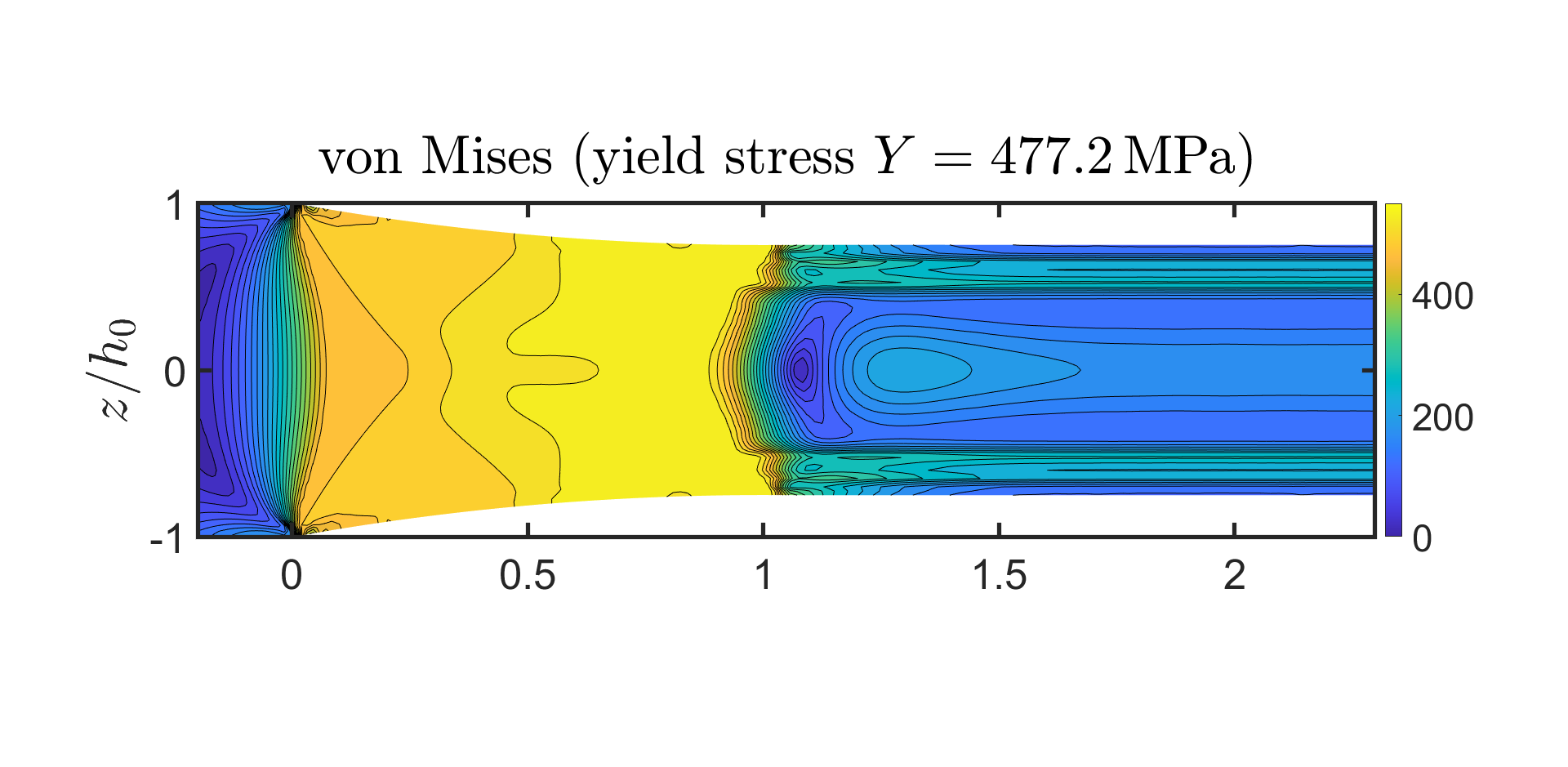}
   \vspace{-0.1\linewidth}
   \caption{}
   \label{fig:symmetric_vonMises_sliplines} 
\end{subfigure}
\begin{subfigure}[b]{\linewidth}
   \centering
   \includegraphics[width=\linewidth]{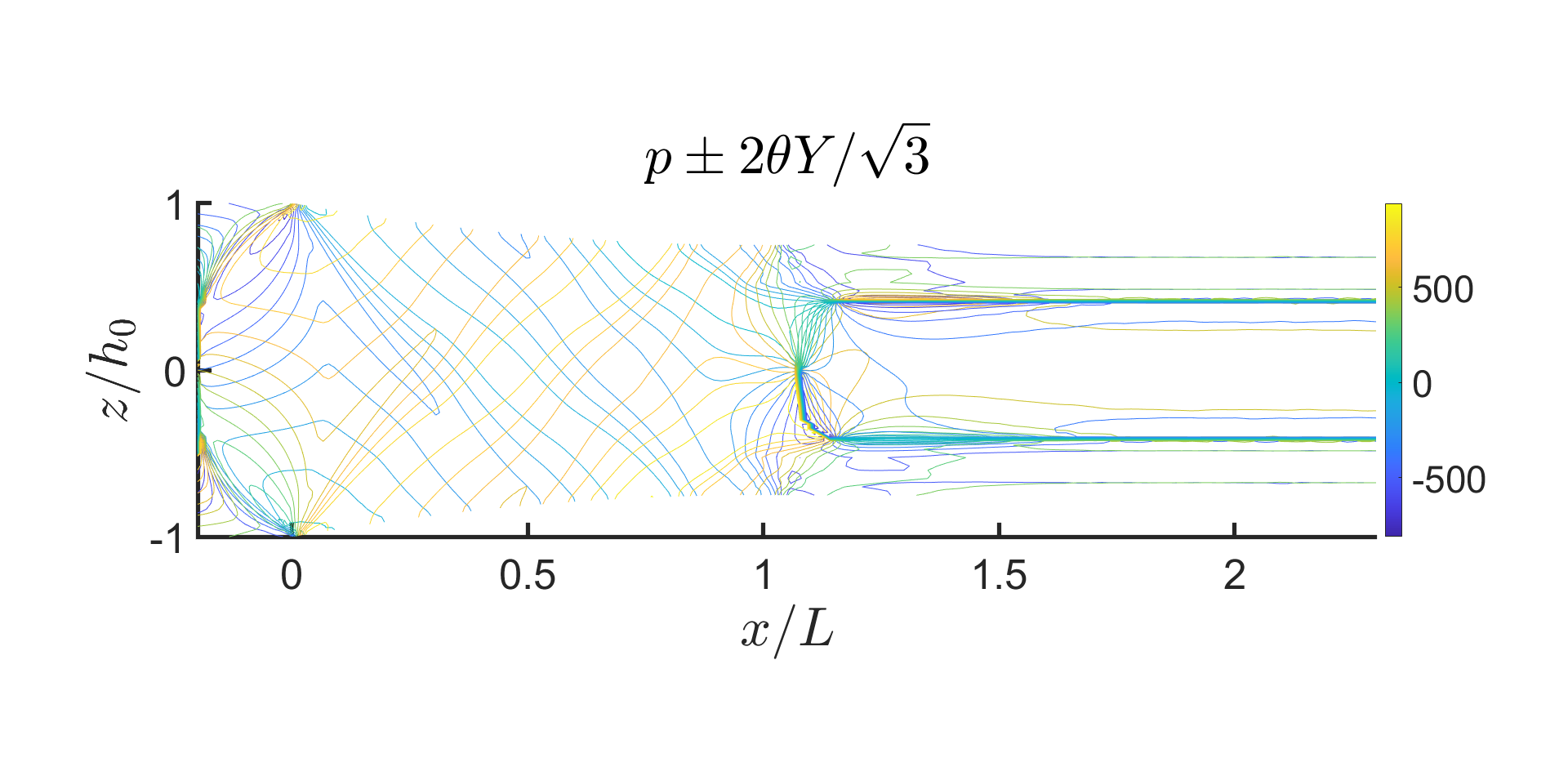}
   \vspace{-0.1\linewidth}
   \caption{}
   \label{fig:symmetric_sliplines}
\end{subfigure}
\caption{Contour plots of (a) von Mises and (b) $p\pm 2\theta Y/\sqrt{3}$, all given in $\si{MPa}$, from a symmetric implicit simulation with $N_e=30$ CPE4R elements, where the symmetry axis is located at $z/h_0=0$ (see Figure~\ref{fig:rolling_pic}).
The radius is $R=32\,\si{mm}$ which gives $\varepsilon = 0.354$.
The tangential velocity of the rollers is $R \Omega = 1287.25\,\si{mm \,s^{-1}}$.}
\label{fig:symmetric_residual stress}
\end{figure}%
The through-thickness variation of the stress inside the roll gap illustrated in \figref{fig:yield_plots_part1} has a knock-on effect on residual stress after rolling.
In \figref{fig:symmetric_residual stress} we consider a symmetric simulation with a roller radius $R=32\,\si{mm}$, which leads to $\varepsilon = 0.354$, and all other parameters are the same as described in Section~\ref{sim_details}.
The length--height ratio of these contour plots is set to the true ratio used in the simulation, for ease of interpretation.
We measure the von Mises stress and the quantities $p\pm 2\theta Y/\sqrt{3}$, where $p=-0.5\left(\sigma_{xx}+\sigma_{zz}\right)$ is the hydrostatic pressure and $\theta$ is the anticlockwise rotation angle of the stress element from the positive $x$-axis for which the stress is in a state of pure shear. 
The von Mises stress in the sheet after it passes through the roll gap can be used to give an indication of residual stress.
For this reason the plotted domain is extended beyond the roll gap ($x/L>1$).
The quantities $p\pm 2\theta Y/\sqrt{3}$ are of interest because these quantities are constant along two orthogonal sets of characteristic lines for plastic deformation, and the direction of these characteristic lines coincide with the lines of maximum shear stress (slip lines), but only when plastic deformation is occurring (see~\ref{sliplines_app} for more information).

Here, we are concerned with residual stress, seen for $x/L>1.5$ in \figref{fig:symmetric_vonMises_sliplines}, where the material is far from the external loading (i.e., the rollers). 
We observe distinct zones in the centre and at the edges; cross-checking against the individual stress components reveals that the central zone experiences lengthwise compression while the outer zones are under longitudinal tension.
This is consistent with longitudinal stress results reported by \citet{tadic2023analysis}.
The residual von Mises stress is highest in narrow horizontal bands in between the sheet surface and the sheet centre.
In the portions of the sheet that are at yield in \figref{fig:symmetric_vonMises_sliplines}, the contour lines of $p\pm 2\theta Y/\sqrt{3}$ in \figref{fig:symmetric_sliplines} meet at $90^{\circ}$, as expected theoretically~\citep{johnson2013plane}. These are the slip lines.  In the sub-yield zone, the contour lines are not slip lines and do not meet at $90^{\circ}$.
The $p\pm 2\theta Y/\sqrt{3}$ contour lines oscillate through the roll gap in a similar manner to the shear lobes discussed earlier, which is unsurprising given they line up with the directions of maximum shear stress during plastic flow.
After oscillating through the roll gap, the $p\pm 2\theta Y/\sqrt{3}$ contour lines also accumulate within a thin layer that roughly lines up with the regions of largest residual von Mises stress noted above.
This indicates that a strong relationship exists between the oscillations in shear stress in the roll gap and residual stress outside the roll gap.

\subsection{Asymmetric rolling}
\label{asymmetric_rolling}
We hypothesise
that the connection between shear stress in the roll gap and residual stress significantly far away from the roll gap has implications
for curvature prediction in asymmetric rolling.
Here, by way of a preliminary investigation,
we consider an asymmetric simulation with rollers of
radius $R=32\,\si{mm}$, where the top roller rotates 2.5\% faster than the bottom roller.
The bottom roller rotates with the same tangential velocity as before ($R \Omega = 1287.25\,\si{mm \,s^{-1}}$), and all other parameters are the same as described in Section~\ref{sim_details}.

\begin{figure}%
\centering%
\begin{subfigure}[b]{\linewidth}%
   \centering%
   \includegraphics[width=\linewidth]{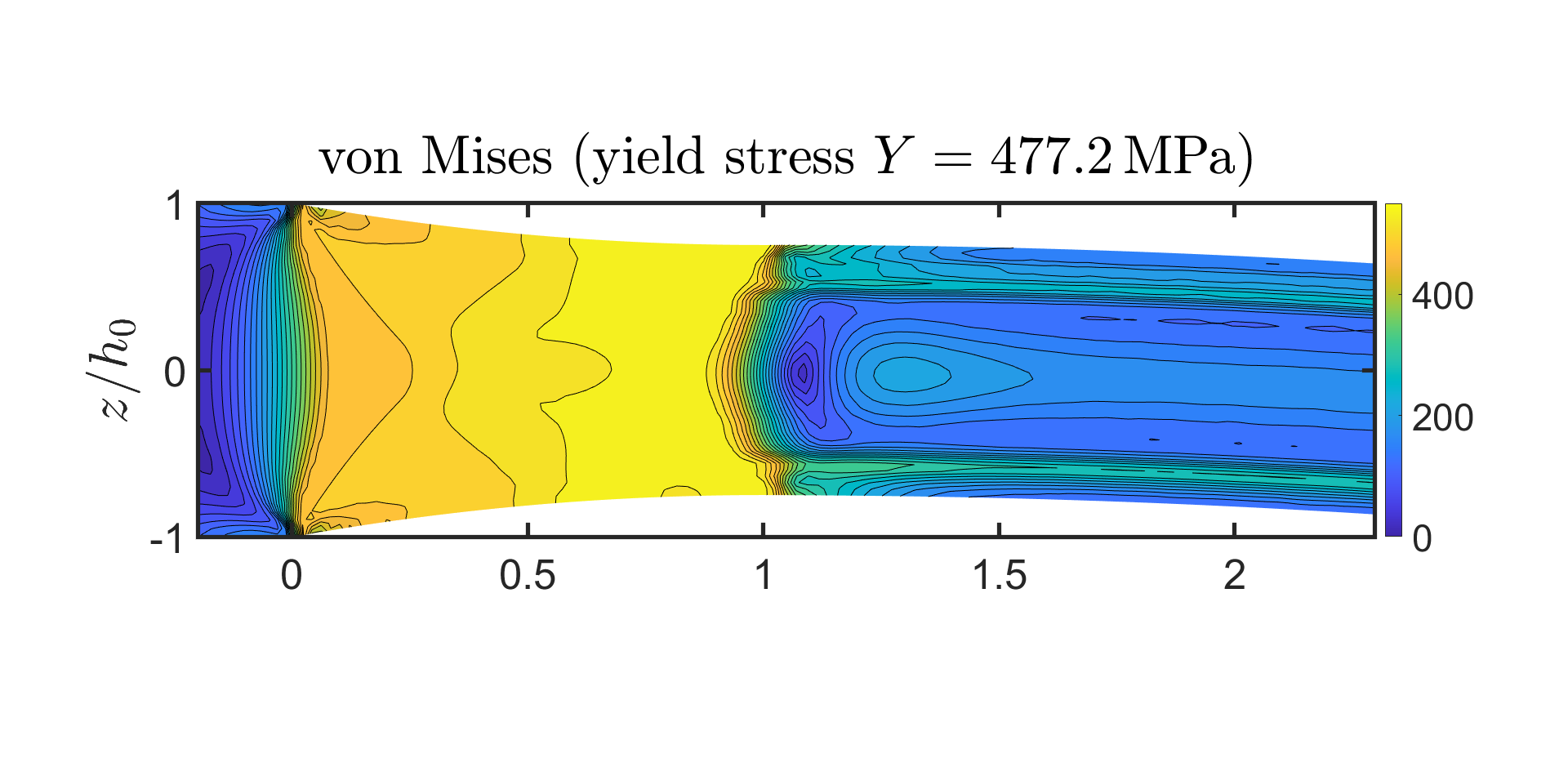}%
   \vspace{-0.1\linewidth}%
   \caption{}%
   \label{fig:vonMises_sliplines}%
\end{subfigure}
\begin{subfigure}[b]{\linewidth}
   \centering
   \includegraphics[width=\linewidth]{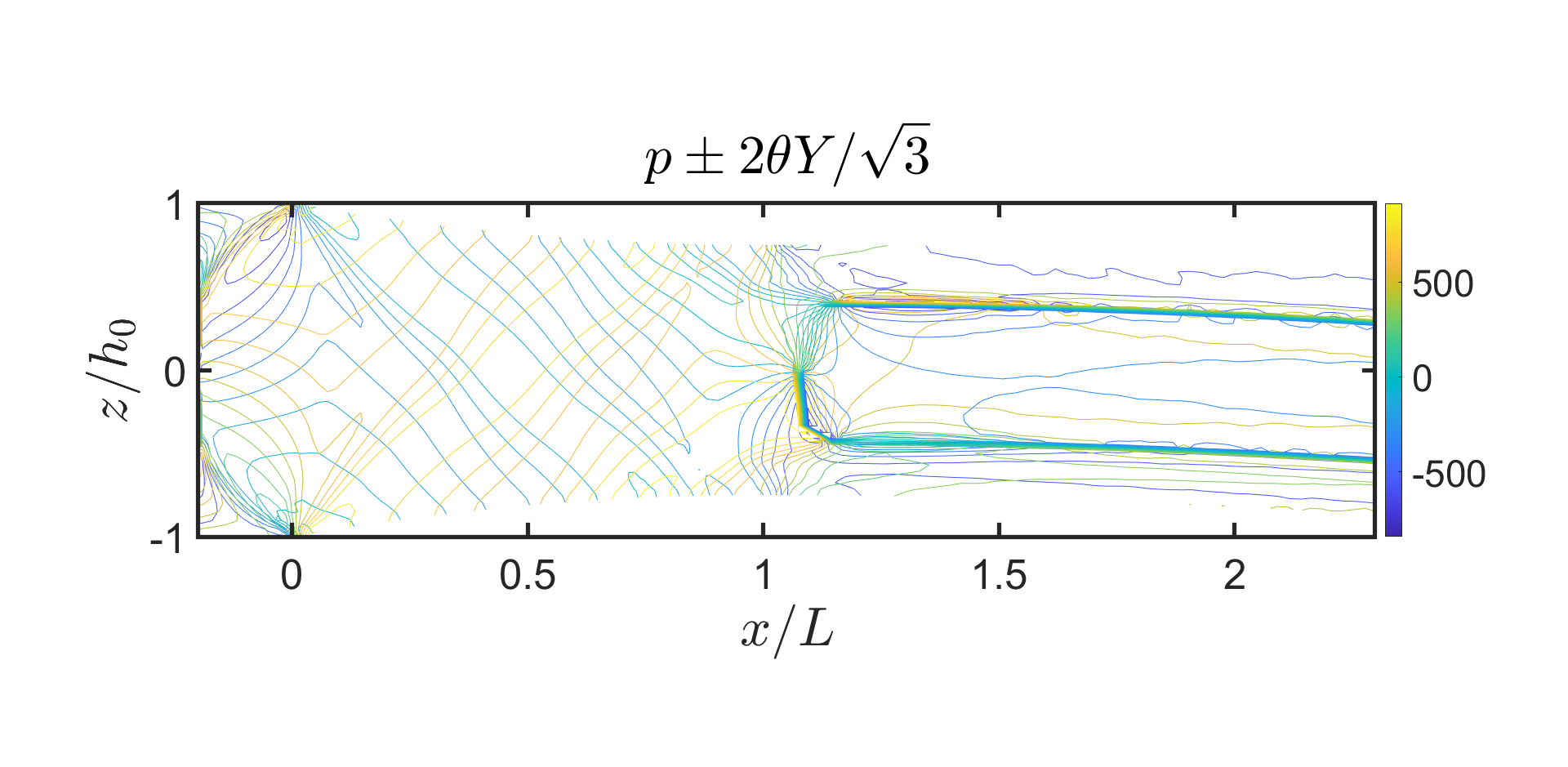}
   \vspace{-0.1\linewidth}
   \caption{}
   \label{fig:sliplines}
\end{subfigure}
\caption{Contour plots of (a) von Mises and (b) $p\pm 2\theta Y/\sqrt{3}$, given in $\si{MPa}$, from an implicit simulation with $2N_e=40$ CPE4R elements and asymmetric roller speeds. 
The radius is $R=32\,\si{mm}$ which gives $\varepsilon = 0.354$.
The top roller rotates 2.5\% faster than the bottom roller.
The tangential velocity of the bottom roller is $R \Omega = 1287.25\,\si{mm \,s^{-1}}$.} \label{fig:asymmetric}
\end{figure}%
The asymmetric results shown in \figref{fig:asymmetric} are obtained via a simulation using $40$ through-thickness elements%
\footnote{The mesh density comparable to the symmetric case above would have required 60 elements through thickness, which was too computationally expensive to be achievable for this asymmetric case. }.
Large amounts of residual stress are observed in the asymmetric von Mises results in \figref{fig:vonMises_sliplines}, similar to what was noted in the symmetric case (\figref{fig:symmetric_vonMises_sliplines}).
Once again the contour lines of $p\pm 2\theta Y/\sqrt{3}$ accumulate within a thin layer that lines up with these regions of largest residual stress.
The magnitude of the residual stress in the sheet for $x/L>1$ is greater towards the bottom of the sheet, and the sheet's curvature is towards the slower roller for this specific set of parameters.
We do not suggest that curvature is always towards the slower roller because, as pointed out by \citet{minton2017meta}, many authors have given contradicting prediction rules for curvature during asymmetric rolling. 
Rather we hypothesise that the oscillatory pattern in stress and strain quantities presented here, which is largely absent in the literature, has a strong influence on the accurate prediction of the residual stress for $x/L >1$, which may have a significant impact on predicting curvature in asymmetric rolling. 
Indeed, the shifting of the oscillatory pattern as parameters are varied likely explains the inconsistent and poorly-understood trends in curvature reported in the literature~\citep{minton2017meta}.  
This is an important aspect of modelling rolling since
unintentional asymmetries cause turn-up and turn-down in sheet rolling which can halt the process or even damage the rolling table and mills and since asymmetries are intentionally exploited in asymmetrical rolling to improve process efficiency
and produce curved products \citep{minton2017meta}.

\section{Further validation}
\label{results.2}
Many of the numerical observations presented in Section~\ref{results} are unexpected or unintuitive, so we provide further validation in this section by comparing with a new analytical method developed using the method of multiple scales by~\citet{Mozhdeh_journ}.
This new model yields a wave-like solution for the stresses in the sheet, where information travels through the sheet along diagonal slip lines and is reflected at the sheet surface, yielding a characteristic zig-zag pattern as illustrated in Figures~\ref{fig:7d}--\ref{fig:7f}.
This multiple-scales model is valid in the limit where the roll gap is very long and thin, i.e.\ for $\varepsilon \to 0$.  We compare our FE predictions of shear stress with the multiple-scales prediction by comparing the left and right columns of \figref{fig:shear_FE_analytical}.
As \citet{Mozhdeh_journ} do not consider deformations outside the roll gap, our attention here is restricted to values of $x/L$ between 0 and 1 (see \figref{fig:rolling_pic} for the domain definition).
Both methods predict repeating lobes of high and low shear stress, as outlined above.
The number and position of lobes predicted by the two methods are in agreement for each value of $\varepsilon$, and the magnitude of shear stress at various through-thickness positions is, in general, comparable.
There are some differences between the predictions from the two methods, which we outline and explain now.

\subsection{Through-thickness shear profile}
One notable difference between the FE and analytic models is in the shear profiles, which can be seen between the neutral point and the end of the roll gap.
In \figref{fig:shear_FE_analytical} the neutral point is observed in the range $x\approx0.6L\text{--}0.75L$ in all contour plots as the $x$-position at which the shear changes sign on the surface of the sheet.
In the FE results, shear stress on the surface varies in a smooth fashion while the analytical model demonstrates a distinct discontinuity in shear stress.
The discontinuous change in shear stress seen in the analytical model is due to the slipping Coulomb friction law that is employed.
In reality, material sticks to the roller around the neutral point (as shown in \figref{fig:vel_E} above) and shear stress changes sign smoothly \citep{domanti1995two,cawthorn}.
This sticking region around the neutral point is predicted in our FE results, but not in the rigid-plastic
analytical model.

\subsection{Stick--slip region}
\begin{figure}
\centering
\includegraphics[width=\linewidth]{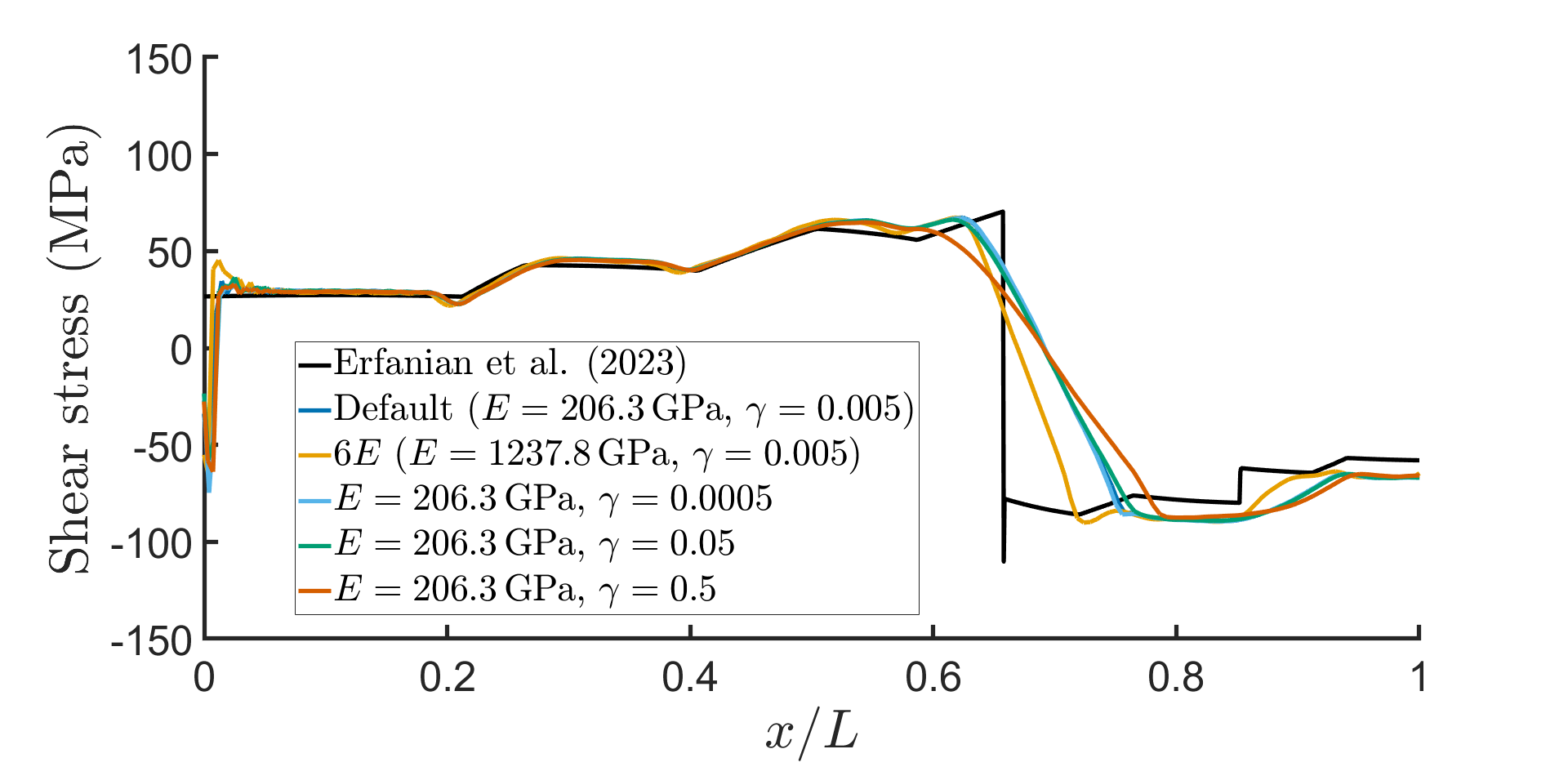}
\caption{Shear stress along the top surface of the deformed sheet from the analytical model by \citet{Mozhdeh_journ} and from simulations with different Young's modulus ($E$) and elastic slip tolerance ($\gamma$) values. The default values are $E=206.3\,\si{GPa}$ and $\gamma=0.005$. All simulations have radius $R=257.45\,\si{mm}$ and $\varepsilon=0.125$.} \label{fig:surface_shear}
\end{figure}

Coulomb friction is employed in our FE model, but frictional constraints are enforced with the stiffness (penalty) method.
The penalty method permits some relative surface motion (an “elastic slip”) when  sticking should occur (i.e., when the shear stress is below the critical magnitude for sliding). 
While the surfaces are ``sticking" the magnitude of sliding is limited to this elastic slip.
Elastic slip is controlled computationally by a slip tolerance factor $\gamma$.
In \figref{fig:surface_shear} we investigate the effect that varying $\gamma$ has on the surface shear stress results, and observe that the gradient in surface shear steepens with decreasing $\gamma$, but to a finite limiting value that is far from the infinite slope predicted analytically. 

\figref{fig:surface_shear} also shows that increasing the Young's modulus by a factor of six has a significant effect on the numerically-predicted shear stress on the sheet's surface: the gradient near the neutral point steepens and the position of the neutral point is closer to the position predicted analytically.
This is expected as these material parameters more closely resemble rigid-plastic behaviour.
This suggests that the friction rule alone is not responsible for the discrepancies between the FE and analytical results in \figref{fig:shear_FE_analytical}, and that the rigid-plastic (i.e., no elastic effects) assumption in the analytical model is also an important contributing factor.

\section{Conclusion}
\label{conclusion}
In this paper several FE models have been presented and compared to find an accurate FE model for rolling. 
FE rolling models in the existing literature are typically limited in their resolution of through-thickness variations.
This is important because through-thickness variation of stress and strain can have consequences for microstructure evolution and residual stress.
FE specifications such as element types, solver type (implicit versus explicit), and mesh density were investigated here to arrive at the final FE model.
This final FE model is a static, implicit model with $N_e=30$ CPE4R elements employed through the half-thickness of the sheet, significantly more than the $N_e=5\text{--}10$ typically used in common practice today.
These particular choices of element type, solver type and element size allow for a FE model that accurately predicts through-thickness variations in stresses and strains during steady cold rolling.
These through-thickness predictions are largely absent from the existing FE literature, probably due to the insufficient through-thickness discretisation used in previous studies.

We show that our through-thickness predictions compare well with a rigid-perfectly-plastic analytical model~\citep{Mozhdeh_journ}.
Excellent agreement in shear stress was found when comparing the analytical and FE models.
Both models captured the oscillatory nature of shear during rolling.
The relatively small discrepancies between the shear results of the two models can be attributed to the size of the roll-gap aspect ratio, $1/\varepsilon$ and the differences in the material and friction models, with the analytic model not allowing for a sticking region near the neutral point. 
In the limits as $\varepsilon \rightarrow 0$, $\gamma \rightarrow 0$ and $E \rightarrow \infty$, we hypothesise that the agreement between the FE and analytical results would be even further improved.

Without the ability to compare to an analytical model, previous studies have often relied on the convergence of roll force and roll torque to assess the quality of numerical rolling simulations.  Here, we show that roll force and roll torque are poor indicators of simulation quality. 
This is because these averaged quantities often converge well even for under-resolved simulations (see $N_e=5$ results in \figref{fig:roll_force_torque}) that fail to accurately predict other quantities of interest such as the through-thickness roll-gap shear distribution (see $N_e=5$ results in \figref{fig:shear_vertical}). 
Comparison to a steady-state analytical model does however require that time-domain FE simulations converge to a steady state. 
Here, instead of starting our simulations with the sheet in front of the rollers and pushing the sheet into the roll gap, as is common practice, we started with the sheet in the roll gap but with the rollers removed and then squeezed the rollers into the sheet in a preliminary bite step before rolling (as described in subsection \ref{subsec:general_fe_conditions}).
This was found to improve convergence to a steady state of rolling.
We confirm that a steady state has been reached via stagnant through-thickness stress and strain quantities, not via roll force and torque, as conducted by previous authors. Although the roll force and torque are necessarily constant during the steady state of the rolling process, constant roll force and torque are not sufficient for identifying steady-state behaviour on their own, as shown here.

Unexpectedly, our results show the existence of a region near the neutral point where the sheet and rollers are sticking, rather than slipping. 
This is not commonly described in previous studies, even in studies of extreme cases such as foil rolling where many different regions within the roll gap are detailed and analysed (e.g., \citet{fleck1992cold}, Figure 3c, zones E and F). 
The assumption that only slipping occurs at the neutral point is a common one, and indeed this assumption is the cause of the analytic model deviating from the FE model near the netural point.
While care has been taken here to ensure that the sticking region near the neutral point is not a numerical artefact, it is a consequence of Coulomb friction, which may be an overly-simplified model of the friction that practically occurs in metal rolling.  Comparison with the analytical model shows that the effects of this sticking region on the stress oscillation pattern through-thickness are important beyond the neutral point and affects the shear stresses at the roll-gap exit.

We also investigated the effect of the oscillatory pattern in shear stress on residual stress.
We found that a strong relationship exists between regions of large residual stress on the exit side of the roll gap and the $p\pm 2\theta Y/\sqrt{3}$ contour lines, whose direction coincides with the directions of maximum shear during plastic flow.
Since the oscillatory pattern is largely absent from the existing analytical and numerical literature, this pattern provides a promising route to curvature predictions in asymmetric rolling.

All of the analysis here, and many of the previous studies in Table~\ref{table:FE_settings}, assume plane strain to reduce the problem to 2D.
This is valid sufficiently far from the edge of the sheet provided the sheet is sufficiently wide and the rolls are perfectly aligned. 
However, in order to study edge effects, misalignment of the rolls, or less-wide sheets, a fully 3D study would be needed. 
A consequence of the analysis here is that $2N_e = 60$ elements through-thickness were needed to accurately resolve through-thickness behaviour, and this level of detail would be extremely computationally-expensive in a 3D simulation, especially since, to avoid elements with a large aspect ratio, a comparably large number of across-width elements would presumably also be needed. 
It may be possible for very narrow strips to instead assume a plane stress condition and again reduce the problem to a (different) 2D model, although this has not been investigated further here.

Future work could include employing second-order elements to discretise the metal sheet or incorporating the elastic deformation of the work roller.
Elastic roll deformation would be particularly important for foil rolling, where the roll-gap aspect ratio is very large and the through-thickness oscillatory pattern will repeat many times; the consequences of these oscillations in foil rolling are currently unknown. 
Extending the preliminary study in subsection \ref{asymmetric_rolling} of curvature induced by asymmetric rolling might allow for the prediction and control of curvature; the lack of through-thickness resolution may be what has been preventing previous FE studies from correctly predicting curvature from asymmetric rolling \citep{minton2017meta}. 
While in this study the FE software ABAQUS was used, we believe the principles described here are applicable to all FE studies, and this could be verified and best-practice refined by performing similar studies using other commonly-used software. 
Finally, given the detailed preparation of this particular FE model, we are confident that the final FE model described in this paper can help to guide the formulation of simpler, faster mathematical models with wave-like solutions.

\section*{Acknowledgements}
This publication has emanated from research conducted with the financial support of Science Foundation Ireland under grant number 18/CRT/6049. For the purpose of open access, the authors have applied a CC BY public copyright licence to any Author Accepted Manuscript version arising from this submission.

ANOC is supported by the European Union, Science Foundation of Ireland and Lero, the Science Foundation Ireland Research Centre for Software, grants \#101028291 \#13/RC/2094 and \#SOWP2-TP0023, respectively.
ME gratefully acknowledges the funding of a University of Warwick Chancellor’s Scholarship. 
EJB is grateful for the UKRI Future Leaders’ Fellowship funding (MR/V02261X/1) supporting this work.
DOK is grateful for funding from Science Foundation Ireland (21/FFP-P/10160).

\appendix
\section{ABAQUS formulation}
\label{app:abaqus_formulation_app}
We present here some relevant background information on the FE modelling technique for rolling.
The metal rolling process is governed by differential equations describing yield conditions, constitutive relationships and conservation of mass and momentum.
In FE analysis, the differential equations are approximately solved between a discrete set of nodes, using shape functions that assume a particular interpolation field between the nodes.
This approach discretises spatial derivatives, so that the partial differential equations are transformed into a system of ordinary differential equations. The simplified form of the equations of motion are written as 
\begin{equation} \label{eq:FE governing equation}
\mathbf{M}\ddot{\mathbf{u}} = \mathbf{f}^\text{ext} - \mathbf{f}^\text{int},
\end{equation}
where $\mathbf{M}$ is the diagonal mass matrix, $\mathbf{u}$ is the nodal displacement vector, and $\mathbf{f}^\text{ext}$ and $\mathbf{f}^\text{int}$ are the external and internal nodal force vectors respectively~\citep{hadadian2019investigation}.
In metal rolling, inertia effects are negligible and are therefore often ignored \citep{tarnovskii2013deformation,cawthorn,minton2017mathematical}.
The rolling process can be idealised by assuming the stresses in the sheet are in quasi-static equilibrium.
The implicit and explicit solvers offered by ABAQUS are both useful for obtaining solutions to FE problems concerning quasi-static processes.
We now give a brief overview of the formulations of ABAQUS/Standard and ABAQUS/Explicit.

\subsection{ABAQUS/Standard formulation}
\label{abaqus_standard}
ABAQUS/Standard is an implicit solver where the model is updated from time $t$ to time $t+\Delta t$ using information from times $t$ and $t+\Delta t$.
In this work, only static analyses are considered in ABAQUS/Standard.
Therefore, since inertia is ignored, the relevant equations of motion are
\begin{equation} \label{eq:implicit governing equation}
\mathbf{F}(\mathbf{u}) = \mathbf{f}^\text{ext} - \mathbf{f}^\text{int} = \mathbf{0}.
\end{equation}
This large system of equations is solved iteratively in each time increment to obtain the updated nodal displacements, using the Newton–Raphson technique.
The displacements are updated via
\begin{subequations}\begin{align} \label{eq:implicit 1}
{\Delta \mathbf{u}_{\left(n+1\right)}}^{\left(t+\Delta t\right)} &= \left[{\mathbf{K}_{\left(n\right)}}^{\left(t+\Delta t\right)}\right]^{-1} \cdot \left({\mathbf{F}_{\left(n\right)}}^{\left(t+\Delta t\right)}\right),
\\\label{eq:implicit 2}
{\mathbf{u}_{\left(n+1\right)}}^{\left(t+\Delta t\right)} &= {{\mathbf{u}}_{(n)}}^{\left(t+\Delta t\right)} + \Delta {\mathbf{u}_{\left(n+1\right)}}^{\left(t+\Delta t\right)},
\end{align}\end{subequations}
where the superscript $\left(t+\Delta t\right)$ is the time at which the equations are evaluated, the subscripts $\left(n\right)$ and $\left(n+1\right)$ indicate the current and next iterations respectively and $\mathbf{K}_{\left(n\right)} = \left(\partial \mathbf{F} / \partial \mathbf{u}\right)_{\left(n\right)}$ is the stiffness matrix at iteration $n$.
The inversion of the stiffness matrix $\mathbf{K}_{\left(n\right)}$ is computationally expensive.
The convergence of the Newton–Raphson method is measured by ensuring that all entries 
$\mathbf{F}(\mathbf{u})$ and $\Delta \mathbf{u}$ are sufficiently small~\citep{jiang2002elastic}.
At the end of time increment $t$, the next time increment $t+\Delta t$ is attempted, and load or displacement is incremented corresponding to $\Delta t$.
If convergence is not achieved by ABAQUS in a certain time increment, $\Delta t$ is reduced and the Newton-Raphson method is attempted again.
The reader should note that no velocities or accelerations are given by ABAQUS/Standard for static analyses.
These can be manually extracted via postprocessing of the displacements (see~\ref{vel_app} for more details).

\subsection{ABAQUS/Explicit formulation}
\label{abaqus_explicit}
In ABAQUS/Explicit, the equations are solved for future time steps using known information, making it an explicit scheme.
This dynamic solution scheme necessitates the full version of equation~\eqref{eq:FE governing equation} to be solved.
The diagonal mass matrix is inverted to find the vector of nodal accelerations,
\begin{equation} \label{eq:equilibrium timed}
\ddot{\mathbf{u}}^{(i)} = \mathbf{M}^{-1}\left(\mathbf{f}^\text{ext} - \mathbf{f}^\text{int}\right),
\end{equation}
at the beginning of time increment number $i$.
The inversion of the mass matrix is numerically much simpler than inverting the stiffness matrix~\citep{web}.
The acceleration vector $\ddot{\mathbf{u}}^{(i)}$ is assumed to remain constant for a very small time increment and the explicit central-difference integration rule~\citep{mesh_sens} is used to update the mid-increment velocities and subsequently the end-of-increment displacements,
\begin{subequations}\begin{align} \label{eq:explicit 1}
\dot{\mathbf{u}}^{\left(i+\frac{1}{2}\right)} &= \dot{\mathbf{u}}^{\left(i-\frac{1}{2}\right)} + \frac{\Delta t^{\left(i+1\right)} + \Delta t^{\left(i\right)}}{2} \ddot{\mathbf{u}}^{(i)},
\\\label{eq:explicit 2}
{\mathbf{u}}^{\left(i+1\right)} &= {\mathbf{u}}^{\left(i\right)} + \Delta t^{\left(i+1\right)}\dot{\mathbf{u}}^{\left(i+\frac{1}{2}\right)}.
\end{align}\end{subequations}
The explicit dynamics procedure is ideally suited for analysing high-speed dynamic events, but many of the advantages of the explicit procedure also apply to the analysis of slower (quasi-static) processes.
Quasi-static analyses typically have longer timescales which increase the numerical computation time required by FE methods.
The stable time increment for ABAQUS/Explicit simulations is the largest possible time step the solver can take without results becoming unstable.
When the solution becomes unstable, the time-history response of solution variables such as displacements will usually oscillate with increasing amplitudes~\citep{abaqus}.
This stability limit is also known as the Courant–Friedrich–Lewy (CFL) limit~\citep{padhye2023mechanics}, and is estimated as
\begin{equation} \label{eq:stable time inc}
\Delta t_{\text{stable}} \approx \frac{L^e \sqrt{\rho}}{\sqrt{E}},
\end{equation}
where $L^e$ is the smallest element dimension in the FE model, $\rho$ is the material density and $E$ is the Young's modulus of the material.
However, ABAQUS/Explicit enables the use of mass scaling.
Mass scaling uniformly scales the density of the material within the simulation to facilitate longer time steps and reduce computational time~\citep{minton2017mathematical}. 
The limited time increment means that explicit methods are conditionally stable, thus the mass scaling factor (MSF) must be chosen carefully to reduce simulation times without introducing inertial effects. 
The introduction of inertial effects can lead to erroneous and unstable outputs~\citep{erroneous,unstable,hadadian2019investigation}. 
The general rule of thumb to assess inertial effects is to ensure that the ratio between the kinetic energy and the internal energy is small (5--10\%)~\citep{web,abaqus}.

\section{FE velocity calculation}
\label{vel_app}
ABAQUS/Standard does not output velocity components or PEEQ rate values for static analyses. 
Therefore, we manually calculate the velocity and PEEQ rate values.
The velocity and PEEQ rate approximations are conducted via a Lagrangian time derivative of the displacements and the PEEQ values respectively.
Since the procedure is identical for calculating both velocity and PEEQ rate values, we give details of the velocity calculation only.

All stress and strain quantities are outputted to an output database (\textit{.ODB}) file for all ABAQUS simulations.
These quantities are available at different time points throughout the simulation, and it is possible for the user to vary the number and frequency of the time points throughout each analysis step.
To approximate the velocity components $\left(v_x,v_z\right)$ of some arbitrary node in the sheet, at time frame $f$, we use a fourth-order-accurate central differencing operator on the displacements
\begin{subequations}\label{eq:velocity calculation}\begin{align}
        {v}^{(f)}_x&= \frac{\frac{1}{12}{u}^{(f-2)}_x - \frac{2}{3}{u}^{(f-1)}_x + \frac{2}{3}{u}^{(f+1)}_x - \frac{1}{12}{u}^{(f+2)}_x}{\Delta t},
\\
        {v}^{(f)}_z&= \frac{\frac{1}{12}{u}^{(f-2)}_z - \frac{2}{3}{u}^{(f-1)}_z + \frac{2}{3}{u}^{(f+1)}_z - \frac{1}{12}
        {u}^{(f+2)}_z}{\Delta t},
\end{align}\end{subequations}
where $\Delta t$ is constant between all time frames considered. 
In order to allow velocities and all other stress and strain components to be evaluated at the same time frame, all data in the main text is evaluated at a time point that is two time frames before $t=0.1\,\si{s}$, which is the end of the \textit{rolling} step.
The authors ensured that $\Delta t$ is sufficiently small for accurate velocity approximations.

\section{Shear locking and incompatible CPE4I elements}
\label{CPE4I_app}
\begin{figure}
\centering
   \centering
   \includegraphics[width=0.8\linewidth]{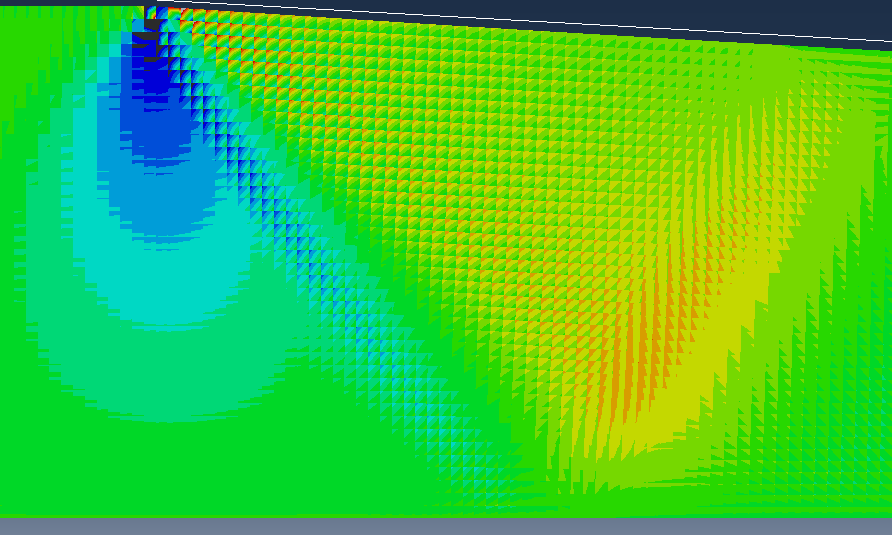}
\caption{Shear stress evaluated at the integration points (i.e., no nodal averaging) for a rolling simulation using $N_e=40$ CPE4 elements through the half-thickness.
There are significant jumps in shear stress across various elements, creating a checkerboard pattern, which is typical of shear-locking behaviour.} \label{fig:CPE4_shear_locking}
\end{figure}

\figref{fig:CPE4_shear_locking} shows shear stress evaluated at the integration points of CPE4 elements during a rolling simulation conducted as outlined in the main text with $N_e=40$ CPE4 elements through the half-thickness of the sheet.
In \figref{fig:CPE4_shear_locking} a clear checkerboard pattern is observed, with more extreme shear stress values observed at the corners of elements. 
This is typical of shear locking behaviour~\citep{abaqus}.

\begin{figure}%
\centering%
   \includegraphics[width=1\linewidth]{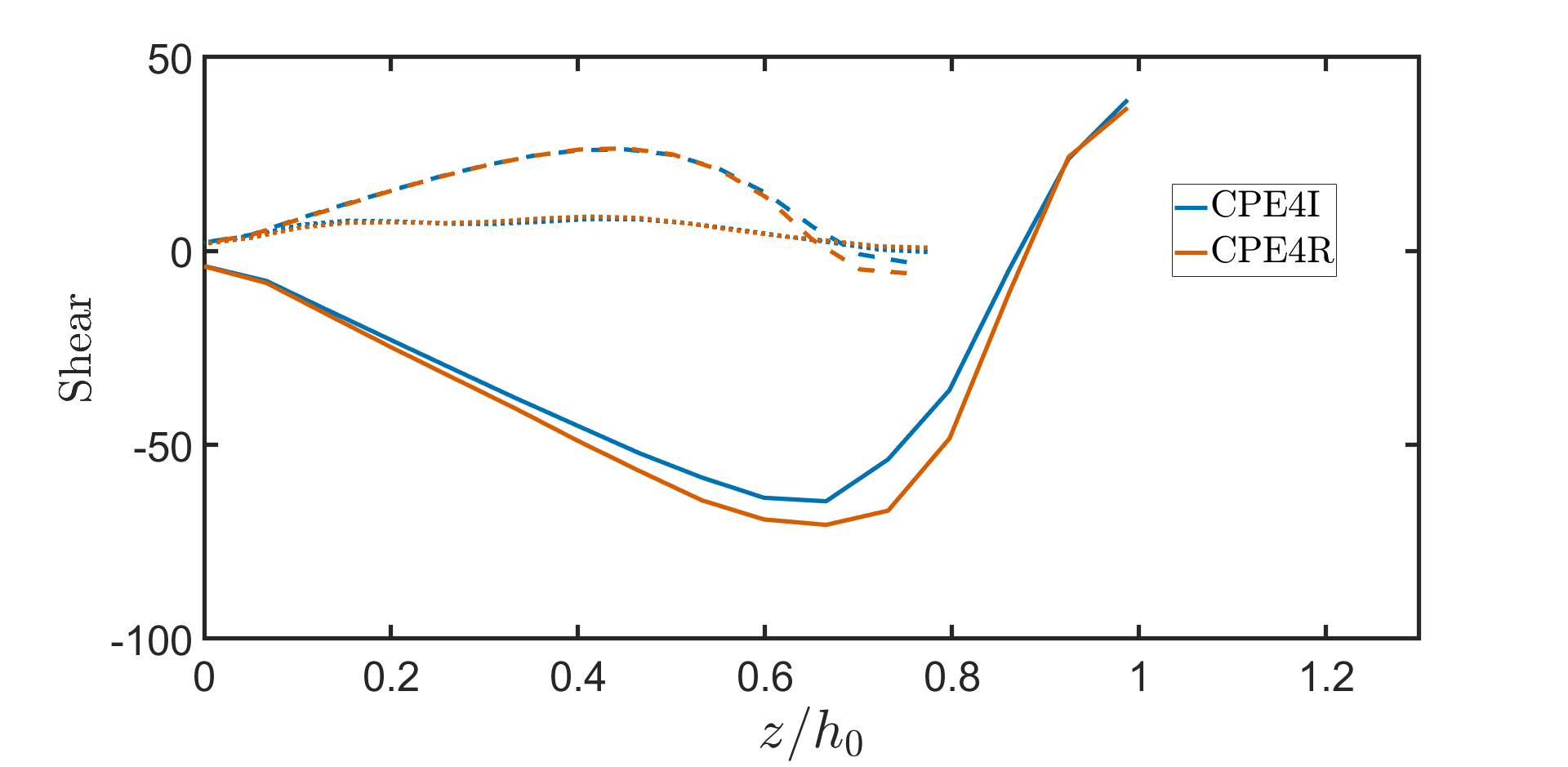}%
\caption{Interpolated results for shear stress ($\si{MPa}$) from simulations with CPE4I elements and CPE4R elements. 
Both simulations employ $N_e=15$ elements, a radius $R=257.45\,\si{mm}$ and $\varepsilon=0.125$. 
The results are almost identical and the CPU time is 17.7 hours for the CPE4I simulation compared to 3.2 hours for the CPE4R simulation.
Readers should refer to \figref{fig:interp_positions} which identifies the horizontal position for the solid, dotted and dashed lines, respectively. 
}\label{fig:CPE4I_CPE4R}%
\end{figure}

Simulations with CPE4I and CPE4R elements employed do not exhibit such shear-locking behaviour. 
\figref{fig:CPE4I_CPE4R} compares interpolated shear stress from two simulations: one with CPE4I elements and the other with CPE4R elements. 
Both simulations employ $N_e=15$ elements through the half-thickness, since simulations with $N_e\geq20$ CPE4I elements required extremely small time increments which gave rise to convergence issues.
The maximum relative error in \figref{fig:CPE4I_CPE4R} is approximately 9\% when compared to the maximum absolute shear stress in the roll gap.
We expect that this error would reduce significantly as the mesh density increases.
Given the expected similar shear stresses with increasing mesh density and that the simulation with CPE4I elements requires 17.7 CPU hours compared to 3.2 CPU hours for the simulation with CPE4R elements, CPE4I elements were not used for the main results in this work.

\section{Relative error: choice of \texorpdfstring{$x/L$}{x/L} locations}
\label{relative_error_appendix}
\begin{figure}
\centering
\begin{subfigure}[b]{\linewidth}
   \centering
   \includegraphics[width=\linewidth,height=0.25\textheight,keepaspectratio]{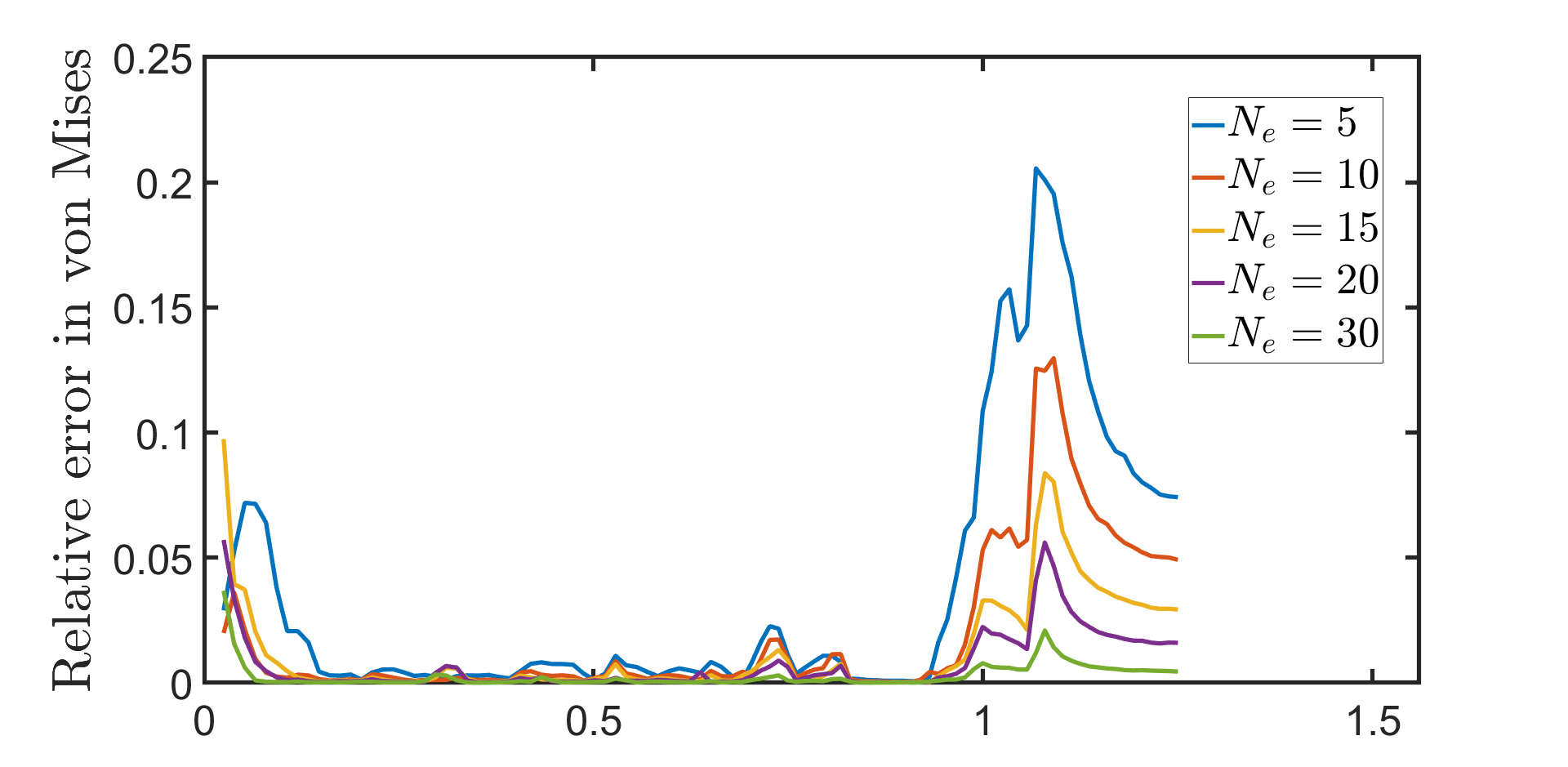}
   \caption{}
   \label{fig:relative_error_vM} 
\end{subfigure}
\begin{subfigure}[b]{\linewidth}
   \centering
   \includegraphics[width=\linewidth,height=0.25\textheight,keepaspectratio]{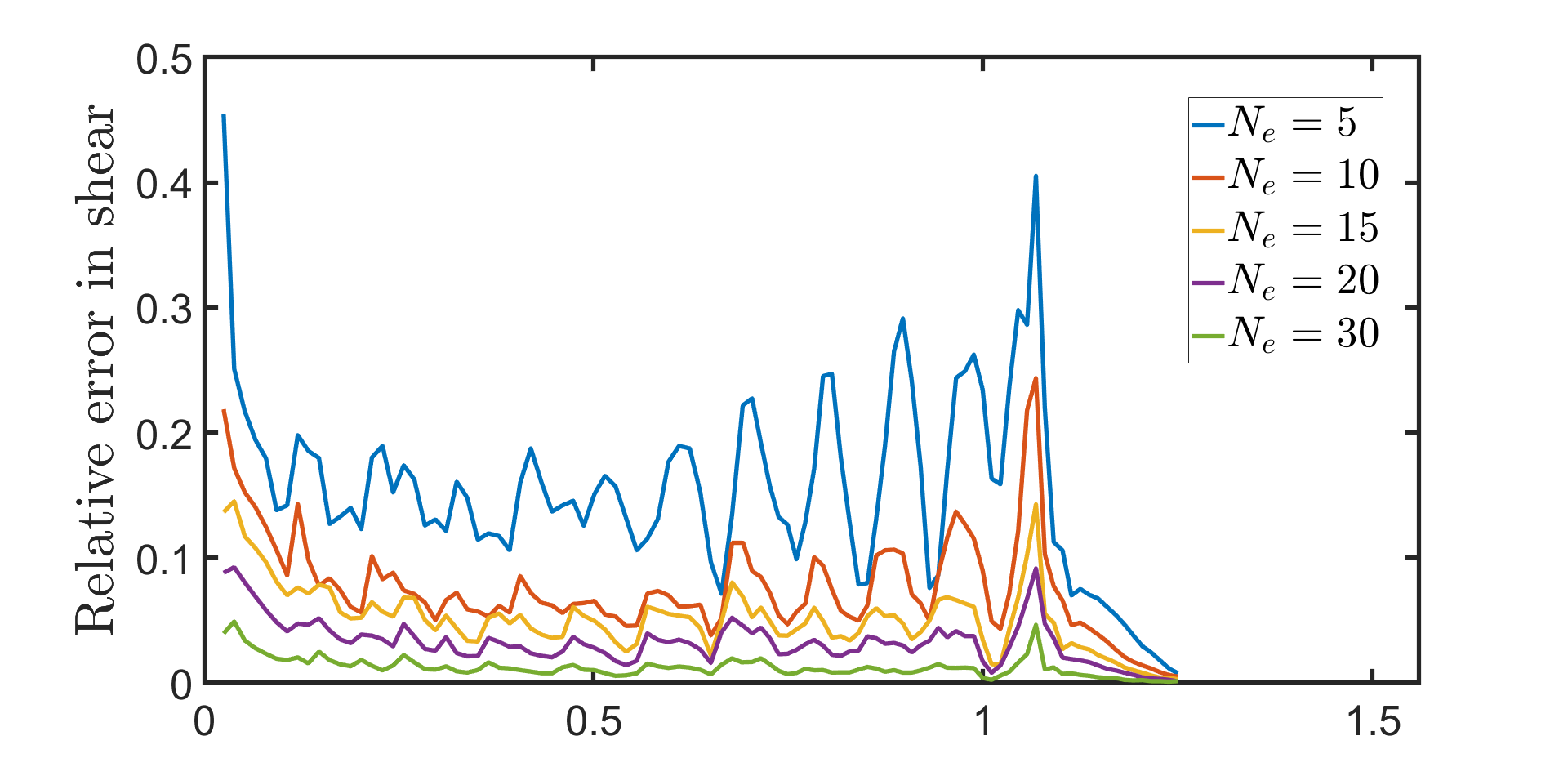}
   \caption{}
   \label{fig:relative_error_shear} 
\end{subfigure}
\begin{subfigure}[b]{\linewidth}
   \centering
   \includegraphics[width=\linewidth,height=0.25\textheight,keepaspectratio]{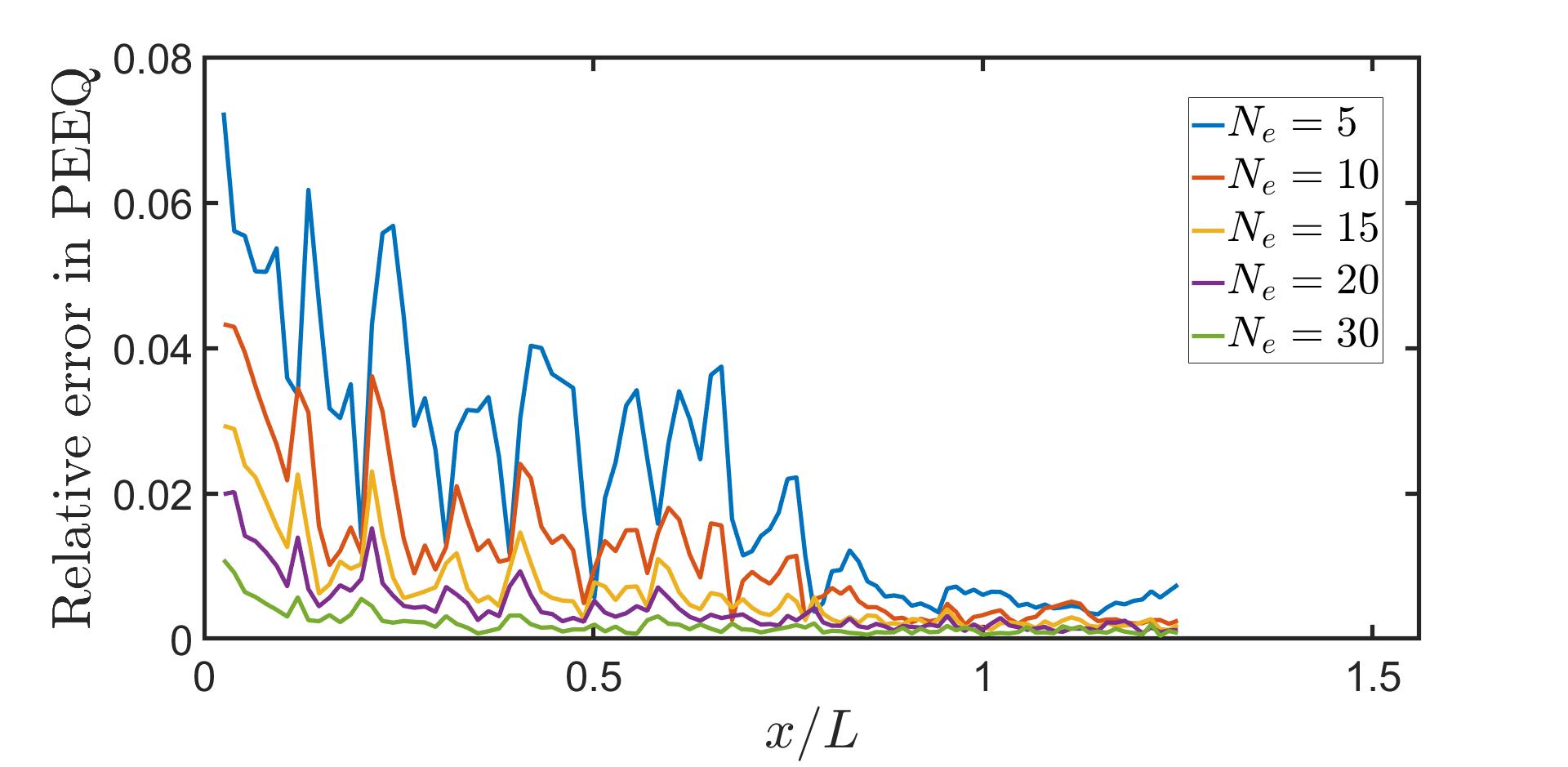}
   \caption{}
   \label{fig:relative_error_PEEQ}
\end{subfigure}
\caption{Relative error in (a) von Mises stress, (b) shear stress and (c) PEEQ (plastic equivalent strain) from simulations with varying mesh densities.
All simulations are compared to a simulation with $N_e=40$ elements through the half-thickness.
The error value at each $x/L$ location represents the maximum error through the thickness of the sheet at that $x/L$ position.
All simulations employ a radius $R=257.45\,\si{mm}$ and $\varepsilon=0.125$.} \label{fig:relative_errors}
\end{figure}

We detail here how the three $x/L$ positions used in the interpolated results throughout this paper were chosen.
These $x/L$ positions represent the positions of maximum relative errors.
To explain how these errors are calculated, let us consider the solid-line shear results in Figure~\ref{fig:shear_vertical}.
The interpolated shear result from the simulation with $N_e=5$ is subsequently interpolated onto a curve with 41 $z/h_0$ points.
The absolute difference between this new curve and the interpolated shear result from the simulation with $N_e=40$ is calculated at each $z/h_0$ point.
Finally, the absolute difference at each point is divided by the maximum absolute shear stress in the entire roll gap, as predicted by the simulation with $N_e=40$.
At each $x/L$ location, the maximum relative error value out of the 41 values is stored in an array.
The same procedure is carried out for $N_e=10$, $N_e=15$, $N_e=20$ and $N_e=30$, and all of these maximum errors are plotted in \figref{fig:relative_errors}.
We considered 100 $x/L$ positions from $x/L=0.025$ up to $x/L=1.25$.
Note that $x/L<0.025$ was not considered. The contact point occurs in the neighbourhood of $x/L=0$ and corresponds to a change from zero to large normal stress, so a small change in the predicted position of the contact point can lead to misleadingly large errors recorded at $x/L=0$.

\figref{fig:relative_errors} shows that $x/L=0.025$ is a position where high relative errors occur, and hence this is one of the $x/L$ positions used for interpolations in this paper.
\figref{fig:relative_error_vM} and \figref{fig:relative_error_shear} also show a spike in relative errors near $x/L=1.09$. This is another $x/L$ position used to calculate the interpolated results.
It is not clear from \figref{fig:relative_error_shear} and \figref{fig:relative_error_PEEQ} which third $x/L$ position to choose. However, \figref{fig:relative_error_vM} shows a small peak in relative error around the neutral point at $x/L = 0.6913$ (the neutral point value of $x/L=0.6913$ is predicted by the implicit simulation with $N_e=40$).
Given this peak and the interesting dynamics around the neutral point in general, $x/L=0.6913$ is the third position used in the interpolated results throughout this paper.

\section{Slip line theory}
\label{sliplines_app}
This theory is only applicable to rigid-perfectly-plastic materials.
In two-dimensional plane-strain analyses, the stress tensor takes the form
\begin{equation}
  \begin{aligned}
    \boldsymbol{\sigma} = \boldsymbol{S} - p\boldsymbol{I}
    &= \setlength{\arraycolsep}{0.5em}\begin{pmatrix}
        S_{xx} &  S_{xz} \\
        S_{xz} &  S_{zz}
    \end{pmatrix} + \setlength{\arraycolsep}{0.5em}\begin{pmatrix}
        -p & 0 \\
        0 & -p
        \end{pmatrix}\\
    &= \setlength{\arraycolsep}{0.5em}\begin{pmatrix}
        S_{xx} - p &  S_{xz} \\
        S_{xz} &  S_{zz} - p
    \end{pmatrix},
  \end{aligned}
\end{equation}
where $\boldsymbol{S}$ is the deviatoric part of the stress tensor and $-p=\left(\sigma_{xx}+\sigma_{zz}\right)/2$ is the two-dimensional hydrostatic pressure~\citep{howell2009applied}.
Under rotation of the coordinate system by $\theta$ degrees from the positive $x$-axis in the anticlockwise direction, the stress state $\boldsymbol{\sigma}$ may be rewritten as 
\begin{equation} \label{eq:rotated stress tensor}
  \begin{aligned}
    \boldsymbol{\sigma}' = \boldsymbol{R} \boldsymbol{\sigma} \boldsymbol{R}^T &= \setlength{\arraycolsep}{0.5em}\begin{pmatrix}
    \cos{\theta} &  \sin{\theta} \\
    -\sin{\theta} &  \cos{\theta}
    \end{pmatrix} \setlength{\arraycolsep}{0.5em}\begin{pmatrix}
        S_{xx}-p &  S_{xz} \\
        S_{xz} &  S_{zz}-p
        \end{pmatrix} \setlength{\arraycolsep}{0.5em}\begin{pmatrix}
    \cos{\theta} &  -\sin{\theta} \\
    \sin{\theta} &  \cos{\theta}
    \end{pmatrix}\\
    &= \begin{pmatrix}
        S_{xx}\cos{2\theta} + S_{xz}\sin{2\theta}- p &   S_{xz}\cos{2\theta} - S_{xx}\sin{2\theta} \\
        S_{xz}\cos{2\theta} - S_{xx}\sin{2\theta}&   -S_{xx}\cos{2\theta} - S_{xz}\sin{2\theta}- p
    \end{pmatrix},
  \end{aligned}
\end{equation}
where both $\boldsymbol{\sigma}$ and $\boldsymbol{\sigma}'$ represent the same stress state in different coordinate systems, and $\boldsymbol{R}$ is a rotation matrix \citep{huang2023critical}. 
It is possible to write the von Mises yield criterion in terms of the deviatoric stress components as $S_{ij}S_{ij} = 2Y^2/3$, where $Y$ is the yield stress in tension or compression of the material.
For some rotation angle $\theta$, the stress state $\boldsymbol{\sigma}'$ is in a state of pure shear,
\begin{equation} \label{eq:pure shear stress tensor}
  \begin{alignedat}{2}
    \boldsymbol{\sigma}' & = \renewcommand{\arraystretch}{1.5}\setlength{\arraycolsep}{0.5em}\begin{pmatrix}
        -p &   \frac{Y}{\sqrt{3}} \\
        \frac{Y}{\sqrt{3}} &   -p
        \end{pmatrix}, \\
  \end{alignedat}
\end{equation}
and the stress state $\boldsymbol{\sigma}$ in the original coordinate system can be written as
\begin{equation} \label{eq:pure shear original stress tensor}
  \begin{alignedat}{2}
    \boldsymbol{\sigma} & = \renewcommand{\arraystretch}{1.5}\setlength{\arraycolsep}{0.5em}\begin{pmatrix}
        p - \frac{Y}{\sqrt{3}} \sin{2\theta}&   \frac{Y}{\sqrt{3}}\cos{2\theta} \\
        \frac{Y}{\sqrt{3}}\cos{2\theta} &   p + \frac{Y}{\sqrt{3}} \sin{2\theta}
        \end{pmatrix}.
  \end{alignedat}
\end{equation}
We can use this definition for $\boldsymbol{\sigma}$ in the two-dimensional stress equilibrium equations to write 
\begin{subequations}\begin{align} \label{eq:stress 1}
0 = \pdv{\sigma_{xx}}{x} + \pdv{\sigma_{xz}}{z} &= -\pdv{p}{x} - \frac{2Y}{\sqrt{3}}\!\left(\!\cos{2\theta} \pdv{\theta}{x} +  \sin{2\theta} \pdv{\theta}{z}\right)\!,
\\\label{eq:stress 2}
0 = \pdv{\sigma_{xz}}{x} + \pdv{\sigma_{zz}}{z} &= -\pdv{p}{z} - \frac{2Y}{\sqrt{3}}\!\left(\!\sin{2\theta} \pdv{\theta}{x} -  \cos{2\theta} \pdv{\theta}{z}\right)\!.
\end{align}\end{subequations}
Equations \ref{eq:stress 1} and \ref{eq:stress 2} are hyperbolic PDEs and can be solved by the method of characteristics, (see \citet{johnson2013plane} for a detailed description).
The characteristic directions of each of these equations are given by
\label{eq:characteristic direction}\begin{align}
\dfrac{dz}{dx} &= \tan{\theta} &
&\text{and}&
\dfrac{dz}{dx} &= \tan{\left(\theta +\frac{\pi}{2}\right)}
\end{align}
respectively, and it is clear that the families of characteristics form an orthogonal network.
The family of characteristics given by the $\theta$ parameter are called the $\alpha$-lines and those given by the $\theta+\pi/2$ parameter are called $\beta$-lines.
Both sets of lines coincide with the lines of maximum shear stress.
It can be shown that $p+ 2\theta Y/\sqrt{3}$ is constant along the $\alpha$-lines and $p- 2\theta Y/\sqrt{3}$ is constant along the $\beta$-lines.
However, these quantities are only constant for material that has yielded plastically \citep{johnson2013plane}.
Similarly, the lines will not be orthogonal in regions where the material is not deforming plastically.

\bibliography{IJMS}

\end{document}